\documentclass{aa}  

\usepackage{multirow} 
\usepackage{graphicx}
\usepackage{txfonts}
\usepackage{lipsum}
\usepackage{subcaption}         
\usepackage{lscape}             
\usepackage{placeins}           
\usepackage{xcolor}

\usepackage{hyperref}
\hypersetup{
	colorlinks=true,
	linkcolor={red!50!black},
	citecolor={blue!70!black},
	urlcolor={blue!80!black}
}

\begin{document}

   \title{Exceptional Gamma-Ray Flaring Activity of the Blazar S4~0954+65 in Early 2025}
   
%

   \author{
             Zhihao Ouyang\inst{1}
        \and Jingyu Wu\inst{1}
        \and Hubing Xiao\inst{1}\fnmsep\thanks{Corresponding author: Hubing Xiao (hubing.xiao@shnu.edu.cn)}
        \and Lili Yang\inst{2}\fnmsep\thanks{Corresponding author: Lili Yang (yanglli5@sysu.edu.cn)}
        \and Shangchun Xie\inst{1}
        \and Shaohua Zhang\inst{1}
        \and Zhijian Luo\inst{1}
        \and Jianzhen Chen\inst{1}
        \and Jianguo Wang\inst{3}
        \and Mingjun Liu \inst{4}
        \and Qinyu Wu \inst{4, 5}
        \and Huaqing Cheng \inst{4}
        \and Wenda Zhang \inst{4, 5}
        \and Junhui Fan\inst{6, 7, 8}
        }

   \institute{
             Shanghai Key Lab for Astrophysics, Shanghai Normal University, Shanghai 200234, People's Republic of China
        \and School of Physics and Astronomy, Sun Yat-sen University, Zhuhai 519082, People’s Republic of China
        \and Yunnan Observatories, Kunming, 650216, People's Republic of China
        \and National Astronomical Observatories, Chinese Academy of Sciences, 20A Datun Road, Beijing 100101, People's Republic of China
        \and School of Astronomy and Space Science, University of Chinese Academy of Sciences, 19A Yuquan Road, Beijing 100049, People's Republic of China
        \and Center for Astrophysics, Guangzhou University, Guangzhou 510006, People's Republic of China
        \and Key Laboratory for Astronomical Observation and Technology of Guangzhou, Guangzhou 510006, People's Republic of China
        \and Astronomy Science and Technology Research Laboratory of Department of Education of Guangdong Province, Guangzhou 510006, People's Republic of China
    }

   \date{Received September 30, 20XX}

 
  \abstract
    {
    S4~0954+65 (4FGL~J0958.7+6534) is a TeV-detected blazar at a redshift of $z = 0.3694 \pm 0.0011$, classified as an intermediate-synchrotron-peaked BL Lac object.
    In early 2025, it entered an exceptional $\gamma$-ray high state, exceeding the flux of the 2015 outburst, during which contemporaneous MAGIC TeV observations were obtained.
    This event offers an opportunity to investigate the radiation mechanisms, particle acceleration, and jet dynamics.
    }
    {
    We aim to investigate the origin and physical properties of the exceptional 2025 flare, and to constrain the emission processes responsible for this flaring activity.
    }
    {
    We performed a multi-wavelength analysis using $\gamma$-ray, X-ray, optical/UV, radio, and 43~GHz Very Long Baseline Array (VLBA) observations.
    We examined multi-band correlations with the z-transformed discrete correlation function (zDCF), analyzed the spectral evolution and parsec-scale jet kinematics, and modeled the broadband spectral energy distributions (SEDs).
    }
    {
    The $\gamma$-ray variations lead the optical and radio emission by 3.22~days and possibly 18.26~days, respectively, while no significant correlation is found between the $\gamma$-ray and X-ray emission.
    The $\gamma$-ray spectra show a harder-when-brighter behavior, consistent with enhanced particle acceleration during the active state.
    A similar spectral trend is also observed in the X-ray band, where the photon index is anti-correlated with flux.
    During the major $\gamma$-ray flare, VLBA images reveal the emergence of a new superluminal radio knot from the core, whose extrapolated ejection time is consistent with the peak of the $\gamma$-ray flare.
    }
    {
    The temporal and structural evolution during the 2025 flare is consistent with a shock-in-jet scenario, in which a newly emerging disturbance propagates downstream and drives the flare.
    Broadband SED modeling shows that a leptonic synchrotron self-Compton plus external Compton model with dusty torus seed photons reproduces the multi-wavelength observations with physically plausible parameters.
    A lepto-hadronic interpretation remains possible but requires substantially higher jet power.
    }


   \maketitle
\nolinenumbers

\section{Introduction}

Blazars constitute the most extreme subclass of active galactic nuclei (AGNs), characterized by relativistic jets oriented close to the line of sight, which leads to strong Doppler boosting and rapid multi-wavelength variability \citep[e.g.,][]{Urry1995PASP107, Padovani2017A&ARv25, Blandford2019ARA&A57}.
Notably, many of them exhibit extreme flares in which their electromagnetic flux can rise by several orders of magnitude over periods ranging from minutes to a few days \citep[e.g.,][]{Aharonian2007ApJ664, Albert2007ApJ669, Abdo2011ApJ733, Ackermann2016ApJ824, Wang2024ApJS270, Kouch2026A&A708_A382}.
They are commonly divided into BL Lacertae objects (BL Lacs) and flat spectrum radio quasars (FSRQs), according to the equivalent width of their optical emission lines \citep{Stickel1991ApJ374, Urry1995PASP107, Scarpa1997A&A325}. 
Their broadband spectral energy distributions (SEDs) typically show two broad components: a low-energy hump attributed to synchrotron radiation from relativistic electrons in the jet, and a high-energy hump whose origin remains under debate. 
In leptonic scenarios, the high-energy emission is produced by inverse Compton scattering of synchrotron or external photons by relativistic electrons, while hadronic models invoke interactions involving relativistic protons, such as proton synchrotron radiation or cascades initiated by photo-hadronic processes \citep[e.g.,][]{Tavecchio1998ApJ509, Ghisellini2009MNRAS397, Mucke2003AP18, Bottcher2009ApJ703, Gao2019NatureAstr3, Cerruti2019MNRAS483, Xue2019ApJ871, Xue2022PRD106, Xue2025EPJC85, Ouyang2025ApJ980,  Zhao-Xie2026ChPhL43, Xiao2026ApJ1005}. 
Although leptonic models successfully reproduce many observed SEDs with relatively modest jet powers, hadronic scenarios are attractive because they can link blazars to ultra-high-energy cosmic rays and neutrinos \citep[e.g.,][]{Gao2019NatureAstr3, Cerruti2019MNRAS483}. 
Distinguishing between these two classes of models remains one of the challenges in high-energy astrophysics and requires detailed, time-resolved, multi-wavelength observations, especially during major flaring events.

S4~0954+65 is a well-known object that was originally identified as a radio source, with its optical counterpart reported by \citet{Cohen1977MmRAS84}. 
It was classified as a BL Lac object by \citet{Walsh1984MNRAS211}, and was later detected in the $\gamma$-ray band by the Energetic Gamma Ray Experiment Telescope (EGRET) \citep{Mukherjee1995ApJ445}. 
Recent optical spectroscopy gives a redshift of $z=0.3694\pm0.0011$ \citep{BecerraGonzalez2021MNRAS504}. 
Although S4~0954+65 is usually classified as an intermediate-synchrotron-peaked BL Lac (IBL) object, its weak but detectable Mg~II emission line and inferred disk luminosity suggest that it may represent a transitional object between BL Lac objects and FSRQs \citep{Ghisellini2011MNRAS414, BecerraGonzalez2021MNRAS504}. 
S4~0954+65 also belongs to the relatively small group of IBLs detected at very-high-energy (VHE; $E>100$~GeV) $\gamma$-rays. 
The source was first detected at VHE by the Major Atmospheric Gamma Imaging Cherenkov Telescopes (MAGIC) during an exceptionally bright optical state in early 2015 \citep{Mirzoyan2015TheAstrono6999, MAGIC2018A&A617}. 
This flare was accompanied by strong activity across multiple wavelengths and the emergence of a new radio jet component, pointing to a connection between high-energy emission and parsec-scale jet dynamics \citep{Tanaka2016PASJ68, MAGIC2018A&A617}. 
Subsequent multi-wavelength studies have revealed extreme variability on timescales from minutes to years, large polarization rotations, and complex spectral behavior, making S4~0954+65 an excellent laboratory for studying particle acceleration and radiation processes in relativistic jets \citep{Pandey2023A&A679}.
More recently, Imaging X-ray Polarimetry Explorer (IXPE) observations have provided constraints on the X-ray polarization of S4~0954+65 during an optical--X-ray flare, offering an additional diagnostic for distinguishing between leptonic and hadronic emission scenarios \citep{Kouch2025A&A695}.

In early 2025, S4~0954+65 entered an exceptionally bright high state, with its $\gamma$-ray flux reaching a record level \citep{Bartolini2025ATel17042, Bernard2025ATel16994}. 
The source became brighter than during the 2015 outburst \citep{Ojha2015ATel7093, MAGIC2018A&A617}, and also surpassed the major activity reported in May 2022 \citep{Rani2022ATel15375, Raiteri2023MNRAS526}. 
This event, therefore, represents one of the strongest outbursts ever recorded from S4~0954+65. 
It is particularly important because it was accompanied by multi-wavelength coverage, including $\gamma$-ray, X-ray, optical/UV, and radio observations, together with the monitoring of the parsec-scale jet using very long baseline interferometry (VLBI). 
Such combined observations provide an opportunity to connect high-energy variability with the structural evolution of the inner jet.

In this work, we present a comprehensive multi-wavelength study of the 2025 flaring activity of S4~0954+65. 
We analyze the temporal variability from $\gamma$-ray to radio bands, investigate the cross-correlations between different bands, examine the $\gamma$-ray spectral evolution, and study the parsec-scale jet kinematics using 43~GHz Very Long Baseline Array (VLBA) observations. 
We further construct quasi-simultaneous broadband SEDs and model them in both leptonic and lepto-hadronic frameworks, aiming to constrain the radiation mechanism and the physical connection between the flare and the jet evolution. 
Throughout this work, we adopt the $\Lambda$CDM cosmological parameters from \citet{Planck2020A&A641}, with $H_{0} = 67.7$~km~s$^{-1}$~Mpc$^{-1}$ and $\Omega_{\rm m} = 0.310$.

\section{Observations and Data Analysis}

\subsection{Gamma-ray observations}

We performed a $\gamma$-ray analysis of S4~0954+65 using \emph{Fermi} Large Area Telescope (\emph{Fermi}-LAT) Pass 8 data between Modified Julian Date (MJD) 60593 and 60857 (10 October 2024 to 1 June 2025) in the energy range 0.1--300 GeV.  

Photon events were selected within a $15^\circ$ radius region of interest (ROI) centered on the source.
To ensure data quality, we applied the standard event selection criteria {\ttfamily `evclass=128'} and {\ttfamily `evtype=3'}. 
In addition, we excluded events with zenith angles larger than $90^\circ$ to suppress contamination from $\gamma$-rays produced in the Earth's limb, which arise from cosmic-ray interactions in the upper atmosphere and constitute a strong background for the LAT. 
We further restricted the analysis to good time intervals using the filter {\ttfamily `(DATA\_QUAL$\geqslant$0)\&\&(LAT\_CONFIG==1)'}.
The analysis was carried out with {\ttfamily Fermitools} \citep[v2.2.0;][]{Fermitools2019ascl} and the {\ttfamily P8R3\_SOURCE\_V3} instrument response functions. 
The source model was constructed using the {\ttfamily LATSourceModel}\footnote{https://github.com/physicsranger/make4FGLxml} package and included all sources from the \emph{Fermi}-LAT Fourth Source catalog \citep[4FGL-DR4;][]{Abdollahi2022ApJS260, Ballet2023arXiv230712546} within $25^\circ$ of S4~0954+65, as well as the Galactic diffuse emission ({\ttfamily gll\_iem\_v07.fits}) and isotropic background component ({\ttfamily iso\_P8R3\_SOURCE\_V3\_v1.txt}). 
Sources only have free spectral parameters if they were found in the catalog with $12\sigma$ average significance and are within $5^\circ$ of the ROI center.
We also freed the parameters of sources within $10^\circ$ that have a variability index $\geqslant$24.725 \citep{Abdollahi2022ApJS260}. 
For S4~0954+65, a power-law model was adopted in the analysis.
An unbinned likelihood analysis was performed for the data. 
Based on the best-fit model, we generated $\gamma$-ray light curves with one-day and seven-day time bins. 
Only flux points with TS$\geqslant$9\footnote{
The test statistic (TS) is defined as 
${\rm TS} = -2 \ln \left( \frac{\mathcal{L}_{\rm max, ~ 0}}{\mathcal{L}_{\rm max, ~ 1}} \right)$, where $\mathcal{L}_{\rm max, ~ 0}$ is the maximum likelihood value of the model without the target source (i.e., the null hypothesis), and $\mathcal{L}_{\rm max, ~ 1}$ is the maximum likelihood value of the model including the target source at the specified position. 
A larger TS value indicates a higher significance of the source detection \citep{Mattox1996ApJ461}.}
were considered significant detections and included; otherwise, 95\% confidence level upper limits were derived using the {\ttfamily UpperLimits}\footnote{https://fermi.gsfc.nasa.gov/ssc/data/analysis/scitools/upper\_limits.html} tool.
The light curve is shown in Figure~\ref{Fig:LC_J0958}. 
For the spectral energy distribution analysis, we adopted the same source model while fixing the spectral indices to the best-fit values obtained from the fit over the full energy range.

\subsection{X-ray and optical observations}

\subsubsection{Swift}

The X-Ray Telescope \citep[XRT;][]{Burrows2004_SwiftXRT} onboard the Neil Gehrels \emph{Swift} Observatory carried out four observations of S4~0954+65 between MJD 60593 and 60857 in photon counting (PC) mode. 
The XRT spectra were generated using the online data products provided by the UK \emph{Swift} Science Data Centre at the University of Leicester \citep{Evans2009MNRAS397}. 
Events with grades 0--12 were selected, and pile-up corrections were applied following the standard procedures described by the UK \emph{Swift} Science Data Centre when the count rate exceeded 0.5 counts s$^{-1}$ in PC mode. 
The spectra were grouped using the {\ttfamily grppha} tool to ensure at least one count per bin, allowing the use of Cash statistics \citep{Cash1979ApJ228}. 
Then, the grouped spectra were fitted in {\ttfamily XSPEC} \citep{Arnaud1996} with an absorbed power-law model. 
The solar abundances were set to {\ttfamily wilm} \citep{Wilms2000ApJ542}, and the photoelectric absorption cross-sections were adopted from {\ttfamily vern} \citep{Verner1996ApJ465}. 
The Galactic hydrogen column density was fixed at $N_{\rm H} = 4.68 \times 10^{20}\ {\rm cm}^{-2}$, as reported by \citet{HI4PICollaboration2016A&A594}. 
The fitted parameter uncertainties correspond to the 68.3\% confidence level ($1 \sigma$ for one parameter of interest). 
The best-fit spectral parameters are listed in Table~\ref{Tab:xrt_spec} in the Appendix. 
The integral energy fluxes in the 0.5--4 keV band were derived from the best-fit model, and the resulting light curve is shown in Figure~\ref{Fig:LC_J0958}. 
The XRT spectra used in the subsequent broadband SED analysis were corrected for Galactic absorption using the cross-sections of \citet{Morrison1983ApJ270}.

For the Ultraviolet/Optical Telescope \citep[UVOT;][]{Roming2005SSR120} data onboard the \emph{Swift} Observatory, we performed aperture photometry using the {\ttfamily uvotmaghist} task with the standard calibration (version 20211108). 
Source counts were extracted from a circular region with a radius of $5''$, while the background was estimated from a nearby source-free region with a radius of $20''$. 
All the UVOT data were checked for the small-scale sensitivity inhomogeneities, which occur when the source falls within the small areas of low sensitivity, and the photometric data in which the source was saturated were excluded.
Galactic extinction was corrected for the observed magnitude with a value of $E(B-V) = 0.1030$ \citep{Schlafly2011ApJ737} using the interstellar extinction law with $R_{V} = 3.1$ \citep{Fitzpatrick1999PASP111}. 
Finally, the corrected magnitudes were converted into flux densities using the standard zero points from \citet{Breeveld2011J1358}.
The UVOT photometric flux densities are given in Table~\ref{Tab:uvot_phot}, and the resulting light curve is shown in Figure~\ref{Fig:LC_J0958}. 

However, the \emph{Swift} observations lacked simultaneous coverage of the $\gamma$-ray peak (around MJD 60722--60724).

\subsubsection{Einstein Probe}

The Wide-field X-ray Telescope (WXT) on board \emph{Einstein Probe} \citep[\emph{EP};][]{Yuan2022, Yuan2025} employs a very large field-of-view (3600 square degrees) to perform the sky survey in high cadences. 
The sky region of S4 0954+65 has been observed with \emph{EP}-WXT 761 times from MJD 60593 to 60681. 
The X-ray events data were selected and calibrated using the WXT Data Analysis Software ({\ttfamily WXTDAS}) and calibration database (CALDB) designed for WXT (Liu et al. in prep.). 
The CALDB is generated based on the results of the on-ground calibration experiments \citep{Cheng2025}. 
To obtain high-significance detections, we merged all these observations into longer time intervals (e.g., 3 days). 
The merged events and spectra data were produced by the {\ttfamily wxtmerge} tool in {\ttfamily WXTDAS} and the {\ttfamily addascaspec} tool in {\ttfamily HEASoft}. 
The {\ttfamily XSPEC} tool was employed to analyze the spectra, in which the spectra were grouped using the {\ttfamily grppha} tool to ensure at least 20 counts per bin.
The fitting setup was applied following the same setup used in \emph{Swift}-XRT data.
The \emph{EP}-WXT light curve is shown in Figure~\ref{Fig:LC_J0958}.

\subsubsection{ZTF}

We collected optical data from the Zwicky Transient Facility (ZTF), which surveys the northern sky in the $g$, $r$, and $i$ bands \citep{Bellm2019PASP131, Graham2019PASP131}.
The optical light-curve data were obtained from the ZTF Public Data Release 24 (DR24)\footnote{https://irsa.ipac.caltech.edu/data/ZTF/docs/releases/dr24/ztf\_release\\\_notes\_dr24.pdf}, covering observations from March 2018 to October 2025.
We used the `{\ttfamily ztfquery.lightcurve}' package \citep{Rigault2018zndo1345222} to retrieve data at the position of S4~0954+65 (RA=09:58:47.245, Dec=+65:33:54.819) within a radius of $1.5''$.
\citet{Xiong2025ApJS279} showed that the variability behaviors in the $g$ and $r$ bands do not differ significantly. 
Therefore, only the $g$-band data were used in our analysis.
To ensure data quality, we included only measurements with `{\ttfamily catflags=0}'.
The Galactic extinction correction was applied following the same procedure used for \emph{Swift}-UVOT data.
Figure~\ref{Fig:LC_J0958} presents the ZTF $g$-band light curve used in our analysis.

\subsection{Radio observations}

\subsubsection{SMA}
The 230~GHz (1.3 mm) observations of S4~0954+65 were obtained with the Submillimeter Array (SMA), located in Hawaii. 
S4~0954+65 is part of the SMA calibrator monitoring program, which is designed to measure the flux densities of compact extragalactic radio sources that serve as calibrators at millimeter wavelengths \citep{Gurwell2007ASPC375}. 
These observations are performed in scans of a few minutes.
For this study, we made use of the publicly available SMA calibrator data products provided through the SMA archive\footnote{http://sma1.sma.hawaii.edu/callist/callist.html}.
Figure~\ref{Fig:LC_J0958} shows the SMA 230~GHz light curve. 
However, no quasi-simultaneous SMA data are available for the $\gamma$-ray peak (around MJD 60722--60724).

\subsubsection{VLBA}

\begin{figure*}
    \centering
    \includegraphics[width=\hsize]{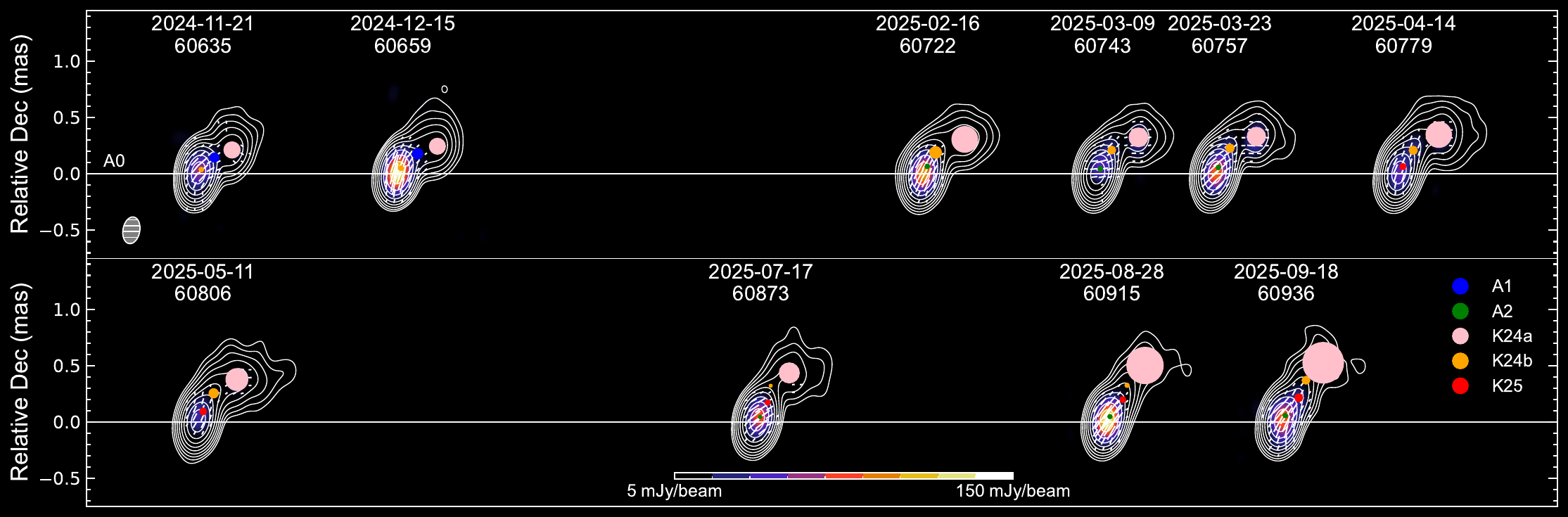}
    \caption{
    Sequence of total (contours) and polarized (color scale) intensity images of S4~0954+65 at 43 GHz, convolved with a beam of 0.24 $\times$ 0.15 mas$^2$ at $PA = -10 ^{\circ}$.
    The global total intensity peak is 2937 mJy/beam, and the global polarized intensity peak is 149 mJy/beam.
    The line segments in each image indicate the polarization direction, and the horizontal line indicates the position of the core, A0.
    The blue and green circles show the locations of stationary knots A1 and A2.
    The pink, orange, and red circles denote the locations of moving knots K24a, K24b, and K25, respectively.
    The detailed parameters of the knots are given in Table~\ref{Tab:VLBA_knot_params}.
    }
    \label{Fig:VLBA_image}%
\end{figure*}

We analyzed 43 GHz (7 mm) VLBA observations of S4~0954+65 obtained from the VLBA-BU-BLAZAR monitoring program\footnote{www.bu.edu/blazars/VLBAproject.html} \citep{Jorstad2016Galax4}. 
The dataset consists of the total-intensity images from 10 epochs, spanning from 21 November 2024 (MJD 60635) to 18 September 2025 (MJD 60936). 
The data had already been reduced and calibrated using the Astronomical Image Processing System (AIPS), as described in \citet{Jorstad2005AJ130, Jorstad2017ApJ846}. 
We performed the subsequent analysis using {\ttfamily DIFMAP}, following the procedures described by \citet{Jorstad2017ApJ846} and \citet{Weaver2022ApJS260}.
To investigate the kinematics of the inner jet, the total-intensity images were modeled with multiple two-dimensional circular Gaussian components using the `{\ttfamily modelfit}' task in {\ttfamily DIFMAP}. 
The modeling procedure began with a Gaussian component representing the brightness distribution of the core, and additional components were subsequently introduced at approximate locations where bright features appeared in the jet. 
The process was terminated when the inclusion of an additional Gaussian component no longer led to a significant improvement in the $\chi^2$ value.
These additional Gaussian components are referred to as ``knots.'' 
The parameters used to characterize each component are: 
(1) flux density, $S$; 
(2) radial distance from the core, $r$; 
(3) relative position angle (PA) with respect to the core, $\Theta$ (measured from north through east); 
and (4) angular size, $a$, corresponding to the full width at half maximum (FWHM) of the circular Gaussian component. 
The observed brightness temperature was then calculated as $T_{\rm b,~ obs} = 7.5 \times 10^{8} S / a^2$ K, as well as the uncertainties of all fitted parameters were estimated following \citet{Weaver2022ApJS260}. 
The fitted parameters of all components are listed in Table~\ref{Tab:VLBA_knot_params} in the Appendix, and the corresponding VLBA maps are shown in Figure~\ref{Fig:VLBA_image}.

\section{Results}

\begin{figure*}
    \centering
    \includegraphics[width=0.95\hsize]{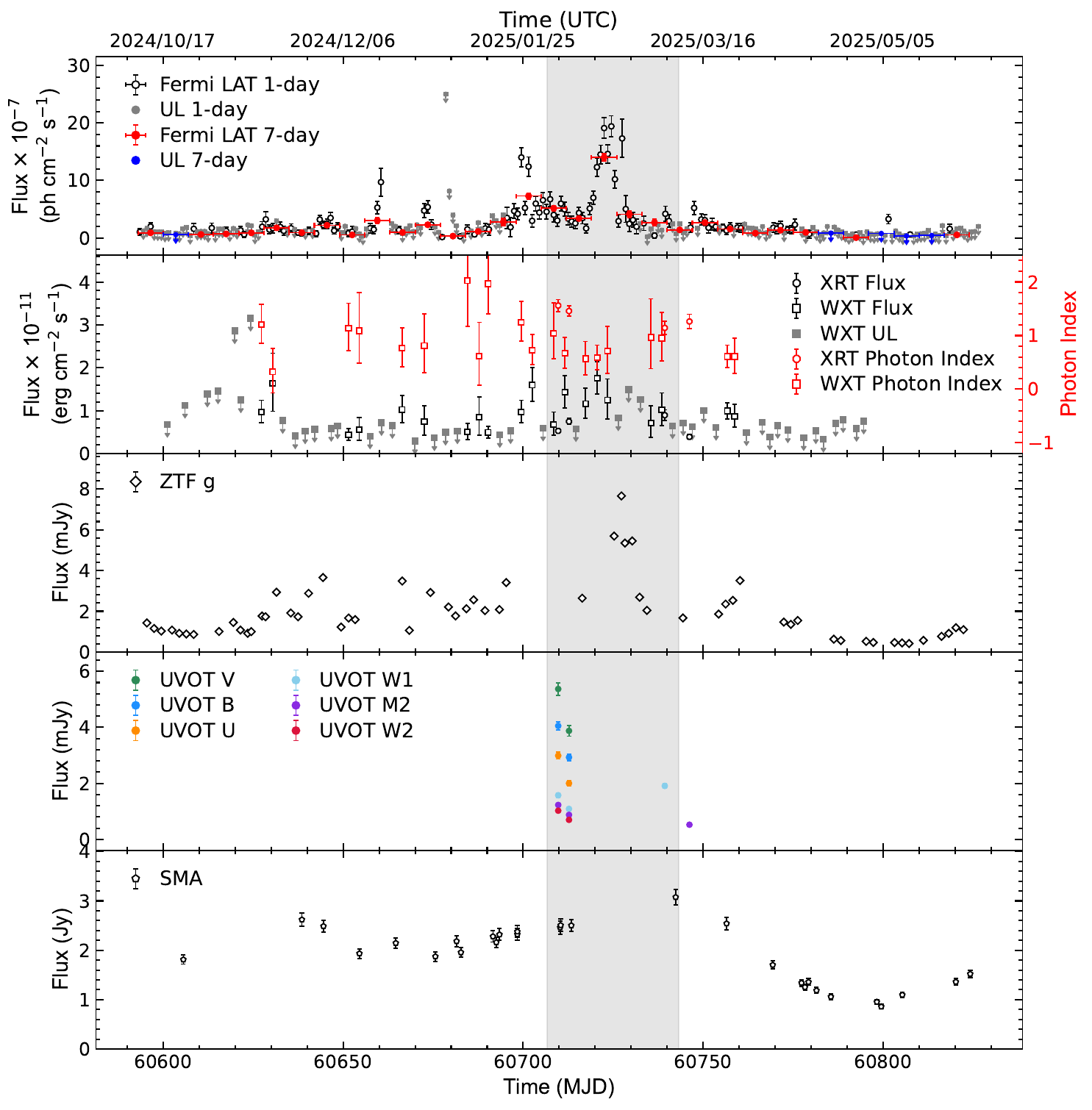}
    \caption{
    Multi-wavelength light curve of S4~0954+65 between MJD 60593 and 60857.
    From the top to the bottom panel:
    \emph{Fermi}-LAT flux and the 95\% confidence level upper limit in 0.1--300 GeV for one-day (black-hollow) and seven-day (red-solid) bins; 
    \emph{Swift}-XRT (circle) and \emph{EP}-WXT (square) fluxes in 0.5--4.0 keV and corresponding photon indices;
    ZTF $g$-band flux density corrected for Galactic extinction;
    \emph{Swift}-UVOT U, B, V, UVW1, UVM2, and UVW2 bands flux density corrected for Galactic extinction;
    and SMA 230 GHz flux density.
    The gray shaded region represents the emergence of a new knot, as detailed in Section~\ref{jet_kine}.
    }
    \label{Fig:LC_J0958}%
\end{figure*}

\subsection{Temporal variability}

Figure~\ref{Fig:LC_J0958} shows the multi-wavelength light curves of S4~0954+65 during MJD 60593--60857, covering the $\gamma$-ray, X-ray, optical/UV, and radio bands.

In the $\gamma$-ray band, the source exhibits strong variability, with a major flaring episode occurring during MJD $\sim$60695--60732. 
The flare reaches its maximum at MJD $\sim$60722--60724, with a peak flux of $\sim2 \times 10^{-6}$ ph cm$^{-2}$ s$^{-1}$. 
The one-day binned light curve reveals pronounced short timescale variability, while the seven-day binned light curve traces the overall evolution of the flare.
We quantified the variability in excess of the measurement errors using the fractional variability amplitude,
\begin{equation}
F_{\rm var}=\sqrt{\frac{S^{2}-\langle \sigma_{\rm err}^{2}\rangle}{\langle x\rangle^{2}}},
\end{equation}
where $\langle x\rangle$ is the mean of measured quantity, $S^{2}$ is the sample variance and $\langle \sigma_{\rm err}^{2}\rangle$ is the mean squared measurement uncertainty \citep{Vaughan2003MNRAS345}.
The uncertainty in $F_{\rm var}$, denoted by $\sigma_{F_{\rm var}}$, was calculated as \citep{Vaughan2003MNRAS345, Poutanen2008MNRAS389} 
\begin{equation}
    \sigma_{F_{\rm var}} = \sqrt{F_{\mathrm{var}}^{2}+\sqrt{\left(\sqrt{\frac{2}{N_{\rm data}}} \cdot \frac{\langle \sigma_{\rm err}^{2}\rangle}{\langle x\rangle^{2}}\right)^{2}+\left(\sqrt{\frac{\langle \sigma_{\rm err}^{2}\rangle}{N_{\rm data}}} \cdot \frac{2 F_{\mathrm{var}}}{\langle x\rangle}\right)^{2}}}-F_{\mathrm{var}} ,
\end{equation}
where $N_{\rm data}$ is the number of data points.
The light curve is considered to show significant variability when $F_{\rm var}/\sigma_{F_{\rm var}}\geqslant 3$.
We obtained the fractional variability of $\gamma$-ray is $F_{\rm var,~\gamma}=1.03\pm0.03$.

The X-ray light curve exhibits mild variability.
Using all available X-ray measurements, we obtain $F_{\rm var,~X}=0.27\pm0.06$, confirming that the flux variability in X-rays was modest compared to the flux variability in $\gamma$-rays.

In the optical band, the ZTF $g$-band light curve shows clear variability, with a flux enhancement during the flaring period and $F_{{\rm var},~g}=0.740\pm0.002$. 
The optical peak appears slightly delayed with respect to the $\gamma$-ray peak, as will be quantified in the following subsection. 
The \emph{Swift}-UVOT multi-band observations (V, B, U, W1, M2, and W2) show a decrease in flux during MJD $\sim$60709--60713, broadly coincident in time with the $\gamma$-ray activity.

At radio frequencies, the SMA 230 GHz light curve shows a smoother and more gradual variation than the higher-energy bands, with $F_{\rm var,~SMA}=0.31\pm0.01$, and a flux increase is observed accompanying the $\gamma$-ray flare, as shown in Figure~\ref{Fig:LC_J0958}.

\subsection{Multi-band correlation}

Multi-band correlation provides a direct way to examine the physical connection between emissions at different wavelengths. 
In blazars, although the broadband radiation is dominated by the relativistic jet, different energy bands may originate from regions with distinct physical conditions and spatial locations along the jet. 
As a result, the presence of temporal correlations and time delays between different bands can place important constraints on the relative locations of emission regions, as well as on particle acceleration and cooling processes.

The time lag between different bands is quantified using the z-transformed Discrete Correlation Function \citep[zDCF;][]{Alexander1997ASSL218, Alexander2013arXiv1302_1508}, which is designed for unevenly sampled light curves.
Compared to the traditional Discrete Correlation Function \citep[DCF;][]{Edelson1988ApJ333}, the zDCF applies equal population binning and Fisher's $z$-transform, providing more robust estimates of delay-dependent multi-band cross-correlations.
This method has been widely used in AGN variability studies to investigate multi-band correlations \citep[e.g.,][]{MAGIC2023A&A670_1ES_0647+250, KrishnaMohana2024MNRAS527, MAGIC2025A&A697_B2_1811+31}.
To assess the significance of the correlation, we employ a Monte Carlo approach by computing the zDCF for ensembles of simulated light curves that share the same power spectral density (PSD) and probability distribution function (PDF) as the observed ones. 
Further details of significance estimation are provided in Appendix \ref{zdcf_significance}.


Figure~\ref{Fig:zdcf} presents the zDCF results for the $\gamma$-ray versus X-ray (upper panel), optical $g$-band (middle panel), and radio 230 GHz (lower panel) light curves.
Positive lag values indicate that the $\gamma$-ray leads those in the comparison band.

For the $\gamma$-ray and X-ray light curves, no zDCF peak exceeds the $2\sigma$ significance level.
The highest peak closest to zero lag occurs at $-1.72$ days but reaches only $1.50\sigma$.
The absence of a statistically significant peak suggests that the present data do not establish a temporal correlation or a reliable lag between these two bands.
For the $\gamma$-ray and optical light curves, the zDCF shows a significant peak at a lag of 3.22 days. 
The peak exceeds the $3\sigma$ significance level with 3.09$\sigma$, supporting the presence of a statistically significant correlation between the two bands. 
This result suggests a close physical connection between the $\gamma$-ray and optical emission regions, with the high-energy emission arising earlier than the optical emission.
For the $\gamma$-ray and radio light curves, the zDCF shows a peak at a lag of 18.26 days. 
Its significance is above $2\sigma$ and approximately 2.17$\sigma$, but below the $3\sigma$ level, suggesting a possible correlation with the $\gamma$-ray variations leading the radio one.
Note that the SMA light curve has a substantial sampling gap around the $\gamma$-ray flare, and the radio maximum may have occurred within this gap and earlier than the observed maximum.
The inferred 18.26 days lag may therefore be overestimated and should be regarded as an indication of a possible radio delay rather than a precise measurement.

\begin{figure}
    \centering
    \includegraphics[width=0.9\hsize]{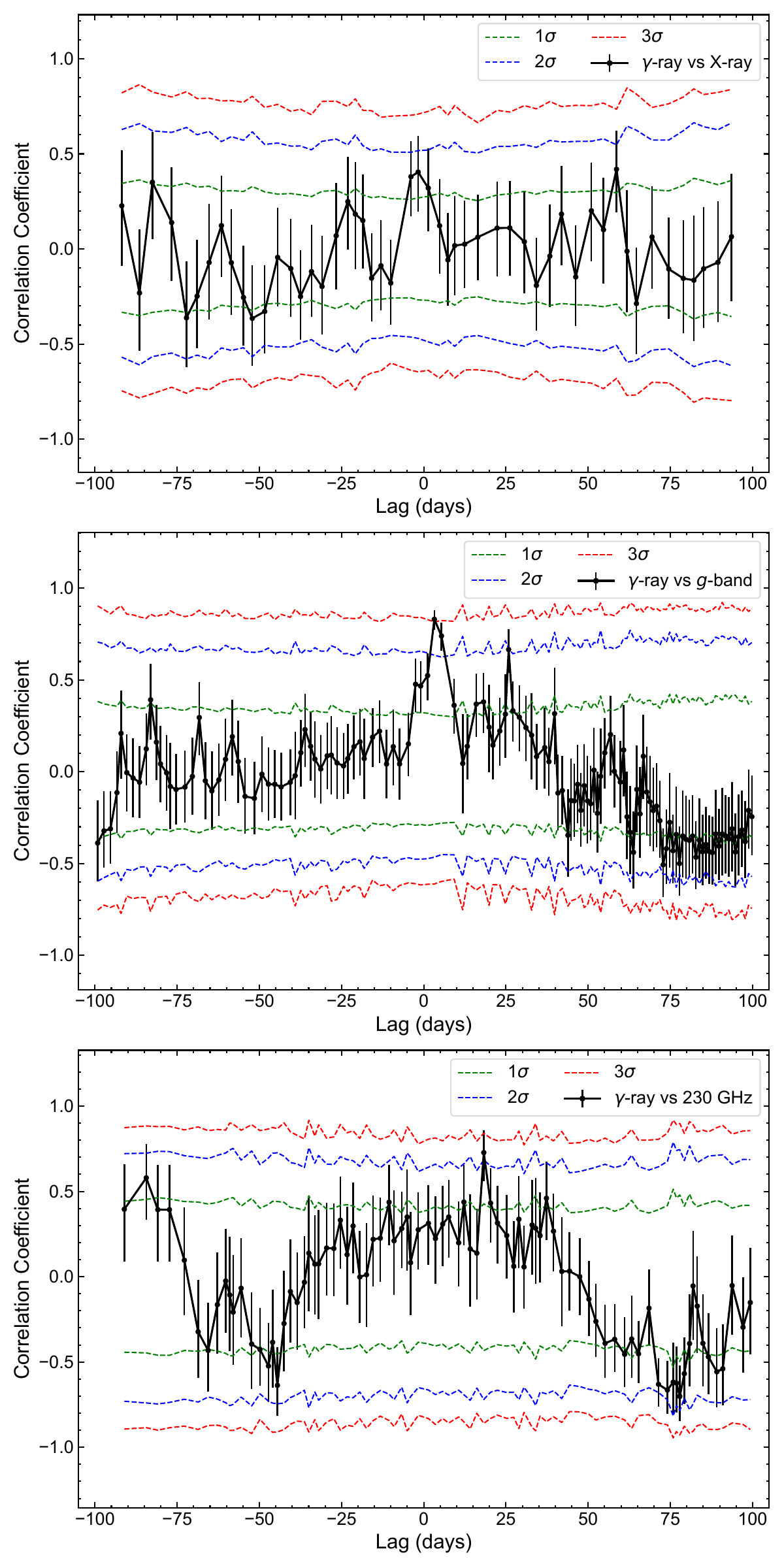}
    \caption{
    Z-transformed discrete correlation function (zDCF) obtained from the correlation between $\gamma$-ray and X-ray (upper panel), the correlation between $\gamma$-ray and ZTF $g$-band (middle panel), and the correlation between $\gamma$-ray and radio SMA 230 GHz (lower panel).
    The positive lag value represents that the $\gamma$-ray is ahead of the X-ray ($g$-band, or 230 GHz) light curve.
    The green, blue, and red dashed lines correspond to 1$\sigma$, 2$\sigma$, and 3$\sigma$ significance levels, respectively.
    }
    \label{Fig:zdcf}%
\end{figure}

\subsection{Jet kinematics}\label{jet_kine}

\begin{figure}
    \centering
    \includegraphics[width=0.95\hsize]{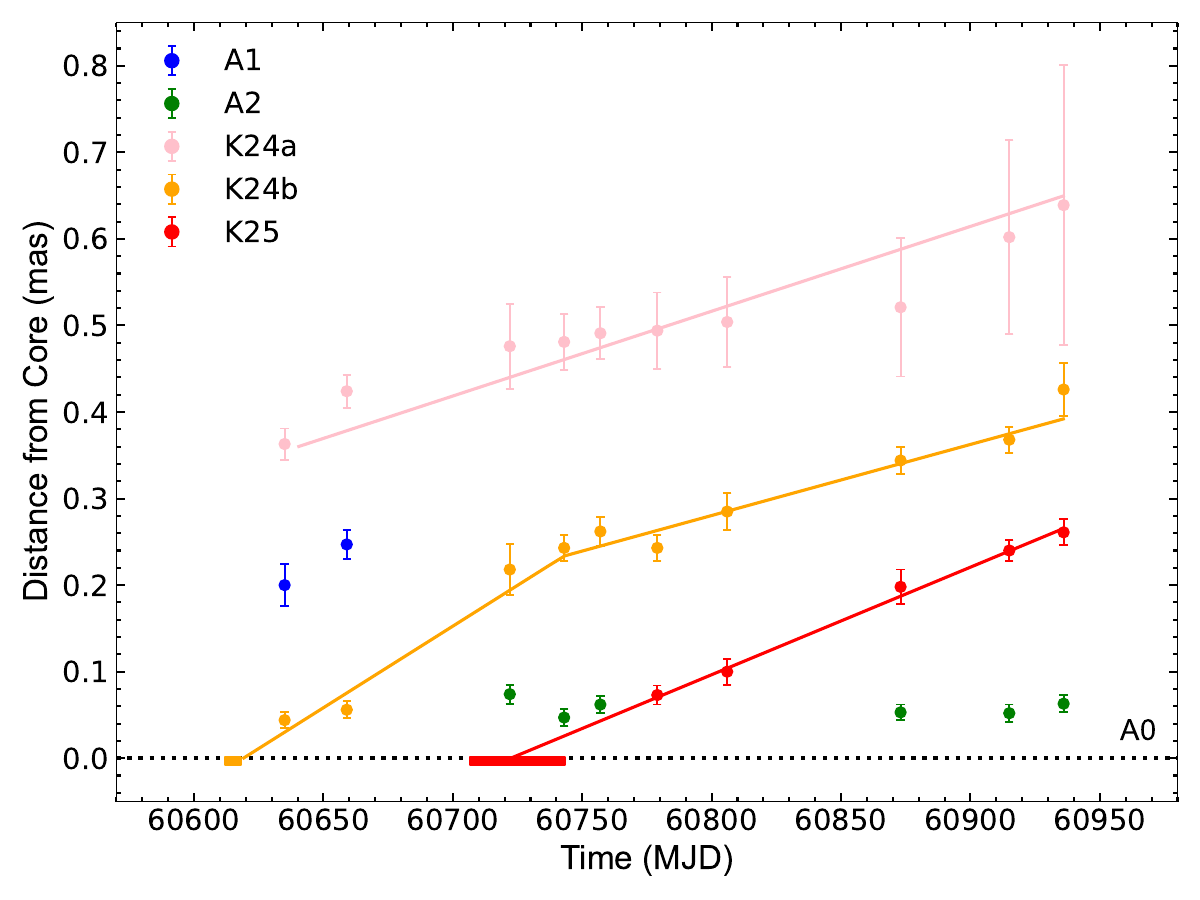}
    \caption{
    Apparent distance from the core A0 as a function of time for the newly emerging (moving) knots K24a, K24b, and K25, as well as for the stationary knots A1 and A2.
    The solid lines are the piecewise fit.
    }
    \label{Fig:VLBA_distance}%
\end{figure}

We analyzed the jet structure using the 43 GHz VLBA images. 
Such an analysis makes it possible to connect the evolution of parsec-scale jet features with multi-wavelength variability, and, in particular, to examine whether major flares are associated either with the ejection of new superluminal knots from the radio core or with the interaction of a moving knot with a recollimation shock.
The temporal evolution of the inner jet is presented in Figure~\ref{Fig:VLBA_image}.
The images reveal some structural evolution during the MJD 60635--60936. 
The brightest feature (core) at the jet center, A0, was assumed to be stationary, and its location is marked by the horizontal line in Figure~\ref{Fig:VLBA_image}. 
In addition to the core A0, we identified two stationary knots (A1 and A2), and three moving knots (K24a, K24b, and K25) propagating downstream of the jet. 
The knots are named according to the year in which they were first identified, and the method of knot identification is followed \citet{Weaver2022ApJS260}.
Figure~\ref{Fig:VLBA_distance} presents the separation of the knots from the core as a function of time. 
The locations of the stationary features A1 and A2 are consistent with those reported in previous studies \citep{Jorstad2017ApJ846, Weaver2022ApJS260}. 
One can analyze the jet kinematics using the knots information and further derive their properties.
On this basis, we followed the procedure reported by \citet[Section 2.2]{Weaver2022ApJS260} and calculated the proper motion, the apparent speed, and the ejection time (i.e., the time at which the ejection passed through the core, derived from the extrapolation of motion) from the piecewise fits. 
The derived parameters are listed in Table~\ref{tab:radio_knots}, and the piecewise fits are shown in Figure~\ref{Fig:VLBA_distance}.

The moving knots exhibit superluminal motion, as commonly observed in blazar jets. 
Among them, K25 is of particular interest because its extrapolated ejection time is consistent, within uncertainties, with the epochs of the major $\gamma$-ray and optical flare peaks, as shown in Figure~\ref{Fig:LC_J0958}. 
In addition, the flux density of the core increased around the emergence of K25 (see Table~\ref{Tab:VLBA_knot_params}).
The inferred knot ejection time overlaps with the flaring activity interval, suggesting a close physical connection between the emergence of the new knot and the broadband flare event. 
Such temporal coincidence supports a scenario in which the flaring activity is produced by a disturbance formed in the inner jet and resolved as a moving knot in the VLBA images, which is consistent with the previous result of this source \citep[e.g.,][]{Morozova2014AJ148, MAGIC2018A&A617}.

\begin{table*}
\centering
\renewcommand{\arraystretch}{1.2}
\setlength{\tabcolsep}{6pt}
\caption{Properties of the emerging knots in the jet of S4~0954+65.}
\label{tab:radio_knots}
\begin{tabular}{lccccccc}
\hline\hline
    Knot 
    & Segment 
    & Mean flux 
    & Mean PA 
    & Mean angular size 
    & Proper motion 
    & Apparent speed 
    & Ejection time \\
    & 
    & (Jy) 
    & (deg) 
    & (mas) 
    & (mas~yr$^{-1}$) 
    & ($c$) 
    & (MJD) \\
    \hline
    
    K24a 
    & -- 
    & $0.154 \pm 0.082$
    & $-44.2 \pm 8.1$
    & $0.206 \pm 0.075$
    & $0.358 \pm 0.013$
    & $8.427 \pm 0.312$
    & -- \\
    
    \multirow{2}{*}{K24b}
    & Seg.~1
    & \multirow{2}{*}{$0.194 \pm 0.217$}
    & \multirow{2}{*}{$-27.3 \pm 3.1$}
    & \multirow{2}{*}{$0.053 \pm 0.028$}
    & $0.689 \pm 0.050$
    & $16.234 \pm 1.174$
    & \multirow{2}{*}{$60615.21 \pm 3.08$} \\
    & Seg.~2
    & 
    & 
    & 
    & $0.300 \pm 0.039$
    & $7.061 \pm 0.910$
    & \\
    
    K25
    & -- 
    & $0.154 \pm 0.106$
    & $-29.0 \pm 5.6$
    & $0.050 \pm 0.009$
    & $0.454 \pm 0.017$
    & $10.685 \pm 0.401$
    & $60724.96 \pm 18.35$ \\
\hline
\end{tabular}
\tablefoot{
For knot K24b, the two kinematic segments are listed separately. 
No ejection time is available for knot K24a, because it was already located relatively far from the core at the epoch of MJD 60635, preventing a reliable estimate of its ejection epoch.
}
\end{table*}

\subsection{Spectral evolution}\label{spec_evo}

\begin{figure}
    \centering
    \includegraphics[width=0.97\hsize]{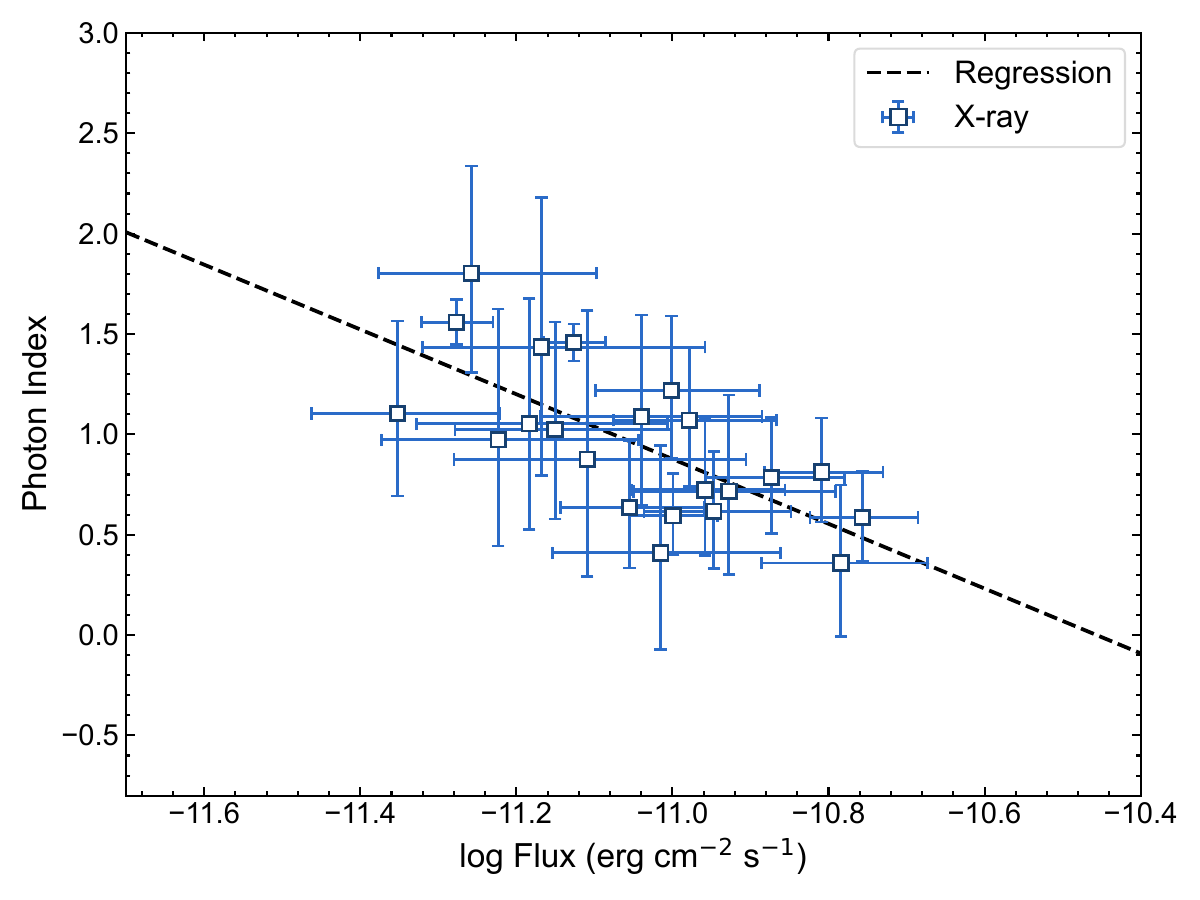}
    \caption{
    X-ray photon index versus X-ray flux. The black line represents the linear regression.
    }
    \label{Fig:flux_vs_index_xray}%
\end{figure}

We explored the spectral evolution in the $\gamma$-ray band over the energy range of 100 MeV--300 GeV. 
Such an analysis provides insight into the underlying particle distribution and radiation processes during different activity states.
Based on the $\gamma$-ray light curve shown in Figure~\ref{Fig:LC_J0958}, we divided the observational period into four distinct intervals: `Pre‑flare', `Flare', `Post‑flare', as well as the `Peak' period. 
The Pre-Flare, Flare, and Post-Flare intervals were defined using the seven-day binned light curve, which traces the long-term evolution of the source. 
The start of the Flare period was identified as the epoch when the seven-day binned flux began to rise systematically above the average level of the preceding low-flux state, while the end of the Flare period was defined as the epoch when the flux declined back to a comparable level.
The intervals preceding and following this Flare period, characterized by relatively low flux levels, were defined as the Pre-Flare and Post-Flare periods, respectively. 
The Peak period was determined from the one-day binned light curve to capture the highest flux state and corresponds to the time range around the maximum flux.

\begin{figure}
    \centering
    \includegraphics[width=0.95\hsize]{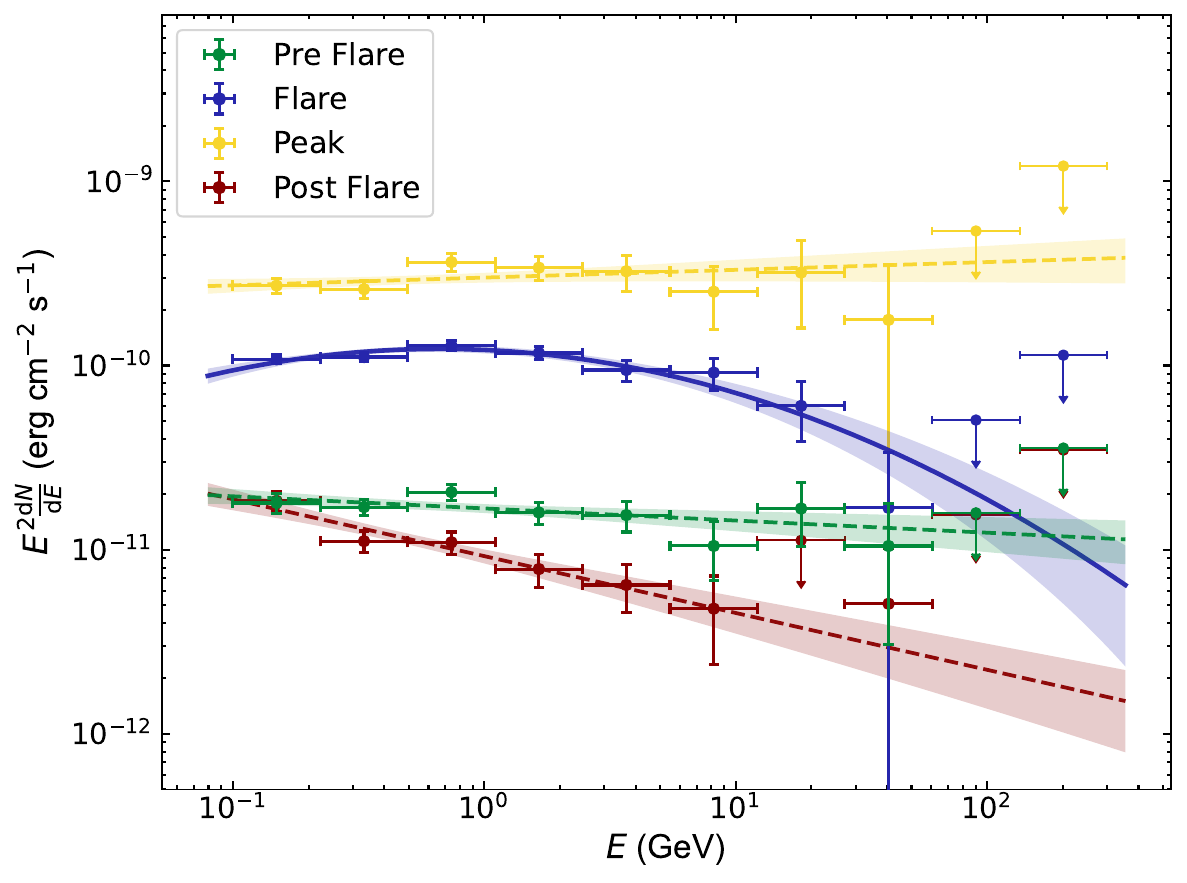}
    \caption{
    The $\gamma$-ray spectra evolution between the Pre-Flare (green), Flare (blue), Peak (yellow), and Post-Flare (red) periods. 
    The dashed line represents the power-law model, and the solid line represents the log-parabola model.
    The fitting parameters are listed in Table~\ref{tab:SED_HE_fit_results}. 
    }
    \label{Fig:SED_HE}%
\end{figure}

For each period, the $\gamma$-ray spectra were constructed and fitted with both a power-law (PL) and a log-parabola (LP) model using the maximum-likelihood fits. 
The models are defined as:
\begin{equation}
    \frac{\mathrm{d} N}{\mathrm{d} E} = N_{0} \left(  \frac{E}{E_{0}} \right)^{-\Gamma_{\gamma}} ,
\end{equation}
and
\begin{equation}
    \frac{\mathrm{d} N}{\mathrm{d} E} = N_{0} \left(  \frac{E}{E_{0}} \right)^{- \left(\alpha_{\gamma} + \beta_{\gamma} \ln (E/E_{0}) \right)} ,
\end{equation}
where $E_{0}$ is the pivot energy, $N_{0}$ is the normalization factor at $E_{0}$, $\Gamma_{\gamma}$ is the photon spectral index, $\alpha_{\gamma}$ represents the spectral index at $E_{0}$, and $\beta_{\gamma}$ describes the curvature of the spectrum.
To quantify the significance of spectral curvature, we adopted the curvature test statistic TS$_{\rm curve}$, defined as ${\rm TS}_{\rm curve} = 2 \left( \ln \mathcal{L}_{\rm LP} - \ln \mathcal{L}_{\rm PL} \right)$ as in \emph{Fermi}-LAT catalog.
Here, $\mathcal{L}_{\rm LP}$ and $\mathcal{L}_{\rm PL}$ denote the maximum likelihood values obtained by fitting the spectra with the log-parabola and power-law models, respectively.
A TS$_{\rm curve} \geqslant 9$, corresponding to a 3$\sigma$ significance level, indicates a stronger preference for a curved spectrum.
The spectra for the four periods are presented in Figure~\ref{Fig:SED_HE}, and the corresponding fitting parameters are listed in Table~\ref{tab:SED_HE_fit_results}. 

During the Pre-Flare period, the spectrum is well described by a simple power-law, with a photon index of $\Gamma_{\gamma} = 2.07 \pm 0.04$ and no significant curvature. 
In the Flare period, the flux increases substantially, and the spectrum shows a noticeable deviation from a power-law shape. 
The LP model is statistically preferred, with TS$_{\rm curve}$ = 25.71, indicating significant spectral curvature. 
At the same time, the spectral index hardens slightly to $\alpha_{\gamma} = 2.01 \pm 0.03$, accompanied by a curvature parameter $\beta_{\gamma} = 0.08 \pm 0.02$.
At the Peak period, the flux reaches its maximum level, and the spectrum becomes harder, with a photon index of $\Gamma_{\gamma} = 1.96 \pm 0.04$. 
In the Post-Flare period, the flux decreases to a lower level, and the spectrum softens significantly, with $\Gamma_{\gamma} = 2.31 \pm 0.07$.

We also examined the X-ray spectral behavior by relating the X-ray flux to the photon index.
Figure~\ref{Fig:flux_vs_index_xray} shows that the photon index is anti-correlated with the X-ray flux, with a Pearson coefficient of $r=-0.70$, where the associated $p$-value gives $p=3.3\times10^{-4}$, suggesting a clear harder-when-brighter trend.
The ordinary least-squares linear regression gives
\begin{equation}
    {\rm Photon~Index} = -(1.61 \pm 0.37) \log {\rm Flux} - (16.87 \pm 4.12). 
\end{equation}

\begin{table*}
\centering
\renewcommand{\arraystretch}{1.3}
\setlength{\tabcolsep}{6pt}
\caption{Gamma-ray SED fitting results for different activity periods.}
\label{tab:SED_HE_fit_results}
\begin{tabular}{lcccccccc}
\hline\hline
    Period
    & Time range
    & Model
    & $N_0$
    & $\Gamma_{\gamma}$ / $\alpha_{\gamma}$
    & $\beta_{\gamma}$
    & $\ln \mathcal{L}$
    & TS$_{\rm curve}$
    & Best model \\

    & (MJD)
    & 
    & (cm$^{-2}$ s$^{-1}$ MeV$^{-1}$)
    &  
    &  
    &  
    & 
    &  \\
    \hline
    
    \multirow{2}{*}{Pre-Flare}
    & \multirow{2}{*}{60593--60695}
    & PL
    & $(2.19 \pm 0.12)\times10^{-8}$
    & $2.07 \pm 0.04$
    & --
    & 202.08
    & \multirow{2}{*}{1.29}
    & \multirow{2}{*}{PL} \\
    &
    & LP
    & $(2.31 \pm 0.16)\times10^{-8}$
    & $2.04 \pm 0.05$
    & $0.04 \pm 0.03$
    & 202.73
    & 
    & \\
    
    \multirow{2}{*}{Flare}
    & \multirow{2}{*}{60695--60732}
    & PL
    & $(1.36 \pm 0.04)\times10^{-7}$
    & $2.08 \pm 0.02$
    & --
    & 179.43
    & \multirow{2}{*}{25.71}
    & \multirow{2}{*}{LP} \\
    &
    & LP
    & $(1.58 \pm 0.06)\times10^{-7}$
    & $2.01 \pm 0.03$
    & $0.08 \pm 0.02$
    & 192.28
    & 
    & \\
    
    \multirow{2}{*}{Peak}
    & \multirow{2}{*}{60722--60725}
    & PL
    & $(3.78 \pm 0.21)\times10^{-7}$
    & $1.96 \pm 0.04$
    & --
    & 177.39
    & \multirow{2}{*}{2.40}
    & \multirow{2}{*}{PL} \\
    &
    & LP
    & $(4.12 \pm 0.31)\times10^{-7}$
    & $1.92 \pm 0.05$
    & $0.05 \pm 0.03$
    & 178.59
    & 
    & \\
    
    \multirow{2}{*}{Post-Flare}
    & \multirow{2}{*}{60732--60827}
    & PL
    & $(1.32 \pm 0.10)\times10^{-8}$
    & $2.31 \pm 0.07$
    & --
    & 180.42
    & \multirow{2}{*}{0.51}
    & \multirow{2}{*}{PL} \\
    &
    & LP
    & $(1.26 \pm 0.12)\times10^{-8}$
    & $2.32 \pm 0.07$
    & $-0.03 \pm 0.04$
    & 180.68
    & 
    & \\
    
\hline
\end{tabular}
\tablefoot{
PL and LP denote the power-law and log-parabola models, respectively.
The pivot energy is set as $E_{0}$ = 698.46 MeV according to the \emph{Fermi}-LAT 4FGL-DR4 catalog \citep{Abdollahi2022ApJS260}.
The preferred best model in each period is determined from the curvature test statistic (TS$_{\rm curve}$) as described in Section~\ref{spec_evo}.
}
\end{table*}

\section{Spectral energy distributions modeling and physical constraints}

\subsection{Construction of the multi-wavelength SEDs}

To investigate the physical conditions in the emission region and interpret the broadband SEDs, we constructed quasi-simultaneous multi-wavelength SEDs for four representative activity periods defined in Section~\ref{spec_evo}.
For the Pre-Flare, Flare, and Post-Flare periods, the SEDs were constructed using multi-wavelength data collected within the corresponding time intervals. 
The $\gamma$-ray spectra were derived from the \emph{Fermi}-LAT data in each period, while the X-ray, ultraviolet, optical, and radio data were selected from observations falling within the same intervals when available. 

The construction of the SED of the Peak period requires additional treatment. 
The Peak period is centered on the maximum of the $\gamma$-ray flare, around MJD 60722--60725, but strictly simultaneous observations are not available in all wavebands within this short interval. 
Therefore, for wavebands without observations exactly during the $\gamma$-ray peak, we adopted the closest available observations around the peak as quasi-simultaneous data. 
The SMA 230 GHz data used in the Peak-period SED were obtained on MJD 60713--60714, while the ZTF $g$-band data were taken from observations on MJD 60725--60726.

\subsection{Model setup}

We model the four quasi-simultaneous SEDs constructed above with a one-zone radiative model.
In this framework, the observed broadband emission is assumed to originate from a single homogeneous emission region, also referred to as the blob, moving relativistically along the jet at a small viewing angle $\theta_{\rm v}$ with respect to our line of sight.
The blob motion is characterized by a bulk Lorentz factor $\Gamma = \left( 1 - \beta_{\Gamma}^{2} \right)^{-1/2}$, where $\beta_{\Gamma} c$ is the velocity of the blob along the jet.
The corresponding Doppler factor is given by $\delta = \left[ \Gamma \left( 1 - \beta_{\Gamma} \cos \theta_{\rm v} \right) \right]^{-1}$.
The emission region is modeled as a spherical region with radius $R_{\rm b}$, filled with non-thermal particles and a tangled magnetic field of strength $B$.
For blazars, the viewing angle is generally expected to be smaller than a few degrees $\theta_{\rm v} \lesssim 1/\Gamma$, under which $\delta \simeq \Gamma$.
In this work, quantities with the superscript ``obs'' are measured in the observer's frame, those with ``AGN'' are measured in the AGN frame, and quantities without any superscript are defined in the jet co-moving frame, unless otherwise stated.

\subsection{Leptonic modeling}

We first consider a leptonic emission model to reproduce the broadband SEDs within a time-dependent framework.
In this framework, the evolution of the particle energy distribution is governed by a continuity equation that accounts for particle injection, radiative cooling, and particle escape.

In leptonic models, the broadband emission is dominated by the relativistic electrons.
The low-energy hump of the broadband SED is dominated by synchrotron emission from relativistic electrons spiraling in the magnetic field.
The high-energy hump is produced by inverse Compton (IC) scattering of soft seed photons.
If the seed photons are the synchrotron photons produced by the same electron population, the process is referred to as synchrotron self-Compton \citep[SSC; e.g.,][]{Maraschi1992ApJ397, Finke2008ApJ686}.
Alternatively, if the seed photons originate from regions external to the jet, such as the accretion disk \citep[AD;][]{Dermer1993ApJ416}, the broad line region \citep[BLR;][]{Sikora1994ApJ421}, or the dusty torus \citep[DT;][]{Blazejowski2000ApJ545}, the process is referred to as external Compton (EC).

The source S4~0954+65 was classified as a BL Lac object due to its rather weak emission lines \citep{Ghisellini2011MNRAS414}.
As such, a widely used framework for blazar SED modeling of the BL Lac subclass is the pure SSC model \citep[e.g.,][]{Maraschi1992ApJ397, Tavecchio1998ApJ509, Finke2008ApJ686, Diwan2023MNRAS524, Xiao2025ApJ991}. 
\citet{Tavecchio1998ApJ509} demonstrated that the physical parameters of the SSC model can be approximately constrained by the peak frequencies ($\nu_{\rm syn,~ peak}$ and $\nu_{\rm IC,~ peak}$) and the peak luminosities ($\nu L_{\nu,~\rm syn,~ peak}$ and $\nu L_{\nu,~\rm IC,~ peak}$) of the synchrotron and IC components.
Following previous work \citep[e.g.,][]{Fan2016ApJS226, Yang2022ApJS262, MAGIC2025MNRAS540_OT081}, we estimated these quantities by fitting the two broad SED humps in log-log space with second-order polynomial functions.
For instance, for the Flare period, we obtained $\nu_{\rm syn,~peak} \sim 2.7 \times 10^{13}$ Hz and $\nu L_{\nu, ~\rm syn,~peak} \sim 4.3 \times 10^{46}$ erg s$^{-1}$ for the synchrotron component, and $\nu_{\rm IC,~peak} \sim 1.4 \times 10^{22}$ Hz and $\nu L_{\nu, ~\rm IC,~peak} \sim 6.9 \times 10^{46}$ erg s$^{-1}$ for the IC component. 
We then applied Equations~(4), (7), and (11) of \citet{Tavecchio1998ApJ509}.
These analytical equations are derived in the framework of a homogeneous one-zone SSC model, in which the synchrotron and IC emission at their peak frequencies is dominantly produced by relativistic electrons with $\gamma_{\rm b}$, where $\gamma_{\rm b}$ is the break Lorentz factor of the electron energy distribution. 
It is further assumed that the same electrons upscatter the produced synchrotron photons in the Thomson regime and that the emission region size is constrained by 
$R_{\rm b} \leqslant c t_{\rm var}\delta/(1+z)$.
Combining these relations, we derived the constraints of $B\delta \sim 3 \times 10^{-2}$ and $B\delta^{3} \sim 336$ for the Flare period SED.
In the constraint, we adopt $t_{\rm var}=5$ days as the characteristic rise time of the main $\gamma$-ray flare from approximately MJD 60717 to its peak around MJD 60722 in the one-day light curve (Figure~\ref{Fig:LC_J0958}).
These relations imply $B \sim 2 \times 10^{-4}$ G and $\delta \sim 115$, which are atypical parameter ranges for blazars.
A similar atypical result is also obtained for the Peak period, with  $B \sim 1 \times 10^{-5}$ G and $\delta \sim 300$. 
In contrast, the Pre-Flare and Post-Flare periods yield more moderate parameters, with $B \sim 2 \times 10^{-3}$ G and $\delta \sim 45$, and $B \sim 6 \times 10^{-2}$ G and $\delta \sim 12$, respectively, which are broadly compatible with typical one-zone SSC expectations.
Therefore, the extremely low magnetic fields and large Doppler factors inferred for the Flare and Peak periods are difficult to accommodate within a one-zone SSC scenario, where $B$ typically ranges from a few mG to $\sim$1 G and $\delta$ is usually below $\sim$40 \citep{Ghisellini2009MNRAS399, Ghisellini2010MNRAS402, Jorstad2017ApJ846, Liodakis2018ApJ866, Lister2018ApJS234, Weaver2022ApJS260, Xiao2025ApJ991}.
Furthermore, these inferred $\delta$ values conflict with the apparent speeds obtained from VLBA images, as listed in Table~\ref{tab:radio_knots}.
This indicates that the observed SED during the high-activity states cannot be satisfactorily explained by a simple one-zone SSC model alone, and that an additional external photon field may be required.

On the other hand, the Mg~II emission line is detected in the optical spectrum, with a FWHM of $\sim 4430~{\rm km~s^{-1}}$ and an equivalent width of $4.9~{\rm \mathring{A}}$ \citep{BecerraGonzalez2021MNRAS504}, suggesting the presence of a BLR in this source. 
Moreover, \citet{BecerraGonzalez2021MNRAS504} reported that S4~0954+65 is a transitional object between BL Lac and FSRQ, and that even if it is classified as a BL Lac, the DT may still be sufficiently powerful to provide external soft photons. 
Therefore, the DT may plausibly supply the soft photon field for EC scattering, making an SSC+EC scenario a more physically reasonable framework for explaining the high-energy emission in S4~0954+65, especially during the high-activity states. 
This interpretation is also consistent with previous studies, which found that an EC component is required to reproduce the high-energy emission of this source \citep{Tanaka2016PASJ68, MAGIC2018A&A617}.

In our modeling, we assume that the primary relativistic electrons are accelerated and are injected into the blob at a constant rate following a power-law with an exponential cut-off (PLEC) distribution, defined as
\begin{equation}
    Q_{\rm e}^{\rm inj}(\gamma_{\rm e}) = Q_{\rm e,~0} \gamma_{\rm e}^{-p_{\rm e}} \exp{\left( - \frac{\gamma_{\rm e}}{ \gamma_{\rm e, ~ max}} \right) }, ~ \gamma_{\rm e} \geqslant \gamma_{\rm e, ~ min} ,
\end{equation}
where $\gamma_{\rm e, ~ min/max}$ represent the electron minimum and maximum Lorentz factor and $p_{\rm e}$ is the electron spectral index.
The normalization $Q_{\rm e, ~ 0}$ are characterized by an electron injection luminosity $L_{\rm e}^{\rm inj}$ through $\int \gamma_{\rm e} m_{\rm e} c^{2} Q_{\rm e}^{\rm inj}(\gamma_{\rm e})  \mathrm{d}\gamma_{\rm e} = 3 L_{\rm e}^{\rm inj} / (4 \pi R_{\rm b}^{3})$.

The temporal evolution of the particle distribution is described by the continuity equation, which is written in the form: 
\begin{equation}\label{kinetic_equation}
    \frac{\partial N(\gamma,t)}{\partial t}
    =
    Q(\gamma,t)
    -
    \frac{\partial}{\partial \gamma}
    \left[ \dot{\gamma}(\gamma,t) ~ N(\gamma,t) \right]
    -
    \frac{N(\gamma,t)}{t_{\rm esc}},
\end{equation}
where 
$Q(\gamma,t)$ is the injection term, $\dot{\gamma}$ represents the energy loss rate (radiative cooling term), and $t_{\rm esc}$ is the escape timescale.
We assume the escape timescale to be comparable to the dynamical timescale.
The $N(\gamma,t)$ in Equation~\eqref{kinetic_equation} denotes the energy distribution of the particle species under consideration. 
For simplicity, the particle-species subscript is omitted in the general form of the continuity equation, and will be specified explicitly in the following.
The radiative cooling includes synchrotron and IC processes \citep[e.g.,][]{Bottcher2012rjag_book}:
\begin{equation}
    \left| \dot{\gamma}_{\rm e} \right|
    =
    \left| \dot{\gamma}_{\rm e, ~ syn} \right|
    +
    \left| \dot{\gamma}_{\rm e, ~ IC} \right|  
    = \frac{4}{3} \frac{\sigma_{\rm T} }{m_{\rm e} c} \gamma_{\rm e}^{2} \left( U_{B}  +  F_{\rm KN} U_{\rm soft} \right)
     ,
\end{equation}
where $\sigma_{\rm T}$ is the Thomson scattering cross section, $U_{B} = \frac{B^2}{8 \pi}$ is the magnetic energy density, $m_{\rm e}$ is electron mass, 
$U_{\rm soft}$ is the total energy density of the target soft photon field,
and $F_{\rm KN}$ is a factor accounting for the Klein-Nishina effect, given by \citep{Jones1968PhRv167, Moderski2005MNRAS363}
\begin{equation}
    F_{\rm KN}  = \frac{m_{\rm e} c^{2}}{ U_{\rm soft} } \int^{\infty}_{0} \frac{9 g(\tilde{b}) }{ \tilde{b}^3} ~ n_{2}(\epsilon, t) ~ \epsilon~ \mathrm{d} \epsilon, 
\end{equation}
where $\epsilon = \frac{h \nu}{m_{\rm e} c^{2}}$ is the dimensionless photon energy with $\nu$ is the photon frequency, $n_{2}(\epsilon, t)$ is the differential number density of the soft photon field,
$\tilde{b} = 4 \gamma_{\rm e} \epsilon$ is a dimensionless parameter that measures the importance of the Klein–Nishina effect, and function $g(\tilde{b})$ is defined in Equation~(C3) of \citet{Moderski2005MNRAS363}.

We follow the treatment adopted by \citet{Tanaka2016PASJ68} and \citet{MAGIC2018A&A617}, in which the EC component is included with external soft photons supplied by the DT. 
The energy density of the DT soft photon field in the jet co-moving frame can be estimated as \citep{Hayashida2012ApJ754}
\begin{equation}
    U_{\mathrm{DT, ~ soft}} = \frac{\xi_{\mathrm{DT}} \Gamma^{2} L_{\mathrm{disk}}}{4 \pi\left(r_{\mathrm{DT}}^{\mathrm{AGN}}\right)^{2} c\left[1+\left(R_{\mathrm{H}} / r_{\mathrm{DT}}^{\mathrm{AGN}}\right)^{4}\right]},
    \label{energy_density_DT}
\end{equation}
where $R_{\mathrm{H}}$ is the distance between the emission region and the central engine, measured along the jet axis, $L_{\mathrm{disk}}$ is the disk luminosity, $\xi_{\mathrm{DT}}=0.1$ is the covering factor of the DT,
and $r_{\mathrm{DT}}^{\mathrm{AGN}}$ is the characteristic radius of the DT in the AGN frame. 
The latter can be estimated as \citep{Ghisellini2009MNRAS397}
\begin{equation}
    r_{\mathrm{DT}}^{\mathrm{AGN}} = 2.5 \times 10^{18} \left( \frac{L_{\mathrm{disk}}}{10^{45}~{\rm erg~s^{-1}}} \right)^{1/2} {\rm cm}.
\end{equation}
We further assume that the DT radiation field can be approximated by an isotropic blackbody spectrum, peaking at $3 \times 10^{13}\Gamma$~Hz in the jet co-moving frame \citep{Cleary2007ApJ660}. 
The resulting DT photon field is used as the external soft photon field for the EC process.

Under a given physical setting, the injected primary relativistic electrons evolve under the competition among injection, radiative cooling, and escape, as described by Equation~\eqref{kinetic_equation}, and eventually reach a steady state when these processes balance each other.

At sufficiently high energies, the emitted $\gamma$-ray photons propagating through the emission region can be absorbed via $\gamma\gamma$ pair production ($\gamma$$\gamma$ absorption, $\gamma + \gamma \rightarrow e^{+} + e^{-}$) in the presence of a soft photon field, and produce secondary electron-positron pairs inside the region.
These secondary particles are then injected into the emission region and evolve under the same transport processes, namely injection, radiative cooling, and escape.
The secondary particle distribution is likewise described by the continuity equation in Equation~\eqref{kinetic_equation}, with the corresponding injection term determined by the differential spectrum of secondary pairs from $\gamma\gamma$ pair-production \citep{Aharonian1983Afz19}
\begin{equation}\label{Q_gg}
\begin{aligned}
    Q^{\gamma\gamma}_{\rm e, ~sec}(\gamma_{\rm e}, t)
    = &
        2 \cdot \frac{3}{32}~c~\sigma_{\mathrm T}
        \int_{\gamma_{\rm e}}^{\infty} \mathrm{d}\epsilon_{1} ~
        \frac{n_{1}(\epsilon_{1},t)}{\epsilon_{1}^{3}}
        \int_{\frac{\epsilon_{1}}{4\gamma_{\rm e}(\epsilon_{1}-\gamma_{\rm e})}}^{\infty} 
        \mathrm{d}\epsilon_{2} ~
        \frac{n_{2}(\epsilon_{2}, t)}{\epsilon_{2}^{2}}
        \\
        & \times \left\{
        \frac{4\epsilon_{1}^{2}}{\gamma_{\rm e}(\epsilon_{1}-\gamma_{\rm e})} ~
        \ln\!\left(\frac{4\epsilon_{2}\gamma_{\rm e}(\epsilon_{1}-\gamma_{\rm e})}{\epsilon_{1}}\right)
        - 8 \epsilon_{1}\epsilon_{2}
        \right.
        \\
        & 
        + \frac{2\epsilon_{1}^{2}(2\epsilon_{1}\epsilon_{2}-1)}
        {\gamma_{\rm e}(\epsilon_{1}-\gamma_{\rm e})}
        - \left(1-\frac{1}{\epsilon_{1}\epsilon_{2}}\right)
        \frac{\epsilon_{1}^{4}}
        {\gamma_{\rm e}^{2}(\epsilon_{1}-\gamma_{\rm e})^{2}}
        \left.\vphantom{\frac{4\epsilon_{1}^{2}}{\gamma_{\rm e}(\epsilon_{1}-\gamma_{\rm e})}}\right\}, 
\end{aligned}
\end{equation}
where $\epsilon_{1}$ represents the dimensionless high-energy photon energy, $\epsilon_{2}$ represents the dimensionless low-energy (soft) photon energy, $n_{1}$ is the differential number density of high-energy photons, and the factor 2 accounts for the production of electrons and positrons \citep{Cerruti2019MNRAS483, Cerruti2021MNRAS502}. 
In the present model, the photon fields ($n_{1}$ and $n_{2}$) are also evolved self-consistently together with the particle distributions. 
Its temporal evolution is described by the continuity equation for photons, 
\begin{equation}\label{photon_kinetic_equation}
    \frac{\partial n_{\rm ph}(\epsilon,t)}{\partial t}
    =
    Q_{\rm ph}(\epsilon,t)
    -
    \frac{n_{\rm ph}(\epsilon,t)}{t_{\gamma\gamma}^{\rm abs}}
    -
    \frac{n_{\rm ph}(\epsilon,t)}{t^{\rm abs}_{\rm SSA}}
    -
    \frac{n_{\rm ph}(\epsilon,t)}{t_{\rm esc}},
\end{equation}
where $n_{\rm ph}(\epsilon,t)$ is the differential photon number density and $Q_{\rm ph}(\epsilon,t)$ is the photon injection term.
The photon injection term includes synchrotron, SSC, and EC radiation from the primary electrons, as well as the radiative contribution from the secondary pairs. 
The remaining terms on the right-hand side describe photon attenuation due to internal $\gamma\gamma$ pair production, synchrotron self-absorption (SSA), and photon escape from the emission region, respectively.
Specifically, the internal $\gamma\gamma$ absorption timescale is given by \citep{Rybicki_Lightman1979rpabook, Dermer2009herb_book}
\begin{equation}
    \left| t_{\gamma\gamma}^{\rm abs} \right|
    =
    \left[
    \frac{2c}{\epsilon_1^{2}}
    \int_{1/\epsilon_1}^{\infty}
    \! \mathrm{d}\epsilon_{2}~
    \frac{n_{2}(\epsilon_{2})}{\epsilon_{2}^{2}}
    \int_{1}^{\epsilon_1\epsilon_{2}}
    \! \mathrm{d}s~\sigma_{\gamma\gamma}(s)~s
    \right]^{-1}, 
\end{equation}
where $\sigma_{\gamma\gamma}(s)$ is the $\gamma\gamma$ pair-production cross section and $s$ is the center-of-momentum-frame Lorentz factor of the produced electron-positron pairs.
The SSA timescale is written as \citep{Rybicki_Lightman1979rpabook, Dermer2009herb_book}
\begin{equation}
    \left| t^{\rm abs}_{\rm SSA} \right|
    = 
    \left[\frac{c}{ 8 \pi m \nu^2} \int_{1}^{\infty} P_{\nu}^{\rm syn}(\gamma_{\rm e}) ~ \gamma_{\rm e}^{2} \frac{\partial }{\partial \gamma_{\rm e}} \left[ \frac{N_{\rm e}(\gamma_{\rm e},t)}{\gamma_{\rm e}^2} \right] \mathrm{d} \gamma_{\rm e}  \right]^{-1},
\end{equation}
where $P_{\nu}^{\rm syn}(\gamma_{\rm e})$ is the synchrotron power emitted by a single electron integrated over solid angle and $N_{\rm e}(\gamma_{\rm e},t)$ is the electron distribution.

The coupled continuity equations for the particle and photon distributions are solved iteratively and self-consistently until a steady-state solution is achieved. 
The final model SED is then obtained by summing all radiative components, including the synchrotron and IC emission from the primary electrons and from the secondary pairs. 
Internal $\gamma\gamma$ absorption and synchrotron self-absorption are treated self-consistently during the modeling.
In addition, the extragalactic background light (EBL) attenuation is also considered in the radiation output using the model of \citet{Dominguez2011MNRAS410}.

Based on the physical setup described above, the one-zone leptonic model is characterized by nine parameters, four describing the injected electron distribution, three specifying the properties of the emission region, and two characterizing the external soft photon field.
Specifically, the disk luminosity is fixed to $L_{\rm disk} = 5.3 \times 10^{44}~{\rm erg~s^{-1}}$, as inferred from the optical spectroscopy \citep{BecerraGonzalez2021MNRAS504}.
The distance between the emission region and the central engine is fixed to $R_{\rm H}=0.5$~pc, consistent with a location outside the BLR and within the region where the DT photon field can still provide an effective target for EC scattering.
The minimum electron Lorentz factor cannot be accurately constrained by the present SEDs because the low-frequency synchrotron emission is strongly affected by SSA, preventing robust constraints on the low-energy part of the electron distribution.
We therefore fix $\gamma_{\rm e,~min}=10^{2}$ to reduce the model degeneracy, and this value is also commonly adopted as a fiducial choice in blazar SED modeling \citep[e.g.,][]{Cerruti2017A&A606, MAGIC2020ApJS247, Wang2024ApJS271}.

\begin{figure*}
    \centering
    \includegraphics[width=0.9\hsize]{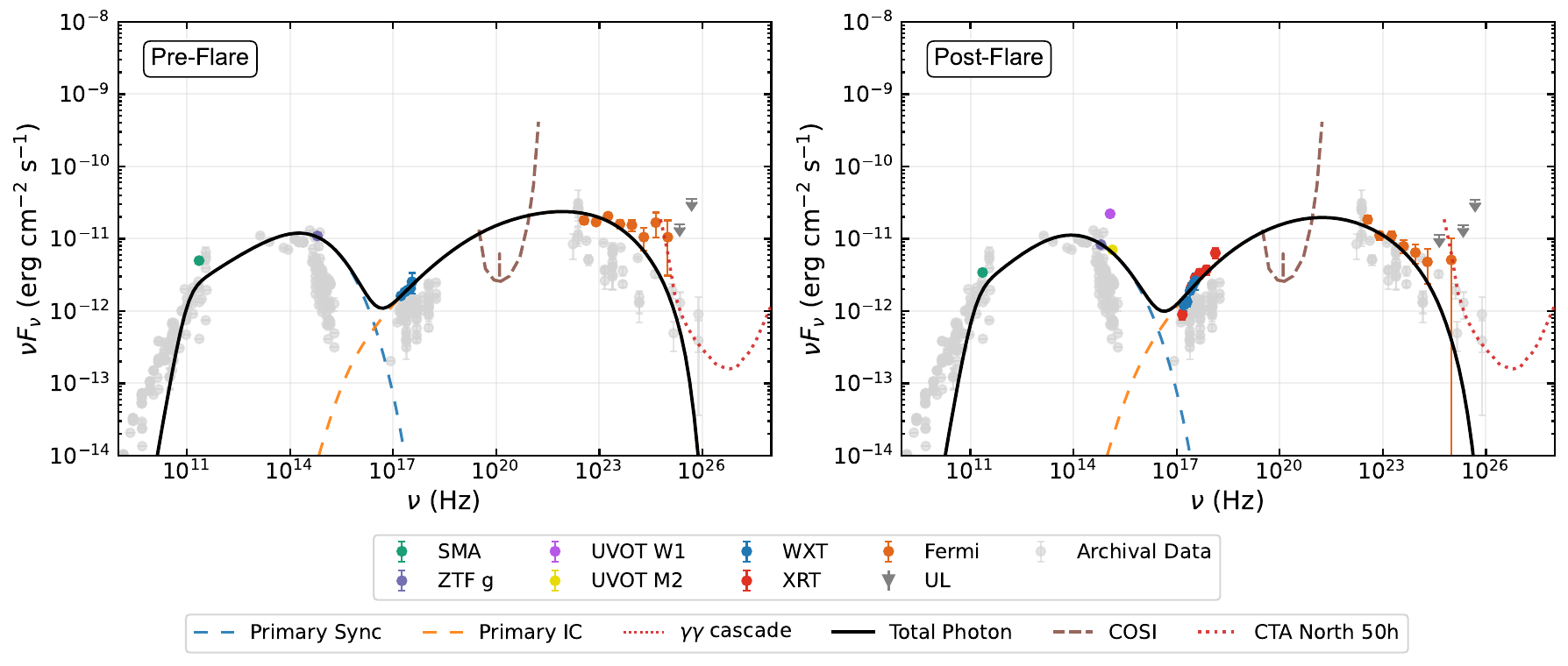}
    \caption{
    Observed spectral energy distribution (SED) modeled with a one-zone pure synchrotron self-Compton (SSC) scenario. 
    The extragalactic background light (EBL) attenuation is treated using the model of \citet{Dominguez2011MNRAS410}. 
    Different colored symbols denote quasi-simultaneous multi-wavelength observations from different telescopes, while the light-gray circles show archival data from the Space Science Data Center (SSDC) SED Builder. 
    Different colored curves represent the individual radiation components included in the model. 
    The blue long-dashed curve represents the synchrotron emission from the primary electrons, while the orange long-dashed curve represents inverse Compton emission from the primary electrons. 
    The red dotted curve shows the emission from secondary electron--positron pairs produced through $\gamma\gamma$ cascades, which is too weak to be clearly visible. 
    The black solid curve represents the total model emission.
    The brown dashed line represents the sensitivity of the Compton Spectrometer and Imager \citep[COSI;][]{Tomsick2022icrcconfE}
    The red long-dotted line represents the Cherenkov Telescope Array Observatory (CTAO) sensitivity in the northern sky for 50 hours. 
    }
    \label{Fig:SED_SSC}%
\end{figure*}

\begin{figure*}
    \centering
    \includegraphics[width=0.9\hsize]{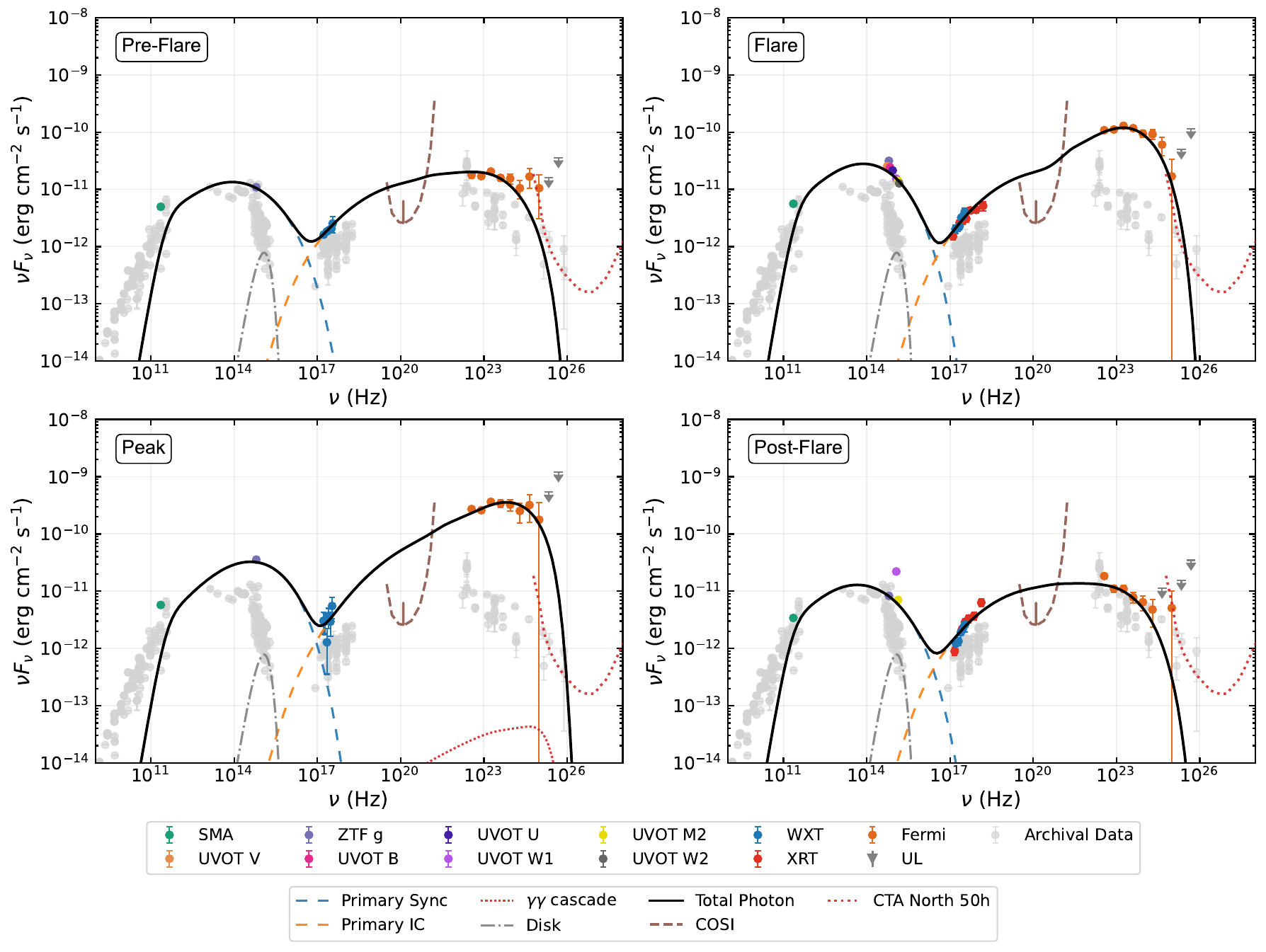}
    \caption{
    Observed spectral energy distribution (SED) modeled with a one-zone leptonic synchrotron self-Compton plus external Compton (SSC+EC) scenario. 
    The attenuation caused by the extragalactic background light (EBL) is treated using the model of \citet{Dominguez2011MNRAS410}. 
    Different colored symbols denote quasi-simultaneous multi-wavelength observations from different telescopes, while the light-gray circles show archival data from the Space Science Data Center (SSDC) SED Builder. 
    Different colored curves represent the individual radiation components included in the model.
    The blue long-dashed curve represents the synchrotron emission from the primary electrons, while the orange long-dashed curve represents inverse Compton emission from the primary electrons, including the SSC and EC components.
    The red dotted curve shows the emission from secondary electron--positron pairs produced through $\gamma\gamma$ cascades, which is too weak to be clearly visible. 
    The gray dash-dotted curve shows the accretion disk emission. 
    The black solid curve represents the total model emission.
    The brown dashed line represents the sensitivity of the COSI \citep{Tomsick2022icrcconfE}
    The red long-dotted line represents the CTAO sensitivity in the northern sky for 50 hours. 
    }
    \label{Fig:SED_SSC+EC}%
\end{figure*}

As mentioned above, the Pre-Flare and Post-Flare periods can be reasonably constrained within a pure SSC framework, whereas the Flare and Peak periods require an additional external photon field. 
We therefore first apply the one-zone SSC model to the two low-activity periods, as shown in Figure~\ref{Fig:SED_SSC}. 
All four periods are then modeled within the SSC+EC framework, as shown in Figure~\ref{Fig:SED_SSC+EC}. 
The corresponding model parameters are listed in Table~\ref{Tab:sed_params_lepton}.

\subsection{Lepto-hadronic modeling}

The association of the blazar TXS~0506+056 with a high-energy neutrino event in 2017 marked the first compelling evidence that blazars can be sources of multi-messenger emission \citep{IceCube2018Sci361}.
This provided the important opportunity to constrain blazar emission using hadronic models. 
Several statistical studies have reported possible spatial or temporal associations between high-energy neutrinos and selected blazar populations \citep[e.g.,][]{Plavin2020ApJ894, Hovatta2021A&A650, Buson2022ApJ933L, Buson2023arXiv230511263}. 
However, the broader association between blazars and neutrinos remains debated, as several studies have shown that the inferred significance is sensitive to the source sample, neutrino localization uncertainties, and statistical methodology \citep[e.g.,][]{Aartsen2017ApJ835, Abbasi2023ApJ954, Bellenghi2023ApJ955L, Abbasi2024ApJ973, Kouch2024A&A690, Lu2025PRD112}.
Motivated by these results, we also explore a lepto-hadronic scenario for S4~0954+65, in order to examine whether the observed broadband SED can be reproduced within a lepto-hadronic framework and to compare the corresponding physical requirements with those inferred from the leptonic model.

In the lepto-hadronic model, relativistic protons interact with the ambient soft photon field in the emission region through the Bethe-Heitler (BH) pair production and photo-pion ($p\gamma$) production processes. 
Here, the target photon field is provided by the internally produced synchrotron and IC radiation and, where relevant, by the external DT photon field. 
Since the efficiency of photohadronic interactions depends sensitively on the density of target photons, the inclusion of an external photon field can provide additional targets for relativistic protons and thus increase the interaction rate. 
As a result, a larger fraction of the proton energy can be converted into secondary particles and radiation, which can reduce the proton power required to reproduce the observed emission \citep{Reimer2019ApJ881}.
The $p\gamma$ process produces pions, which subsequently decay into secondary particles. 
Neutral pions decay into high-energy photons, while charged pions decay into muons and neutrinos, followed by muon decay into electrons/positrons and additional neutrinos. 
The BH process injects secondary electron--positron pairs without producing neutrinos. 
The secondary pairs generated through these processes can subsequently radiate via synchrotron and IC emission. 
All other basic assumptions of the emission region are the same as those adopted in the leptonic model described above.

The primary relativistic protons are also assumed to be accelerated and injected into the blob at a constant rate. 
The injected proton distribution is assumed to follow a power-law with an exponential cutoff (PLEC) as well,
\begin{equation}
    Q_{\rm p}^{\rm inj}(\gamma_{\rm p}) = Q_{\rm p,~0} \gamma_{\rm p}^{-p_{\rm p}}
    \exp\left(-\frac{\gamma_{\rm p}}{\gamma_{\rm p,~max}}\right),
    \qquad \gamma_{\rm p} \geq \gamma_{\rm p,~min},
\end{equation}
where $\gamma_{\rm p,~min/max}$ are the minimum and maximum proton Lorentz factors, $p_{\rm p}$ is the proton spectral index, and $Q_{\rm p,~0}$ is the normalization of the injection function for proton. 
The latter is characterized by the proton injection luminosity $L_{\rm p}^{\rm inj}$ through
$\int \gamma_{\rm p} m_{\rm p} c^{2} Q_{\rm p}^{\rm inj}(\gamma_{\rm p})  \mathrm{d}\gamma_{\rm p} = 3 L_{\rm p}^{\rm inj} / (4 \pi R_{\rm b}^{3})$,
where $m_{\rm p}$ is the proton mass.

The temporal evolution of the proton distribution follows the same continuity equation as in Equation~\eqref{kinetic_equation}, but with the injection and cooling terms appropriate for protons. 
The total proton radiative cooling term is written as
\begin{equation}
    \left| \dot{\gamma}_{\rm p} \right|
    =
    \left| \dot{\gamma}_{\rm p,~{\rm syn}} \right|
    +
    \left| \dot{\gamma}_{\rm p,~{\rm BH}} \right|
    +
    \left| \dot{\gamma}_{{\rm p},~p\gamma} \right|,
\end{equation}
where the individual terms correspond to proton synchrotron cooling, BH pair production, and $p\gamma$ production, respectively.
The proton synchrotron cooling rate is given by \citep[e.g.,][]{Bottcher2012rjag_book}
\begin{equation}
    \left| \dot{\gamma}_{\rm p,~{\rm syn}} \right|
    =
    \frac{4}{3}\frac{\sigma_{\rm T}}{m_{\rm e}c}\gamma_{\rm p}^{2}U_{B}
    \left(\frac{m_{\rm e}}{m_{\rm p}}\right)^{3},
\end{equation}
the BH cooling rate is \citep{Blumenthal1970PRD1, Chodorowski1992ApJ400}
\begin{equation}
    \left| \dot{\gamma}_{\rm p,~{\rm BH}} \right|
    =
    \frac{3}{8\pi}\sigma_{\rm T}\alpha_{f}c\frac{m_{\rm e}}{m_{\rm p}}
    \int_{\epsilon_{r,~{\rm th}}^{\rm BH}}^{\infty}
    \mathrm{d}\epsilon_{r}~
    n_{2}\!\left(\frac{\epsilon_{r}}{2\gamma_{\rm p}}\right)
    \frac{\varphi(\epsilon_{r})}{\epsilon_{r}^{2}},
\end{equation}
and the $p\gamma$ cooling rate is \citep{Begelman1990ApJ362}
\begin{equation}
    \left| \dot{\gamma}_{{\rm p},~p\gamma} \right|
    =
    \frac{c}{2\gamma_{\rm p}}
    \int_{\frac{\epsilon_{r,~{\rm th}}^{p\gamma}}{2\gamma_{\rm p}}}^{\infty}
    \mathrm{d}\epsilon_{2}~
    \frac{n_{2}(\epsilon_{2})}{\epsilon_{2}^{2}}
    \int_{\epsilon_{r,~{\rm th}}^{p\gamma}}^{2\gamma_{\rm p}\epsilon_{2}}
    \mathrm{d}\epsilon_{r}~
    \epsilon_{r}\sigma_{p\gamma}(\epsilon_{r})K_{p\gamma}(\epsilon_{r}).
\end{equation}
Here, $\epsilon_{r}$ is the dimensionless photon energy measured in the proton rest frame, $\epsilon_{r,~{\rm th}}^{\rm BH}\simeq 2$ and $\epsilon_{r,~{\rm th}}^{p\gamma} \simeq 145~{\rm MeV}/(m_{\rm e}c^{2})$ are the corresponding dimensionless threshold photon energies for the BH and $p\gamma$ processes, respectively \citep{Kelner2008PhRvD78}, and $\alpha_{f}\simeq 1/137$ is the fine-structure constant. 
The quantity $n_{2}$ denotes the differential number density of the target soft photon field, $\varphi(\epsilon_{r})$ is given in Equation~(14) of \citet{Blumenthal1970PRD1} and Equation~(3.12) of \citet{Chodorowski1992ApJ400}, $\sigma_{p\gamma}(\epsilon_{r})$ is the cross section of the $p\gamma$ process, and $K_{p\gamma}(\epsilon_{r})$ is the corresponding inelasticity of the collision. 
The values of $\sigma_{p\gamma}(\epsilon_{r})$ and $K_{p\gamma}(\epsilon_{r})$ are taken from the crude fits to the experimental data given by \citet{Begelman1990ApJ362}.

The secondary particles produced through the BH, $p\gamma$, and $\gamma\gamma$ processes are also evolved using the same continuity equation as in Equation~\eqref{kinetic_equation}, with the corresponding injection terms determined by the relevant production rates:
\begin{equation}
    Q_{\rm e, ~sec} (\gamma_{\rm e}, t) = Q^{\gamma\gamma}_{\rm e, ~sec}(\gamma_{\rm e}, t) + Q^{\rm BH}_{\rm e, ~sec}(\gamma_{\rm e}, t) + Q^{p\gamma}_{\rm e, ~sec}(\gamma_{\rm e}, t) .
\end{equation}
The term $Q^{\gamma\gamma}_{\rm e,~sec}$ has been defined above, while $Q^{\rm BH}_{\rm e,~sec}$ and $Q^{p\gamma}_{\rm e,~sec}$ denote the differential injection spectra of secondary pairs from the BH and $p\gamma$ processes, respectively, which are calculated using the analytical expressions of \citet{Kelner2008PhRvD78}.
These secondary pairs mainly cool through synchrotron and IC processes, and their radiative output is included in the total emission.
The photon field is also evolved self-consistently together with the particle distributions, following Equation~\eqref{photon_kinetic_equation}. 
The neutrino emission is produced by the decay of charged pions and muons generated in the $p\gamma$ interactions, and the corresponding production rates are calculated following \citet{Kelner2008PhRvD78}.

Compared with the leptonic model, the lepto-hadronic model introduces additional parameters related to the proton distribution, mainly including the proton injection luminosity, the proton spectral index, and the minimum and maximum proton Lorentz factors. 
In the modeling, we fix $\gamma_{\rm p,~min}=10^{2}$ and adopt $p_{\rm p}=2$. 
The minimum proton Lorentz factor is likewise not constrained by the observed SED, as the predicted emission is essentially insensitive to $\gamma_{\rm p,~min}$. 
This is because the proton-induced radiation is dominated by relativistic protons with Lorentz factors above $\gamma_{\rm p,~min}$.
We therefore fix $\gamma_{\rm p,~min}=10^{2}$, a fiducial value also adopted in previous lepto-hadronic modeling studies \citep[e.g.,][]{Rodrigues2024A&A689, Rodrigues2024A&A681, Rodrigues2026A&A706, Xiao2026ApJ1005}, avoiding the introduction of an unconstrained free parameter.
The value $p_{\rm p}=2$ is consistent with the canonical value expected from Fermi I acceleration, which is one of the most efficient acceleration mechanisms in astrophysical shock waves \citep[e.g.,][]{Longair2011hea_book}.
The maximum proton Lorentz factor is constrained by requiring the acceleration timescale to be shorter than the minimum of the energy-loss and escape timescales. 
The acceleration timescale is \citep[e.g.,][]{Mucke2003AP18}
\begin{equation}
t_{\rm acc}=\alpha\frac{r_{\rm L}}{c}
=\alpha\frac{\gamma_{\rm p}m_{\rm p}c}{eB},
\end{equation}
where $r_{\rm L}$ is the Larmor radius of the relativistic proton and $\alpha \geqslant 1$ is the acceleration efficiency parameter, with $\alpha=1$ corresponding to the Bohm limit, while larger values indicate less efficient particle acceleration. 
For the parameter range considered here, the proton energy-loss timescale is longer than the escape timescale. 
Therefore, the maximum proton Lorentz factor is limited by
\begin{equation}
    \gamma_{\rm p,max} \leqslant \frac{e B R_{\rm b}}{\alpha m_{\rm p} c^{2}}
    \simeq 6.4 \times 10^{7} \left(\frac{\alpha}{50}\right)^{-1} \left(\frac{B}{ 1~{\rm G}}\right) \left(\frac{R_{\rm b}}{ 10^{16}~{\rm cm}}\right).
\end{equation}

\begin{figure*}
    \centering
    \includegraphics[width=0.9\hsize]{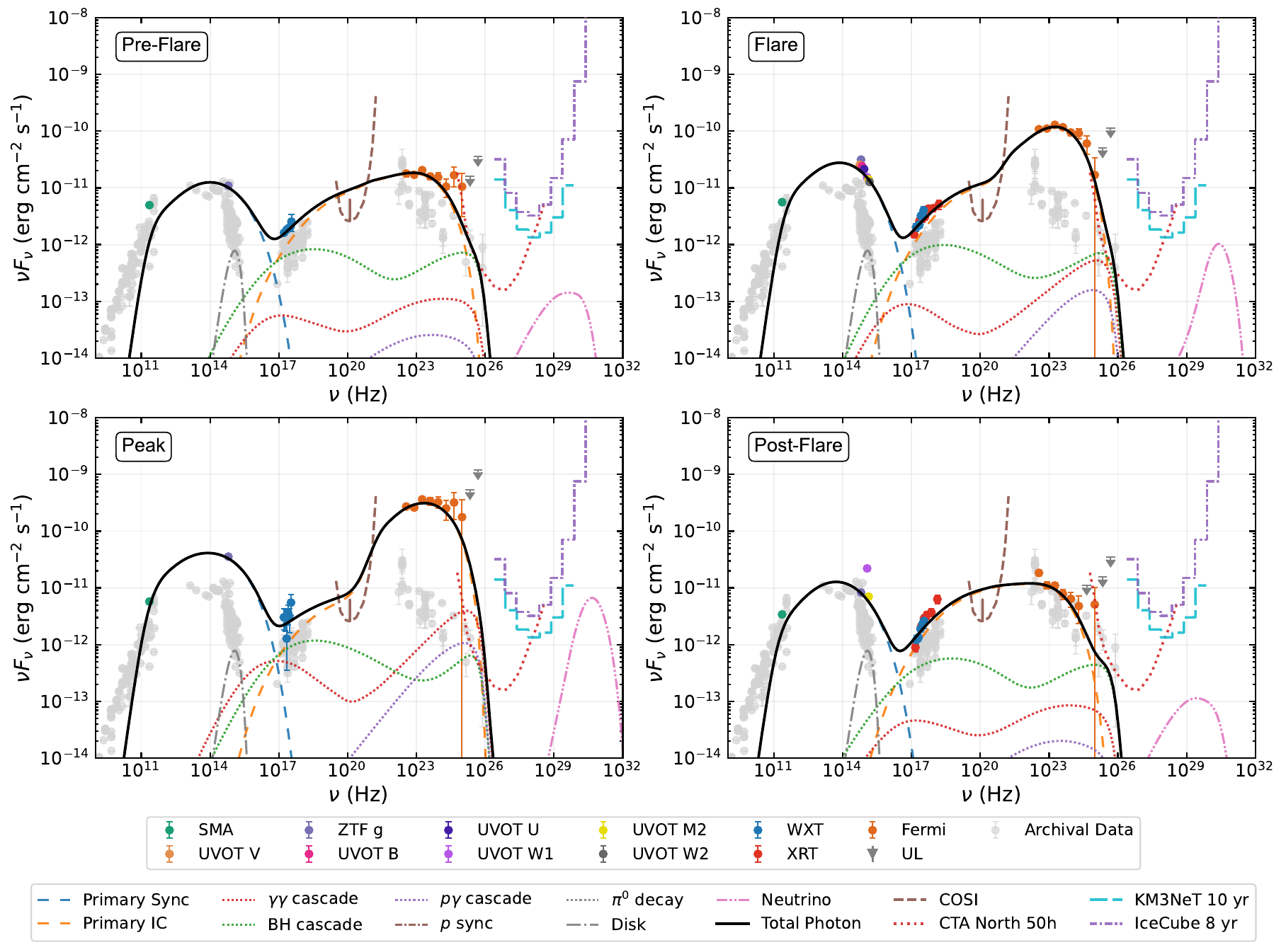}
    \caption{
    Observed spectral energy distributions (SEDs) modeled with a one-zone lepto-hadronic scenario. 
    The attenuation caused by the extragalactic background light (EBL) is treated using the model of \citet{Dominguez2011MNRAS410}. 
    Different colored symbols denote the quasi-simultaneous multi-wavelength observations from different telescopes, while the light-gray circles represent archival data obtained from the Space Science Data Center (SSDC) SED Builder. 
    Different colored curves denote the individual model components. 
    The blue long-dashed curve represents the synchrotron emission from the primary electrons, while the orange long-dashed curve represents inverse Compton emission from the primary electrons, including the SSC and EC components.
    The red dotted curve represents the emission from secondary electron--positron pairs generated through $\gamma\gamma$ cascades. 
    The green dotted curve shows the BH cascade emission, while the purple dotted curve represents the cascade emission associated with $p\gamma$ production. 
    The brown dash-dotted curve denotes the proton synchrotron emission, and the gray dotted curve shows the emission from $\pi^{0}$ decay. 
    The pink dash-dotted curve represents the predicted neutrino emission. 
    The gray dash-dotted curve shows the accretion disk emission. 
    The black solid curve represents the total photon emission.
    The brown dashed line represents the sensitivity of the COSI \citep{Tomsick2022icrcconfE}
    The red long-dotted line represents the CTAO sensitivity in the northern sky for 50 hours. 
    The cyan long-dashed line represents the sensitivity of the Cubic Kilometre Neutrino Telescope (KM3NeT) for 10 years \citep{Aiello2024APh16202990}, and the purple dot-dashed line represents the sensitivity of the IceCube Neutrino Observatory for 8 years \citep{AartsenIceCube2019EPJC79}.
    }
    \label{Fig:SED_SSC+EC_had}%
\end{figure*}

The results of the lepto-hadronic model for four periods are shown in Figure~\ref{Fig:SED_SSC+EC_had}, and the corresponding parameters are listed in Table~\ref{Tab:sed_params_hadron}.

\begin{table*}
\centering
\small
\renewcommand{\arraystretch}{1.4}
\setlength{\tabcolsep}{6pt}
\caption{Parameters of the one-zone leptonic model during different activity periods}
\label{Tab:sed_params_lepton}
\begin{tabular}{lccc|ccccc}
    \hline\hline
    Parameter & Unit
    & \multicolumn{2}{c}{Pure SSC}
    & \multicolumn{5}{c}{SSC+EC} \\
    \cline{3-4} \cline{5-9}
     &
     & Pre-Flare & Post-Flare
     & Pre-Flare & Flare & Peak & Post-Flare & 2015 Feb \\
    \hline

     $B$                       & G
     & 0.06 & 0.17 & 0.47 & 0.32 & 0.23 & 0.45 & 0.34 \\

     $R_{\rm b}$               & cm
     & $7.05\times 10^{16}$
     & $5.30\times 10^{16}$
     & $1.53\times 10^{16}$
     & $1.98\times 10^{16}$
     & $9.29\times 10^{15}$
     & $3.32\times 10^{16}$
     & $8.77\times 10^{15}$ \\

     $\delta$                  &
     & 20.30 & 13.86 & 19.04 & 25.80 & 32.17 & 13.58 & 36.54 \\

     $p_{\rm e}$               &
     & 2.35 & 2.18 & 2.23 & 2.06 & 1.95 & 2.05 & 2.20 \\

     $\gamma_{\rm e,~min}$     &
     & 100 & 100 & 100 & 100 & 100 & 100 & 100 \\

     $\gamma_{\rm e,~max}$     &
     & $2.71\times 10^{4}$
     & $2.22\times 10^{4}$
     & $1.43\times 10^{4}$
     & $7.01\times 10^{3}$
     & $1.36\times 10^{4}$
     & $1.04\times 10^{4}$
     & $1.72\times 10^{4}$ \\

     $L_{\rm e}^{\rm inj}$     & erg~s$^{-1}$
     & $1.70\times 10^{43}$
     & $1.57\times 10^{43}$
     & $4.21\times 10^{42}$
     & $3.95\times 10^{42}$
     & $4.29\times 10^{42}$
     & $8.85\times 10^{42}$
     & $2.73\times 10^{42}$ \\

     $U_{\rm DT,~soft}^{\rm AGN}$ & erg~cm$^{-3}$
     & --
     & --
     & $2.80\times 10^{-5}$
     & $2.80\times 10^{-5}$ 
     & $2.80\times 10^{-5}$ 
     & $2.80\times 10^{-5}$ 
     & $2.80\times 10^{-5}$ \\

    \hline

     $P_{B}$                   & erg~s$^{-1}$
     & $2.61\times 10^{43}$
     & $6.18\times 10^{43}$
     & $7.12\times 10^{43}$
     & $1.01\times 10^{44}$
     & $1.81\times 10^{43}$
     & $1.50\times 10^{44}$
     & $4.51\times 10^{43}$ \\

     $P_{\rm e}$               & erg~s$^{-1}$
     & $5.01\times 10^{45}$
     & $1.83\times 10^{45}$
     & $8.79\times 10^{44}$
     & $1.50\times 10^{45}$
     & $2.47\times 10^{45}$
     & $8.26\times 10^{44}$
     & $2.13\times 10^{45}$ \\

     $P_{\rm p,~cold}$         & erg~s$^{-1}$
     & $3.02\times 10^{46}$
     & $1.07\times 10^{46}$
     & $6.08\times 10^{45}$
     & $9.65\times 10^{45}$
     & $1.23\times 10^{46}$
     & $5.51\times 10^{45}$
     & $1.38\times 10^{46}$ \\

     $P_{\rm rad}$             & erg~s$^{-1}$
     & $9.94\times 10^{43}$
     & $1.76\times 10^{44}$
     & $1.09\times 10^{44}$
     & $1.95\times 10^{44}$
     & $3.47\times 10^{44}$
     & $1.67\times 10^{44}$
     & $2.53\times 10^{44}$ \\

     $P_{\rm tot}$             & erg~s$^{-1}$
     & $3.53\times 10^{46}$
     & $1.28\times 10^{46}$
     & $7.14\times 10^{45}$
     & $1.15\times 10^{46}$
     & $1.51\times 10^{46}$
     & $6.65\times 10^{45}$
     & $1.62\times 10^{46}$ \\

    \hline
\end{tabular}
\end{table*}

\begin{table*}
\centering
\renewcommand{\arraystretch}{1.4}
\setlength{\tabcolsep}{9pt}
\caption{Parameters of the one-zone lepto-hadronic model during different activity periods}
\label{Tab:sed_params_hadron}
\begin{tabular}{lccccc}
    \hline\hline
    Parameter & Unit & Pre-Flare & Flare & Peak & Post-Flare \\
    \hline

     $B$                       & G
     & 0.43 & 0.34 & 0.30 & 0.42 \\

     $R_{\rm b}$               & cm
     & $2.26\times 10^{16}$
     & $2.20\times 10^{16}$
     & $3.72\times 10^{16}$
     & $4.55\times 10^{16}$ \\

     $\delta$                  &
     & 17.34 & 27.57 & 32.87 & 12.51 \\

     $p_{\rm e}$               &
     & 2.08 & 2.01 & 1.97 & 1.88 \\

     $\gamma_{\rm e,~min}$     &
     & 100 & 100 & 100 & 100 \\

     $\gamma_{\rm e,~max}$     &
     & $1.31\times 10^{4}$
     & $6.06\times 10^{3}$
     & $8.12\times 10^{3}$
     & $8.75\times 10^{3}$ \\

     $L_{\rm e}^{\rm inj}$     & erg~s$^{-1}$
     & $4.12\times 10^{42}$
     & $2.53\times 10^{42}$
     & $1.99\times 10^{42}$
     & $9.14\times 10^{42}$ \\

     $p_{\rm p}$               &
     & 2.00 & 2.00 & 2.00 & 2.00 \\

     $\gamma_{\rm p,~min}$     &
     & 100 & 100 & 100 & 100 \\

     $\gamma_{\rm p,~max}$     &
     & $2.41\times 10^{6}$
     & $3.81\times 10^{6}$
     & $9.37\times 10^{6}$
     & $3.26\times 10^{6}$ \\

     $L_{\rm p}^{\rm inj}$     & erg~s$^{-1}$
     & $2.42\times 10^{46}$
     & $2.12\times 10^{45}$
     & $4.28\times 10^{44}$
     & $3.95\times 10^{46}$ \\

     $U_{\rm DT,~soft}^{\rm AGN}$ & erg~cm$^{-3}$
     & $2.80\times 10^{-5}$
     & $2.80\times 10^{-5}$
     & $2.80\times 10^{-5}$
     & $2.80\times 10^{-5}$ \\

    \hline

     $P_{B}$                   & erg~s$^{-1}$
     & $1.05\times 10^{44}$
     & $1.60\times 10^{44}$
     & $5.00\times 10^{44}$
     & $2.10\times 10^{44}$ \\

     $P_{\rm e}$               & erg~s$^{-1}$
     & $6.70\times 10^{44}$
     & $1.07\times 10^{45}$
     & $9.80\times 10^{44}$
     & $6.68\times 10^{44}$ \\

     $P_{\rm p}$               & erg~s$^{-1}$
     & $5.42\times 10^{48}$
     & $1.20\times 10^{48}$
     & $3.45\times 10^{47}$
     & $4.59\times 10^{48}$ \\

     $P_{\rm rad}$             & erg~s$^{-1}$
     & $1.15\times 10^{44}$
     & $1.59\times 10^{44}$
     & $2.69\times 10^{44}$
     & $1.77\times 10^{44}$ \\

     $P_{\rm tot}$             & erg~s$^{-1}$
     & $5.43\times 10^{48}$
     & $1.20\times 10^{48}$
     & $3.47\times 10^{47}$
     & $4.59\times 10^{48}$ \\

    \hline
\end{tabular}
\end{table*}

\subsection{Jet power calculation}
\label{jet_power_calculation}

The physical parameters obtained from the SED modeling can be further used to estimate the power carried by different components of the relativistic jet. 
For a spherical emission region with radius $R_{\rm b}$ moving with bulk Lorentz factor $\Gamma$, the jet power associated with component $i$ can be written as \citep[e.g.,][]{Ghisellini2009MNRAS399, Ghisellini2010MNRAS402, Celotti2008MNRAS385}
\begin{equation}
    P_i = \pi R_{\rm b}^{2} \Gamma^{2} c U_i ,
    \label{eq:jet_power_general}
\end{equation}
where $U_i$ is the corresponding energy density measured in the jet co-moving frame.
In the leptonic scenario, we consider the power carried by the magnetic field, relativistic electrons, cold protons, and radiation.
In the lepto-hadronic scenario, we consider the power carried by the magnetic field, relativistic electrons, relativistic protons, and radiation.
The magnetic energy density is calculated using the expression defined in the previous section.
The energy densities of relativistic electrons and protons are calculated from their steady-state particle distributions as
\begin{equation}
    U_{\rm e} = m_{\rm e} c^2 \int_{\gamma_{\rm e,~min}}^{\gamma_{\rm e,~max}} 
    \gamma_{\rm e} N_{\rm e}(\gamma_{\rm e}) \mathrm{d}\gamma_{\rm e},
\end{equation}
and
\begin{equation}
    U_{\rm p} = m_{\rm p} c^2 \int_{\gamma_{\rm p,~min}}^{\gamma_{\rm p,~max}} 
    \gamma_{\rm p} N_{\rm p}(\gamma_{\rm p}) \mathrm{d}\gamma_{\rm p}.
\end{equation}
For the leptonic model, the cold proton power is estimated by assuming one cold proton per electron.
The radiative energy density is estimated from the observed bolometric luminosity as
\begin{equation}
    U_{\rm r} = \frac{L_{\rm bol}^{\rm obs}}{ 4 \pi R_{\rm b}^{2} c \delta^4}.
\end{equation}
The jet powers for different activity periods are calculated and listed in Table~\ref{Tab:sed_params_lepton} and Table~\ref{Tab:sed_params_hadron}.

\section{Discussion}

The observational and modeling results together provide a coherent picture for the physical origin of the flare activity in S4~0954+65. 
The source exhibits a pronounced multi-wavelength outburst during MJD $\sim$60695--60732, with a strong variability occurring in the $\gamma$-ray band. 
The flare peak reaches a flux of $\sim 2 \times 10^{-6}$ ph cm$^{-2}$ s$^{-1}$ at MJD $\sim$60722--60724, while contemporaneous activity is also present in the X-ray, optical/UV, and radio bands.

The zDCF analysis places direct constraints on the relative timing among different wavebands. 
No significant zDCF peak is detected between the $\gamma$-ray and X-ray light curves, indicating that the correlation does not exist between these bands.
By contrast, the $\gamma$-ray leads the optical ones by 3.22 days with 3.09$\sigma$, while a possible lead of 18.26 days over the 230~GHz radio variations is detected with a significance of 2.17$\sigma$.
The short optical lag suggests that the optical emission region is closely connected to the high-energy dissipation zone.
The possible radio delay is qualitatively consistent with emission from a more extended and partially optically thick downstream region, where opacity and propagation effects become important, but the present sampling does not provide a precise measurement of this delay \citep[e.g.,][]{Chatterjee2008ApJ689, MaxMoerbeck2014MNRAS445}.
Such time lags can also be interpreted in the framework of shock propagation along the jet. 
In this scenario, relativistic electrons are accelerated at the shock front and then cool through synchrotron and IC processes \citep{Marscher1985ApJ298}. 
In the presence of the external soft photon field provided by the DT, the accelerated high-energy electrons can lose energy more efficiently through the EC process than through the SSC process, leading to the early production of $\gamma$-ray emission.
As the electrons cool to lower energies, synchrotron radiation in the optical band becomes more important, which can naturally account for the observed lead of the $\gamma$-ray emission over the optical emission. 
As the shock propagates further downstream and the emission region expands, opacity decreases, and the region gradually becomes optically thin at radio frequencies. 
This provides a possible explanation for delayed SMA 230~GHz emission, although denser radio sampling would be required to test this interpretation and determine the lag precisely.

This propagation picture is further supported by the parsec-scale jet evolution revealed by the 43~GHz VLBA data.
The emergence of the moving knot K25 is temporally consistent, within uncertainties, with the epochs of the major $\gamma$-ray and optical flare peaks, as listed in Table~\ref{tab:radio_knots}. 
At the same time, the flux density of the radio core increases around the emergence of this component. 
These results link the high-energy flare to the propagation of a newly ejected disturbance in the inner jet, as commonly found in VLBI studies of flaring blazars \citep[e.g.,][]{Marscher2010ApJ710, MAGIC2014A&A569, Jorstad2017ApJ846, Weaver2022ApJS260, Kim2022ApJ925}.

The $\gamma$-ray spectral evolution provides evidence for changes in the underlying particle distribution during the outburst. 
From the Pre-Flare to the Flare and Peak periods, the spectrum hardens, followed by notable softening in the Post-Flare period. 
This behavior is consistent with a ``harder-when-brighter'' trend, which is commonly observed in blazars \citep[e.g.,][]{Kraub2016A&A591, Middei2022MNRAS514, Wang2024ApJS270}. 
Such a trend is generally interpreted as evidence for enhanced particle acceleration during high-activity states, where the acceleration timescale becomes comparable to or shorter than the radiative cooling timescale \citep[e.g.,][]{Kirk1998A&A333, Sokolov2004ApJ613, Taylor2026A&A706}.
The X-ray fractional variability is mild, no significant $\gamma$-ray--X-ray zDCF correlation is detected, and the photon index uncertainties do not support a definite temporal softening after the flare.
However, the significant anti-correlation between the X-ray photon index and flux nevertheless suggests that the X-ray spectrum tends to be harder when the source is brighter.
This trend is likely consistent with enhanced particle acceleration, similar to that inferred from the $\gamma$-ray spectral evolution.

Taken together, the multi-wavelength time lags, spectral hardening during the flare, X-ray spectral evolution, and temporal association between the flare and K25 favor a shock-in-jet interpretation. 
In this framework, K25 may trace the propagation of shocked plasma or a newly ejected disturbance in the parsec-scale jet, where efficient particle acceleration and magnetic field compression produce the observed broadband flare \citep{Marscher1985ApJ298, Marscher2010ApJ710, Jorstad2017ApJ846}.

The SED modeling constrains the origin of the high-energy emission. 
A simple one-zone SSC interpretation requires highly atypical values of the magnetic field strength and Doppler factor during the Flare and Peak periods. 
Although the Pre-Flare and Post-Flare SEDs can be formally reproduced by a pure SSC model, as shown in Figure~\ref{Fig:SED_SSC}, the inferred jet powers are higher than those obtained in the SSC+EC scenario (see Table~\ref{Tab:sed_params_lepton}). 
This suggests that even in these lower-activity states, the pure SSC model is less favored from an energetic point of view. 
Therefore, a pure one-zone SSC scenario cannot satisfactorily account for the observed high-energy component.
When an EC component is included, with external soft photons supplied by the DT, the broadband SEDs can be reproduced with more physically plausible parameters. 
Moreover, the total jet powers inferred in the leptonic models remain below the Eddington luminosity, $L_{\rm Edd}\simeq3\times10^{46}~{\rm erg~s^{-1}}$, estimated using the black hole mass of $M_{\rm BH}=2.3\times10^{8}~M_{\odot}$ \citep{BecerraGonzalez2021MNRAS504}. 
This interpretation is supported by the detection of the Mg~II emission line and by the inferred disk luminosity, both of which indicate the presence of a non-negligible external radiation field in this source \citep{BecerraGonzalez2021MNRAS504}. 
An important point is that the visibility of the emission lines in S4~0954+65 appears to depend strongly on the optical state. 
\citet{BecerraGonzalez2021MNRAS504} showed that spectral features, including the broad Mg~II line, are detectable when the source is in a relatively low optical state, whereas the spectrum becomes nearly featureless during high states. 
This state dependence is likely caused by dilution of the line equivalent widths by the strongly beamed non-thermal jet continuum, rather than by the disappearance of the BLR. 
Thus, the BL~Lac classification of S4~0954+65 may be partly affected by continuum dilution, supporting its transitional nature between BL~Lac objects and FSRQs. 
This provides an important physical basis for including an external photon field in the SED modeling.
It is also consistent with previous modeling studies of S4~0954+65, which found that an EC component is required to reproduce the high-energy emission \citep{Tanaka2016PASJ68, MAGIC2018A&A617}. 
Furthermore, several IBLs, including W~Comae \citep{Acciari2009ApJ707}, 3C~66A \citep{Abdo2011ApJ726}, S5~0716+714 \citep{Rani2013A&A552, Ouyang2025ApJ980}, and PKS~0903-57 \citep{Shah2021MNRAS504}, have also been found to require an EC component for a better description of their broadband emission.

We also explored a lepto-hadronic interpretation of the broadband SED. 
In this framework, the high-energy emission is related to proton-induced processes, including proton synchrotron radiation, BH pair production, $p\gamma$ production, and the associated electromagnetic cascades. 
Such models are particularly relevant in the context of multi-messenger astrophysics, especially after the association of TXS~0506+056 with high-energy neutrino emission \citep{IceCube2018Sci361, Gao2019NatureAstr3, Cerruti2019MNRAS483, Cerruti2021MNRAS502}. 
Polarization measurements provide an additional diagnostic for distinguishing leptonic and hadronic scenarios, since synchrotron radiation from primary or secondary particles can produce higher polarization than IC-dominated emission, especially EC emission from nearly unpolarized external seed photons \citep{Zhang2013ApJ774, Peirson2019ApJ885, Krawczynski2012ApJ744}. 
For S4~0954+65, IXPE observations have provided useful constraints on the X-ray polarization during an optical--X-ray flare \citep{Kouch2025A&A695}. 
Motivated by this diagnostic potential, we tested a lepto-hadronic model, as shown in Figure~\ref{Fig:SED_SSC+EC_had}, in which synchrotron emission from secondary pairs contributes to the X-ray band. 
We also considered a case in which proton synchrotron emission dominates the X-ray band, but this scenario fails to reproduce the low-energy emission because it requires a high magnetic field. 
However, the lepto-hadronic scenario generally requires substantially higher proton power and more extreme jet energetics than the leptonic scenario, as shown in Table~\ref{Tab:sed_params_hadron}.
This is consistent with previous studies \citep[e.g.,][]{Bottcher2013ApJ768, Petropoulou2015MNRAS448, Cerruti2019MNRAS483}.
It is also consistent with the recent work of \citet{Kouch2025A&A695}, who combined multi-wavelength polarization observations with SED modeling and found that hadronic models for the X-ray emission are tentatively disfavored.
However, those polarization observations correspond to a different activity epoch and therefore do not directly discriminate between the models for the 2025 flare studied here.
Similarly, the absence of a firm association between S4~0954+65 and a high-energy neutrino event \citep[e.g.,][]{Giommi2020MNRAS497, Buson2023arXiv230511263, Kouch2026A&A708_A383} does not by itself constrain the radiation mechanism of an individual flare.
Consequently, our SED modeling does not rule out the lepto-hadronic framework.
It only shows that, under the assumptions of the present one-zone models, the SSC+EC interpretation reproduces the broadband emission with physically reasonable parameters.
Contemporaneous multi-wavelength polarization measurements and additional multi-messenger constraints would be needed to distinguish the emission scenarios more accurately.

We found that, in our results, the models underpredict several $\gamma$-ray data points during the Pre-Flare period in Figures~\ref{Fig:SED_SSC} and \ref{Fig:SED_SSC+EC}.
The Pre-Flare interval contains moderate $\gamma$-ray enhancements around MJD 60655 and 60660, as shown in Figure~\ref{Fig:LC_J0958}. 
Since the $\gamma$-ray spectrum is averaged over the full interval, these short-term enhancements may increase the interval-averaged $\gamma$-ray flux and elevate the resulting spectrum.
This may explain why the one-zone models underpredict several high-energy data points.

Within the favored leptonic SSC+EC framework, the fitted parameters provide further clues to the energy conversion during the flare. 
The magnetic field strength remains moderate, with $B\simeq0.23$--$0.32$~G during the Flare and Peak periods, while the Doppler factor increases to $\delta\simeq25.80$--$32.17$. 
At the same time, the relativistic electron spectrum hardens to $p_{\rm e}=1.95$ in the Peak period, consistent with more efficient particle acceleration at the highest activity state. 
The magnetic power is lower than the power carried by relativistic electrons and cold protons, indicating that the emission region is particle dominated rather than magnetically dominated. 
This energy partition is consistent with a scenario in which a propagating shock or disturbance converts a fraction of the bulk kinetic energy of the jet into non-thermal electrons, which then radiate efficiently through synchrotron and IC processes \citep{Marscher1985ApJ298, Kirk1998A&A333, Sokolov2004ApJ613}. 
The observed $\gamma$-ray flare is therefore most likely driven by enhanced particle acceleration and an increased Doppler boosting of the EC emission.

\begin{figure}
    \centering
    \includegraphics[width=0.98\hsize]{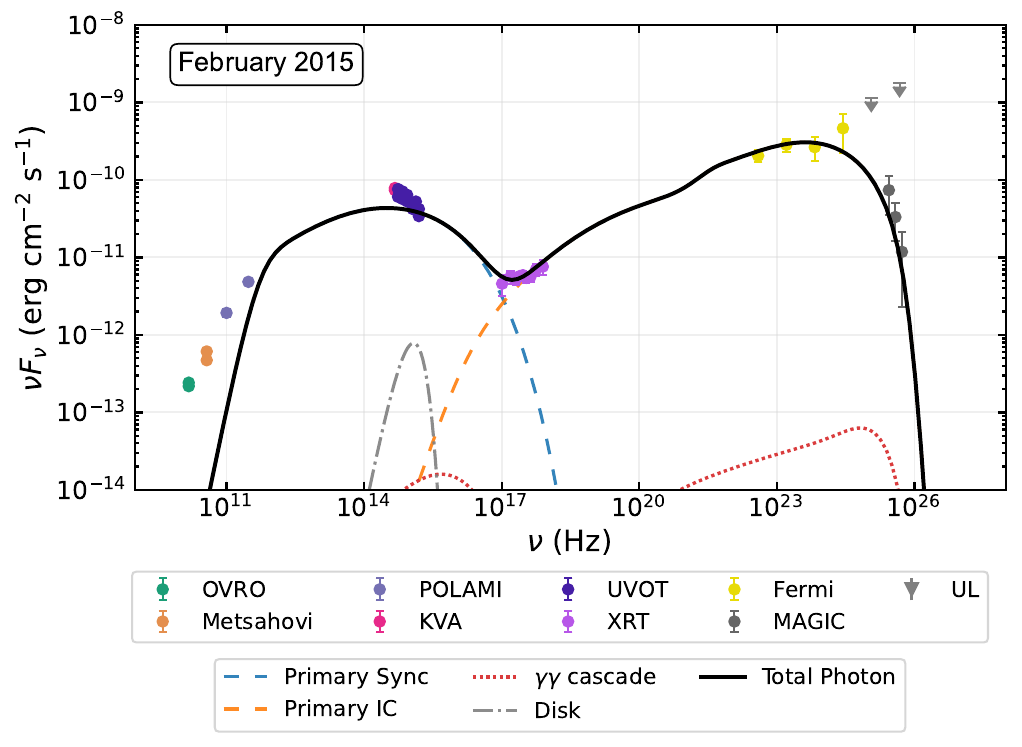}
    \caption{
    Observed spectral energy distribution (SED) for the flaring activity in February 2015, modeled with a one-zone leptonic scenario. 
    The multi-wavelength observational data were obtained from \citet{MAGIC2018A&A617}.
    The attenuation caused by the extragalactic background light (EBL) is treated using the model of \citet{Dominguez2011MNRAS410}. 
    Different colored symbols denote the multi-wavelength observational data from different telescopes, while gray inverted triangles represent upper limits. 
    The line styles and model-component legends are the same as those in Figure~\ref{Fig:SED_SSC+EC}.
    }
    \label{Fig:SED_2015flare}%
\end{figure}

It is also useful to compare the present flaring activity of 2025 with the major outburst observed in February 2015, since the two events reached comparable $\gamma$-ray flux levels \citep{MAGIC2018A&A617}. 
We therefore examined whether similar SED parameters can also reproduce the 2015 flaring state within the same modeling framework. 
As shown in Figure~\ref{Fig:SED_2015flare}, the 2015 SED can be described within the same leptonic SSC+EC framework adopted for the present flare. 
The best-fit parameters are broadly comparable to those obtained for the Peak period in the current activity period.
The corresponding parameter values are provided in Table~\ref{Tab:sed_params_lepton}.
In particular, the 2015 leptonic model gives $B=0.34$~G and $\delta=36.54$, close to the corresponding Peak-period values of $B=0.23$~G and $\delta=32.17$. 
The inferred total jet power, $P_{\rm tot}=1.62\times10^{46}~{\rm erg~s^{-1}}$, is also comparable to that of the present Peak period, $P_{\rm tot}=1.51\times10^{46}~{\rm erg~s^{-1}}$, with a difference of only $\sim$0.03 dex. 
This consistency raises the question of whether the 2015 and 2025 flares represent recurrent episodes of energy dissipation in the jet, in which similar physical conditions and jet power budgets can account for both high-activity episodes.

\section{Conclusion}

In this work, we performed a detailed multi-wavelength study of the blazar S4~0954+65 during MJD 60593--60857, combining $\gamma$-ray, X-ray, optical/UV, radio, and 43~GHz VLBA observations. 
Our main results are summarized as follows.
\begin{enumerate}
    \item The source exhibited a pronounced broadband flaring episode during MJD $\sim$60695--60732, with the strongest activity observed in the $\gamma$-ray band. 
    The flare reached a peak flux of $\sim 2 \times 10^{-6}$ ph cm$^{-2}$ s$^{-1}$ at MJD $\sim$60722--60724.

    \item The zDCF analysis detected no significant correlation between the $\gamma$-ray and X-ray light curves.
    The $\gamma$-ray variations lead the optical ones by 3.22 days with $3.09\sigma$, and indicate a possible lead of 18.26 days over the 230~GHz radio ones with $2.17\sigma$.
    Because the SMA light curve has a substantial gap around the flare, the radio lag may be overestimated and cannot be considered a precise measurement.
    These results favor a compact upstream origin for the high-energy emission, followed by delayed emission at progressively lower frequencies downstream in the jet.

    \item The $\gamma$-ray spectral behaviors show hardening during the active state and softening after the flare, following a harder-when-brighter trend, and are consistent with enhanced particle acceleration during the active state.
    Likewise, the anti-correlation between the X-ray photon index and flux suggests a similar harder-when-brighter behavior.

    \item The 43~GHz VLBA data revealed clear structural evolution in the inner jet.
    The extrapolated ejection time of K25 is consistent, within uncertainties, with the major $\gamma$-ray flare. 
    Together with the multi-band time lags and spectral hardening, this connection may favor a shock-in-jet picture in which a newly emerging disturbance propagates downstream and drives the broadband flare.

    \item A pure one-zone SSC model is insufficient to explain the broadband SEDs with physically reasonable parameters. 
    In our one-zone framework, the leptonic SSC+EC model provides an energetically more economical interpretation, while the lepto-hadronic scenario requires much larger proton power and more demanding jet energetics. 
    However, the lepto-hadronic model cannot be excluded, and contemporaneous polarization or additional multi-messenger constraints are required for a more robust distinction.

\end{enumerate}

\begin{acknowledgements}
The authors thank Yuan Liu and Chichuan Jin for helpful discussions on WXT data processing and flares from blazars. \\
H.B.X. acknowledges the support from National Natural Science Foundation of China (NSFC, grant No. 12203034), the Shanghai Science and Technology Fund (grant No. 22YF1431500), the China Manned Space Project (grant No. CMS-CSST-2025-A07), and the Shanghai Municipal Education Commission regarding artificial intelligence empowered research.
This work is supported by the National Natural Science Foundation of China (NSFC) grants 12261141691.
S.H.Z. acknowledges the support from the National Natural Science Foundation of China (grant Nos. 12173026, 12141302), the Program for Professor of Special Appointment (Eastern Scholar) at Shanghai Institutions of Higher Learning, and from the Shuguang Program (23SG39) of Shanghai Education Development Foundation and Shanghai Municipal Education Commission.
Z.L. acknowledges the support from the National Natural Science Foundation of China (Grant No. 12573009) and the scientific research grants from the China Manned Space Project with Grant No. CMS-CSST-2025-A07 and CMS-CSST-2025-A05.
JHFan’s work is partially supported by the National Natural Science Foundation of China (grant Nos. 12433004) and the National Key Research and Development Program of China, Grant No. 2025YFA1614102.\\

This work is based on the data obtained with the \emph{Einstein Probe} (\emph{EP}), also known as “\emph{Tianguan}”. \emph{EP} is led by the Chinese Academy of Sciences, in collaboration with the European Space Agency, the Max Planck Institute for Extraterrestrial Physics (Germany), and the Centre National d'Études Spatiales (France). This work is supported by the Strategic Priority Research Program of the Chinese Academy of Sciences (Grant No.XDB0550200) and the National Key R\&D Program of China No. 2025YFF0511100.
\\

This work made use of Swift-XRT data supplied by the UK Swift Science Data Centre at the University of Leicester. \\

Based on observations obtained with the Samuel Oschin Telescope 48-inch and the 60-inch Telescope at the Palomar
Observatory as part of the Zwicky Transient Facility project. 
ZTF is supported by the National Science Foundation under Grants No. AST-1440341 and AST-2034437 and a collaboration including current partners Caltech, IPAC, the Oskar Klein Center at Stockholm University, the University of Maryland, University of California, Berkeley, the University of Wisconsin at Milwaukee, University of Warwick, Ruhr University, Cornell University, Northwestern University and Drexel University. 
Operations are conducted by COO, IPAC, and UW. \\

This study makes use of VLBA data from the VLBA-BU Blazar Monitoring Program (BEAM-ME and VLBA-BU-BLAZAR; http://www.bu.edu/blazars/BEAM-ME.html), funded by NASA through the Fermi Guest Investigator Program. 
The VLBA is an instrument of the National Radio Astronomy Observatory.
The National Radio Astronomy Observatory is a facility of the National Science Foundation operated by Associated Universities, Inc.

\end{acknowledgements}

%

\bibliographystyle{aa}
\bibliography{Ref}

@ARTICLE{Xiao2026ApJ1005,
       author = {{Xiao}, Hubing and {Ouyang}, Zhihao and {Yang}, Lili and {Zhu}, Jingtian and {Gu}, Minfeng and {Chen}, Liang and {Zhang}, Shaohua and {Luo}, Zhijian and {Fan}, Junhui},
        title = "{Chasing the Neutrino Blazar Candidates. II. SED Modeling with a Hadronic Model}",
      journal = {\apj},
         year = 2026,
        month = jul,
       volume = {1005},
       number = {1},
          eid = {30},
        pages = {30},
          doi = {10.3847/1538-4357/ae66e9},
archivePrefix = {arXiv},
       eprint = {2605.08775},
 primaryClass = {astro-ph.HE},
       adsurl = {https://ui.adsabs.harvard.edu/abs/2026ApJ..1005...30X}
}

@ARTICLE{Rodrigues2024A&A681,
       author = {{Rodrigues}, X. and {Paliya}, V.~S. and {Garrappa}, S. and {Omeliukh}, A. and {Franckowiak}, A. and {Winter}, W.},
        title = "{Leptohadronic multi-messenger modeling of 324 gamma-ray blazars}",
      journal = {\aap},
         year = 2024,
        month = jan,
       volume = {681},
          eid = {A119},
        pages = {A119},
          doi = {10.1051/0004-6361/202347540},
archivePrefix = {arXiv},
       eprint = {2307.13024},
 primaryClass = {astro-ph.HE},
       adsurl = {https://ui.adsabs.harvard.edu/abs/2024A&A...681A.119R}
}

@ARTICLE{Rodrigues2026A&A706,
       author = {{Rodrigues}, X. and {Rieger}, F. and {Bohdan}, A. and {Padovani}, P.},
        title = "{Hillas meets Eddington: The case for blazars as ultra-high-energy neutrino sources}",
      journal = {\aap},
         year = 2026,
        month = feb,
       volume = {706},
          eid = {A351},
        pages = {A351},
          doi = {10.1051/0004-6361/202556986},
archivePrefix = {arXiv},
       eprint = {2508.18345},
 primaryClass = {astro-ph.HE},
       adsurl = {https://ui.adsabs.harvard.edu/abs/2026A&A...706A.351R}
}

@ARTICLE{Rodrigues2024A&A689,
       author = {{Rodrigues}, X. and {Karl}, M. and {Padovani}, P. and {Giommi}, P. and {Paiano}, S. and {Falomo}, R. and {Petropoulou}, M. and {Oikonomou}, F.},
        title = "{The Spectra of IceCube Neutrino (SIN) candidate sources: V. Modeling and interpretation of multiwavelength and neutrino data}",
      journal = {\aap},
         year = 2024,
        month = sep,
       volume = {689},
          eid = {A147},
        pages = {A147},
          doi = {10.1051/0004-6361/202450592},
archivePrefix = {arXiv},
       eprint = {2406.06667},
 primaryClass = {astro-ph.HE},
       adsurl = {https://ui.adsabs.harvard.edu/abs/2024A&A...689A.147R}
}

@ARTICLE{MAGIC2020ApJS247,
       author = {{Acciari}, V.~A. and {Ansoldi}, S. and {Antonelli}, L.~A. and {Engels}, A. Arbet and {Asano}, K. and {Baack}, D. and {Babi{\'c}}, A. and {Banerjee}, B. and {Barres de Almeida}, U. and {Barrio}, J.~A. and {Becerra Gonz{\'a}lez}, J. and {Bednarek}, W. and {Bellizzi}, L. and {Bernardini}, E. and {Berti}, A. and {Besenrieder}, J. and {Bhattacharyya}, W. and {Bigongiari}, C. and {Biland}, A. and {Blanch}, O. and {Bonnoli}, G. and {Bo{\v{s}}njak}, {\v{Z}}. and {Busetto}, G. and {Carosi}, R. and {Ceribella}, G. and {Cerruti}, M. and {Chai}, Y. and {Chilingaryan}, A. and {Cikota}, S. and {Colak}, S.~M. and {Colin}, U. and {Colombo}, E. and {Contreras}, J.~L. and {Cortina}, J. and {Covino}, S. and {D'Elia}, V. and {Da Vela}, P. and {Dazzi}, F. and {De Angelis}, A. and {De Lotto}, B. and {Delfino}, M. and {Delgado}, J. and {Depaoli}, D. and {Di Pierro}, F. and {Di Venere}, L. and {Do Souto Espi{\~n}eira}, E. and {Dominis Prester}, D. and {Donini}, A. and {Dorner}, D. and {Doro}, M. and {Elsaesser}, D. and {Ramazani}, V. Fallah and {Fattorini}, A. and {Ferrara}, G. and {Fidalgo}, D. and {Foffano}, L. and {Fonseca}, M.~V. and {Font}, L. and {Fruck}, C. and {Fukami}, S. and {Garc{\'\i}a L{\'o}pez}, R.~J. and {Garczarczyk}, M. and {Gasparyan}, S. and {Gaug}, M. and {Giglietto}, N. and {Giordano}, F. and {Godinovi{\'c}}, N. and {Green}, D. and {Guberman}, D. and {Hadasch}, D. and {Hahn}, A. and {Herrera}, J. and {Hoang}, J. and {Hrupec}, D. and {H{\"u}tten}, M. and {Inada}, T. and {Inoue}, S. and {Ishio}, K. and {Iwamura}, Y. and {Jouvin}, L. and {Kerszberg}, D. and {Kubo}, H. and {Kushida}, J. and {Lamastra}, A. and {Lelas}, D. and {Leone}, F. and {Lindfors}, E. and {Lombardi}, S. and {Longo}, F. and {L{\'o}pez}, M. and {L{\'o}pez-Coto}, R. and {L{\'o}pez-Oramas}, A. and {Loporchio}, S. and {Machado de Oliveira Fraga}, B. and {Maggio}, C. and {Majumdar}, P. and {Makariev}, M. and {Mallamaci}, M. and {Maneva}, G. and {Manganaro}, M. and {Mannheim}, K. and {Maraschi}, L. and {Mariotti}, M. and {Mart{\'\i}nez}, M. and {Mazin}, D. and {Mi{\'c}anovi{\'c}}, S. and {Miceli}, D. and {Minev}, M. and {Miranda}, J.~M. and {Mirzoyan}, R. and {Molina}, E. and {Moralejo}, A. and {Morcuende}, D. and {Moreno}, V. and {Moretti}, E. and {Munar-Adrover}, P. and {Neustroev}, V. and {Nigro}, C. and {Nilsson}, K. and {Ninci}, D. and {Nishijima}, K. and {Noda}, K. and {Nogu{\'e}s}, L. and {Nozaki}, S. and {Paiano}, S. and {Palatiello}, M. and {Paneque}, D. and {Paoletti}, R. and {Paredes}, J.~M. and {Pe{\~n}il}, P. and {Peresano}, M. and {Persic}, M. and {Prada Moroni}, P.~G. and {Prandini}, E. and {Puljak}, I. and {Rhode}, W. and {Rib{\'o}}, M. and {Rico}, J. and {Righi}, C. and {Rugliancich}, A. and {Saha}, L. and {Sahakyan}, N. and {Saito}, T. and {Sakurai}, S. and {Satalecka}, K. and {Schmidt}, K. and {Schweizer}, T. and {Sitarek}, J. and {{\v{S}}nidari{\'c}}, I. and {Sobczynska}, D. and {Somero}, A. and {Stamerra}, A. and {Strom}, D. and {Strzys}, M. and {Suda}, Y. and {Suri{\'c}}, T. and {Takahashi}, M. and {Tavecchio}, F. and {Temnikov}, P. and {Terzi{\'c}}, T. and {Teshima}, M. and {Torres-Alb{\`a}}, N. and {Tosti}, L. and {Vagelli}, V. and {van Scherpenberg}, J. and {Vanzo}, G. and {Vazquez Acosta}, M. and {Vigorito}, C.~F. and {Vitale}, V. and {Vovk}, I. and {Will}, M. and {Zari{\'c}}, D. and {Arcaro}, C. and {Carosi}, A. and {D'Ammando}, F. and {Tombesi}, F. and {Lohfink}, A.},
        title = "{New Hard-TeV Extreme Blazars Detected with the MAGIC Telescopes}",
      journal = {\apjs},
         year = 2020,
        month = mar,
       volume = {247},
       number = {1},
          eid = {16},
        pages = {16},
          doi = {10.3847/1538-4365/ab5b98},
archivePrefix = {arXiv},
       eprint = {1911.06680},
 primaryClass = {astro-ph.HE},
       adsurl = {https://ui.adsabs.harvard.edu/abs/2020ApJS..247...16A}
}

@ARTICLE{Wang2024ApJS271,
       author = {{Wang}, Ze-Rui and {Xue}, Rui and {Xiong}, Dingrong and {Wang}, Hai-Qin and {Sun}, Lu-Ming and {Peng}, Fang-Kun and {Mao}, Jirong},
        title = "{Broadband Multiwavelength Study of LHAASO-detected Active Galactic Nuclei}",
      journal = {\apjs},
         year = 2024,
        month = mar,
       volume = {271},
       number = {1},
          eid = {10},
        pages = {10},
          doi = {10.3847/1538-4365/ad168c},
archivePrefix = {arXiv},
       eprint = {2308.10200},
 primaryClass = {astro-ph.HE},
       adsurl = {https://ui.adsabs.harvard.edu/abs/2024ApJS..271...10W}
}

@ARTICLE{Cerruti2017A&A606,
       author = {{Cerruti}, M. and {Benbow}, W. and {Chen}, X. and {Dumm}, J.~P. and {Fortson}, L.~F. and {Shahinyan}, K.},
        title = "{Luminous and high-frequency peaked blazars: the origin of the {\ensuremath{\gamma}}-ray emission from PKS 1424+240}",
      journal = {\aap},
         year = 2017,
        month = oct,
       volume = {606},
          eid = {A68},
        pages = {A68},
          doi = {10.1051/0004-6361/201730799},
archivePrefix = {arXiv},
       eprint = {1707.00804},
 primaryClass = {astro-ph.HE},
       adsurl = {https://ui.adsabs.harvard.edu/abs/2017A&A...606A..68C}
}

@ARTICLE{Lu2025PRD112,
       author = {{Lu}, Ming-Xuan and {Liang}, Yun-Feng and {Ouyang}, Xuerui and {Li}, Rong-Lan and {Wang}, Xiang-Gao},
        title = "{Investigating the correlations between IceCube high-energy neutrinos and Fermi-LAT {\ensuremath{\gamma}}-ray observations. II}",
      journal = {\prd},
         year = 2025,
        month = nov,
       volume = {112},
       number = {10},
          eid = {103013},
        pages = {103013},
          doi = {10.1103/wm72-25tq},
archivePrefix = {arXiv},
       eprint = {2404.19730},
 primaryClass = {astro-ph.HE},
       adsurl = {https://ui.adsabs.harvard.edu/abs/2025PhRvD.112j3013L}
}

@ARTICLE{Bellenghi2023ApJ955L,
       author = {{Bellenghi}, Chiara and {Padovani}, Paolo and {Resconi}, Elisa and {Giommi}, Paolo},
        title = "{Correlating High-energy IceCube Neutrinos with 5BZCAT Blazars and RFC Sources}",
      journal = {\apjl},
         year = 2023,
        month = oct,
       volume = {955},
       number = {2},
          eid = {L32},
        pages = {L32},
          doi = {10.3847/2041-8213/acf711},
archivePrefix = {arXiv},
       eprint = {2309.03115},
 primaryClass = {astro-ph.HE},
       adsurl = {https://ui.adsabs.harvard.edu/abs/2023ApJ...955L..32B}
}

@ARTICLE{Abbasi2023ApJ954,
       author = {{Abbasi}, R. and {Ackermann}, M. and {Adams}, J. and {Agarwalla}, S.~K. and {Aguilar}, J.~A. and {Ahlers}, M. and {Alameddine}, J.~M. and {Amin}, N.~M. and {Andeen}, K. and {Anton}, G. and {Arg{\"u}elles}, C. and {Ashida}, Y. and {Athanasiadou}, S. and {Axani}, S.~N. and {Bai}, X. and {Balagopal}, V.~A. and {Baricevic}, M. and {Barwick}, S.~W. and {Basu}, V. and {Bay}, R. and {Beatty}, J.~J. and {Becker}, K.-H. and {Becker Tjus}, J. and {Beise}, J. and {Bellenghi}, C. and {Benning}, C. and {BenZvi}, S. and {Berley}, D. and {Bernardini}, E. and {Besson}, D.~Z. and {Binder}, G. and {Blaufuss}, E. and {Blot}, S. and {Bontempo}, F. and {Book}, J.~Y. and {Meneguolo}, C. Boscolo and {B{\"o}ser}, S. and {Botner}, O. and {B{\"o}ttcher}, J. and {Bourbeau}, E. and {Braun}, J. and {Brinson}, B. and {Brostean-Kaiser}, J. and {Burley}, R.~T. and {Busse}, R.~S. and {Butterfield}, D. and {Campana}, M.~A. and {Carloni}, K. and {Carnie-Bronca}, E.~G. and {Chattopadhyay}, S. and {Chau}, N. and {Chen}, C. and {Chen}, Z. and {Chirkin}, D. and {Choi}, S. and {Clark}, B.~A. and {Classen}, L. and {Coleman}, A. and {Collin}, G.~H. and {Connolly}, A. and {Conrad}, J.~M. and {Coppin}, P. and {Correa}, P. and {Countryman}, S. and {Cowen}, D.~F. and {Dave}, P. and {De Clercq}, C. and {DeLaunay}, J.~J. and {Delgado}, D. and {Dembinski}, H. and {Deng}, S. and {Deoskar}, K. and {Desai}, A. and {Desiati}, P. and {de Vries}, K.~D. and {de Wasseige}, G. and {DeYoung}, T. and {Diaz}, A. and {D{\'\i}az-V{\'e}lez}, J.~C. and {Dittmer}, M. and {Domi}, A. and {Dujmovic}, H. and {DuVernois}, M.~A. and {Ehrhardt}, T. and {Eller}, P. and {El Mentawi}, S. and {Engel}, R. and {Erpenbeck}, H. and {Evans}, J. and {Evenson}, P.~A. and {Fan}, K.~L. and {Fang}, K. and {Farrag}, K. and {Fazely}, A.~R. and {Fedynitch}, A. and {Feigl}, N. and {Fiedlschuster}, S. and {Finley}, C. and {Fischer}, L. and {Fox}, D. and {Franckowiak}, A. and {Friedman}, E. and {Fritz}, A. and {F{\"u}rst}, P. and {Gaisser}, T.~K. and {Gallagher}, J. and {Ganster}, E. and {Garcia}, A. and {Gerhardt}, L. and {Ghadimi}, A. and {Glaser}, C. and {Glauch}, T. and {Gl{\"u}senkamp}, T. and {Goehlke}, N. and {Gonzalez}, J.~G. and {Goswami}, S. and {Grant}, D. and {Gray}, S.~J. and {Gries}, O. and {Griffin}, S. and {Griswold}, S. and {G{\"u}nther}, C. and {Gutjahr}, P. and {Haack}, C. and {Hallgren}, A. and {Halliday}, R. and {Halve}, L. and {Halzen}, F. and {Hamdaoui}, H. and {Minh}, M. Ha and {Hanson}, K. and {Hardin}, J. and {Harnisch}, A.~A. and {Hatch}, P. and {Haungs}, A. and {Helbing}, K. and {Hellrung}, J. and {Henningsen}, F. and {Heuermann}, L. and {Heyer}, N. and {Hickford}, S. and {Hidvegi}, A. and {Hill}, C. and {Hill}, G.~C. and {Hoffman}, K.~D. and {Hori}, S. and {Hoshina}, K. and {Hou}, W. and {Huber}, T. and {Hultqvist}, K. and {H{\"u}nnefeld}, M. and {Hussain}, R. and {Hymon}, K. and {In}, S. and {Ishihara}, A. and {Jacquart}, M. and {Janik}, O. and {Jansson}, M. and {Japaridze}, G.~S. and {Jayakumar}, K. and {Jeong}, M. and {Jin}, M. and {Jones}, B.~J.~P. and {Kang}, D. and {Kang}, W. and {Kang}, X. and {Kappes}, A. and {Kappesser}, D. and {Kardum}, L. and {Karg}, T. and {Karl}, M. and {Karle}, A. and {Katz}, U. and {Kauer}, M. and {Kelley}, J.~L. and {Zathul}, A. Khatee and {Kheirandish}, A. and {Kiryluk}, J. and {Klein}, S.~R. and {Kochocki}, A. and {Koirala}, R. and {Kolanoski}, H. and {Kontrimas}, T. and {K{\"o}pke}, L. and {Kopper}, C. and {Koskinen}, D.~J. and {Koundal}, P. and {Kovacevich}, M. and {Kowalski}, M. and {Kozynets}, T. and {Kruiswijk}, K. and {Krupczak}, E. and {Kumar}, A. and {Kun}, E. and {Kurahashi}, N. and {Lad}, N. and {Lagunas Gualda}, C. and {Lamoureux}, M. and {Larson}, M.~J. and {Latseva}, S.},
        title = "{Search for Correlations of High-energy Neutrinos Detected in IceCube with Radio-bright AGN and Gamma-Ray Emission from Blazars}",
      journal = {\apj},
         year = 2023,
        month = sep,
       volume = {954},
       number = {1},
          eid = {75},
        pages = {75},
          doi = {10.3847/1538-4357/acdfcb},
archivePrefix = {arXiv},
       eprint = {2304.12675},
 primaryClass = {astro-ph.HE},
       adsurl = {https://ui.adsabs.harvard.edu/abs/2023ApJ...954...75A}
}

@ARTICLE{Aartsen2017ApJ835,
       author = {{Aartsen}, M.~G. and {Abraham}, K. and {Ackermann}, M. and {Adams}, J. and {Aguilar}, J.~A. and {Ahlers}, M. and {Ahrens}, M. and {Altmann}, D. and {Andeen}, K. and {Anderson}, T. and {Ansseau}, I. and {Anton}, G. and {Archinger}, M. and {Arguelles}, C. and {Arlen}, T.~C. and {Auffenberg}, J. and {Axani}, S. and {Bai}, X. and {Barwick}, S.~W. and {Baum}, V. and {Bay}, R. and {Beatty}, J.~J. and {Becker Tjus}, J. and {Becker}, K.-H. and {BenZvi}, S. and {Berghaus}, P. and {Berley}, D. and {Bernardini}, E. and {Bernhard}, A. and {Besson}, D.~Z. and {Binder}, G. and {Bindig}, D. and {Bissok}, M. and {Blaufuss}, E. and {Blot}, S. and {Boersma}, D.~J. and {Bohm}, C. and {B{\"o}rner}, M. and {Bos}, F. and {Bose}, D. and {B{\"o}ser}, S. and {Botner}, O. and {Braun}, J. and {Brayeur}, L. and {Bretz}, H.-P. and {Burgman}, A. and {Casey}, J. and {Casier}, M. and {Cheung}, E. and {Chirkin}, D. and {Christov}, A. and {Clark}, K. and {Classen}, L. and {Coenders}, S. and {Collin}, G.~H. and {Conrad}, J.~M. and {Cowen}, D.~F. and {Cruz Silva}, A.~H. and {Daughhetee}, J. and {Davis}, J.~C. and {Day}, M. and {de Andr{\'e}}, J.~P.~A.~M. and {De Clercq}, C. and {del Pino Rosendo}, E. and {Dembinski}, H. and {De Ridder}, S. and {Desiati}, P. and {de Vries}, K.~D. and {de Wasseige}, G. and {de With}, M. and {DeYoung}, T. and {D{\'\i}az-V{\'e}lez}, J.~C. and {di Lorenzo}, V. and {Dujmovic}, H. and {Dumm}, J.~P. and {Dunkman}, M. and {Eberhardt}, B. and {Ehrhardt}, T. and {Eichmann}, B. and {Euler}, S. and {Evenson}, P.~A. and {Fahey}, S. and {Fazely}, A.~R. and {Feintzeig}, J. and {Felde}, J. and {Filimonov}, K. and {Finley}, C. and {Flis}, S. and {F{\"o}sig}, C.-C. and {Franckowiak}, A. and {Fuchs}, T. and {Gaisser}, T.~K. and {Gaior}, R. and {Gallagher}, J. and {Gerhardt}, L. and {Ghorbani}, K. and {Giang}, W. and {Gladstone}, L. and {Glagla}, M. and {Gl{\"u}senkamp}, T. and {Goldschmidt}, A. and {Golup}, G. and {Gonzalez}, J.~G. and {G{\'o}ra}, D. and {Grant}, D. and {Griffith}, Z. and {Haack}, C. and {Haj Ismail}, A. and {Hallgren}, A. and {Halzen}, F. and {Hansen}, E. and {Hansmann}, B. and {Hansmann}, T. and {Hanson}, K. and {Hebecker}, D. and {Heereman}, D. and {Helbing}, K. and {Hellauer}, R. and {Hickford}, S. and {Hignight}, J. and {Hill}, G.~C. and {Hoffman}, K.~D. and {Hoffmann}, R. and {Holzapfel}, K. and {Homeier}, A. and {Hoshina}, K. and {Huang}, F. and {Huber}, M. and {Huelsnitz}, W. and {Hultqvist}, K. and {In}, S. and {Ishihara}, A. and {Jacobi}, E. and {Japaridze}, G.~S. and {Jeong}, M. and {Jero}, K. and {Jones}, B.~J.~P. and {Jurkovic}, M. and {Kappes}, A. and {Karg}, T. and {Karle}, A. and {Katz}, U. and {Kauer}, M. and {Keivani}, A. and {Kelley}, J.~L. and {Kemp}, J. and {Kheirandish}, A. and {Kim}, M. and {Kintscher}, T. and {Kiryluk}, J. and {Kittler}, T. and {Klein}, S.~R. and {Kohnen}, G. and {Koirala}, R. and {Kolanoski}, H. and {Konietz}, R. and {K{\"o}pke}, L. and {Kopper}, C. and {Kopper}, S. and {Koskinen}, D.~J. and {Kowalski}, M. and {Krings}, K. and {Kroll}, M. and {Kr{\"u}ckl}, G. and {Kr{\"u}ger}, C. and {Kunnen}, J. and {Kunwar}, S. and {Kurahashi}, N. and {Kuwabara}, T. and {Labare}, M. and {Lanfranchi}, J.~L. and {Larson}, M.~J. and {Lennarz}, D. and {Lesiak-Bzdak}, M. and {Leuermann}, M. and {Leuner}, J. and {Lu}, L. and {L{\"u}nemann}, J. and {Madsen}, J. and {Maggi}, G. and {Mahn}, K.~B.~M. and {Mancina}, S. and {Mandelartz}, M. and {Maruyama}, R. and {Mase}, K. and {Maunu}, R. and {McNally}, F. and {Meagher}, K. and {Medici}, M. and {Meier}, M. and {Meli}, A. and {Menne}, T. and {Merino}, G. and {Meures}, T. and {Miarecki}, S. and {Middell}, E. and {Mohrmann}, L. and {Montaruli}, T. and {Moulai}, M. and {Nahnhauer}, R.},
        title = "{The Contribution of Fermi-2LAC Blazars to Diffuse TeV-PeV Neutrino Flux}",
      journal = {\apj},
         year = 2017,
        month = jan,
       volume = {835},
       number = {1},
          eid = {45},
        pages = {45},
          doi = {10.3847/1538-4357/835/1/45},
archivePrefix = {arXiv},
       eprint = {1611.03874},
 primaryClass = {astro-ph.HE},
       adsurl = {https://ui.adsabs.harvard.edu/abs/2017ApJ...835...45A}
}

@ARTICLE{Hovatta2021A&A650,
       author = {{Hovatta}, T. and {Lindfors}, E. and {Kiehlmann}, S. and {Max-Moerbeck}, W. and {Hodges}, M. and {Liodakis}, I. and {L{\"a}hteem{\"a}ki}, A. and {Pearson}, T.~J. and {Readhead}, A.~C.~S. and {Reeves}, R.~A. and {Suutarinen}, S. and {Tammi}, J. and {Tornikoski}, M.},
        title = "{Association of IceCube neutrinos with radio sources observed at Owens Valley and Mets{\"a}hovi Radio Observatories}",
      journal = {\aap},
         year = 2021,
        month = jun,
       volume = {650},
          eid = {A83},
        pages = {A83},
          doi = {10.1051/0004-6361/202039481},
archivePrefix = {arXiv},
       eprint = {2009.10523},
 primaryClass = {astro-ph.HE},
       adsurl = {https://ui.adsabs.harvard.edu/abs/2021A&A...650A..83H}
}

@ARTICLE{Plavin2020ApJ894,
       author = {{Plavin}, Alexander and {Kovalev}, Yuri Y. and {Kovalev}, Yuri A. and {Troitsky}, Sergey},
        title = "{Observational Evidence for the Origin of High-energy Neutrinos in Parsec-scale Nuclei of Radio-bright Active Galaxies}",
      journal = {\apj},
         year = 2020,
        month = may,
       volume = {894},
       number = {2},
          eid = {101},
        pages = {101},
          doi = {10.3847/1538-4357/ab86bd},
archivePrefix = {arXiv},
       eprint = {2001.00930},
 primaryClass = {astro-ph.HE},
       adsurl = {https://ui.adsabs.harvard.edu/abs/2020ApJ...894..101P}
}

@ARTICLE{Kouch2024A&A690,
       author = {{Kouch}, Pouya M. and {Lindfors}, Elina and {Hovatta}, Talvikki and {Liodakis}, Ioannis and {Koljonen}, Karri I.~I. and {Nilsson}, Kari and {Kiehlmann}, Sebastian and {Max-Moerbeck}, Walter and {Readhead}, Anthony C.~S. and {Reeves}, Rodrigo A. and {Pearson}, Timothy J. and {Jormanainen}, Jenni and {Ramazani}, Vandad Fallah and {Graham}, Matthew J.},
        title = "{Association of the IceCube neutrinos with blazars in the CGRaBS sample}",
      journal = {\aap},
         year = 2024,
        month = oct,
       volume = {690},
          eid = {A111},
        pages = {A111},
          doi = {10.1051/0004-6361/202347624},
archivePrefix = {arXiv},
       eprint = {2407.07153},
 primaryClass = {astro-ph.HE},
       adsurl = {https://ui.adsabs.harvard.edu/abs/2024A&A...690A.111K}
}

@ARTICLE{Abbasi2024ApJ973,
       author = {{Abbasi}, R. and {Ackermann}, M. and {Adams}, J. and {Agarwalla}, S.~K. and {Aguilar}, J.~A. and {Ahlers}, M. and {Alameddine}, J.~M. and {Amin}, N.~M. and {Andeen}, K. and {Arg{\"u}elles}, C. and {Ashida}, Y. and {Athanasiadou}, S. and {Ausborm}, L. and {Axani}, S.~N. and {Bai}, X. and {Balagopal}, V.~A. and {Baricevic}, M. and {Barwick}, S.~W. and {Bash}, S. and {Basu}, V. and {Bay}, R. and {Beatty}, J.~J. and {Becker Tjus}, J. and {Beise}, J. and {Bellenghi}, C. and {Benning}, C. and {BenZvi}, S. and {Berley}, D. and {Bernardini}, E. and {Besson}, D.~Z. and {Blaufuss}, E. and {Bloom}, L. and {Blot}, S. and {Bontempo}, F. and {Book Motzkin}, J.~Y. and {Boscolo Meneguolo}, C. and {B{\"o}ser}, S. and {Botner}, O. and {B{\"o}ttcher}, J. and {Braun}, J. and {Brinson}, B. and {Brostean-Kaiser}, J. and {Brusa}, L. and {Burley}, R.~T. and {Butterfield}, D. and {Campana}, M.~A. and {Caracas}, I. and {Carloni}, K. and {Carpio}, J. and {Chattopadhyay}, S. and {Chau}, N. and {Chen}, Z. and {Chirkin}, D. and {Choi}, S. and {Clark}, B.~A. and {Coleman}, A. and {Collin}, G.~H. and {Connolly}, A. and {Conrad}, J.~M. and {Corley}, R. and {Cowen}, D.~F. and {Dave}, P. and {De Clercq}, C. and {DeLaunay}, J.~J. and {Delgado}, D. and {Deng}, S. and {Desai}, A. and {Desiati}, P. and {de Vries}, K.~D. and {de Wasseige}, G. and {DeYoung}, T. and {Diaz}, A. and {D{\'\i}az-V{\'e}lez}, J.~C. and {Dierichs}, P. and {Dittmer}, M. and {Domi}, A. and {Draper}, L. and {Dujmovic}, H. and {Durnford}, D. and {Dutta}, K. and {DuVernois}, M.~A. and {Ehrhardt}, T. and {Eidenschink}, L. and {Eimer}, A. and {Eller}, P. and {Ellinger}, E. and {El Mentawi}, S. and {Els{\"a}sser}, D. and {Engel}, R. and {Erpenbeck}, H. and {Evans}, J. and {Evenson}, P.~A. and {Fan}, K.~L. and {Fang}, K. and {Farrag}, K. and {Fazely}, A.~R. and {Fedynitch}, A. and {Feigl}, N. and {Fiedlschuster}, S. and {Finley}, C. and {Fischer}, L. and {Fox}, D. and {Franckowiak}, A. and {Fukami}, S. and {F{\"u}rst}, P. and {Gallagher}, J. and {Ganster}, E. and {Garcia}, A. and {Garcia}, M. and {Garg}, G. and {Genton}, E. and {Gerhardt}, L. and {Ghadimi}, A. and {Girard-Carillo}, C. and {Glaser}, C. and {Gl{\"u}senkamp}, T. and {Gonzalez}, J.~G. and {Goswami}, S. and {Granados}, A. and {Grant}, D. and {Gray}, S.~J. and {Gries}, O. and {Griffin}, S. and {Griswold}, S. and {Groth}, K.~M. and {Guevel}, D. and {G{\"u}nther}, C. and {Gutjahr}, P. and {Ha}, C. and {Haack}, C. and {Hallgren}, A. and {Halve}, L. and {Halzen}, F. and {Hamdaoui}, H. and {Minh}, M. Ha and {Handt}, M. and {Hanson}, K. and {Hardin}, J. and {Harnisch}, A.~A. and {Hatch}, P. and {Haungs}, A. and {H{\"a}u{\ss}ler}, J. and {Helbing}, K. and {Hellrung}, J. and {Hermannsgabner}, J. and {Heuermann}, L. and {Heyer}, N. and {Hickford}, S. and {Hidvegi}, A. and {Hill}, C. and {Hill}, G.~C. and {Hoffman}, K.~D. and {Hori}, S. and {Hoshina}, K. and {Hostert}, M. and {Hou}, W. and {Huber}, T. and {Hultqvist}, K. and {H{\"u}nnefeld}, M. and {Hussain}, R. and {Hymon}, K. and {Ishihara}, A. and {Iwakiri}, W. and {Jacquart}, M. and {Jain}, S. and {Janik}, O. and {Jansson}, M. and {Japaridze}, G.~S. and {Jeong}, M. and {Jin}, M. and {Jones}, B.~J.~P. and {Kamp}, N. and {Kang}, D. and {Kang}, W. and {Kang}, X. and {Kappes}, A. and {Kappesser}, D. and {Kardum}, L. and {Karg}, T. and {Karl}, M. and {Karle}, A. and {Katil}, A. and {Katz}, U. and {Kauer}, M. and {Kelley}, J.~L. and {Khanal}, M. and {Khatee Zathul}, A. and {Kheirandish}, A. and {Kiryluk}, J. and {Klein}, S.~R. and {Kochocki}, A. and {Koirala}, R. and {Kolanoski}, H. and {Kontrimas}, T. and {K{\"o}pke}, L. and {Kopper}, C. and {Koskinen}, D.~J. and {Koundal}, P. and {Kovacevich}, M. and {Kowalski}, M.},
        title = "{Probing the Connection between IceCube Neutrinos and MOJAVE AGN}",
      journal = {\apj},
         year = 2024,
        month = oct,
       volume = {973},
       number = {2},
          eid = {97},
        pages = {97},
          doi = {10.3847/1538-4357/ad643d},
archivePrefix = {arXiv},
       eprint = {2407.01351},
 primaryClass = {astro-ph.HE},
       adsurl = {https://ui.adsabs.harvard.edu/abs/2024ApJ...973...97A}
}

@ARTICLE{Ackermann2016ApJ824,
       author = {{Ackermann}, M. and {Anantua}, R. and {Asano}, K. and {Baldini}, L. and {Barbiellini}, G. and {Bastieri}, D. and {Becerra Gonzalez}, J. and {Bellazzini}, R. and {Bissaldi}, E. and {Blandford}, R.~D. and {Bloom}, E.~D. and {Bonino}, R. and {Bottacini}, E. and {Bruel}, P. and {Buehler}, R. and {Caliandro}, G.~A. and {Cameron}, R.~A. and {Caragiulo}, M. and {Caraveo}, P.~A. and {Cavazzuti}, E. and {Cecchi}, C. and {Cheung}, C.~C. and {Chiang}, J. and {Chiaro}, G. and {Ciprini}, S. and {Cohen-Tanugi}, J. and {Costanza}, F. and {Cutini}, S. and {D'Ammando}, F. and {de Palma}, F. and {Desiante}, R. and {Digel}, S.~W. and {Di Lalla}, N. and {Di Mauro}, M. and {Di Venere}, L. and {Drell}, P.~S. and {Favuzzi}, C. and {Fegan}, S.~J. and {Ferrara}, E.~C. and {Fukazawa}, Y. and {Funk}, S. and {Fusco}, P. and {Gargano}, F. and {Gasparrini}, D. and {Giglietto}, N. and {Giordano}, F. and {Giroletti}, M. and {Grenier}, I.~A. and {Guillemot}, L. and {Guiriec}, S. and {Hayashida}, M. and {Hays}, E. and {Horan}, D. and {J{\'o}hannesson}, G. and {Kensei}, S. and {Kocevski}, D. and {Kuss}, M. and {La Mura}, G. and {Larsson}, S. and {Latronico}, L. and {Li}, J. and {Longo}, F. and {Loparco}, F. and {Lott}, B. and {Lovellette}, M.~N. and {Lubrano}, P. and {Madejski}, G.~M. and {Magill}, J.~D. and {Maldera}, S. and {Manfreda}, A. and {Mayer}, M. and {Mazziotta}, M.~N. and {Michelson}, P.~F. and {Mirabal}, N. and {Mizuno}, T. and {Monzani}, M.~E. and {Morselli}, A. and {Moskalenko}, I.~V. and {Nalewajko}, K. and {Negro}, M. and {Nuss}, E. and {Ohsugi}, T. and {Orlando}, E. and {Paneque}, D. and {Perkins}, J.~S. and {Pesce-Rollins}, M. and {Piron}, F. and {Pivato}, G. and {Porter}, T.~A. and {Principe}, G. and {Rando}, R. and {Razzano}, M. and {Razzaque}, S. and {Reimer}, A. and {Scargle}, J.~D. and {Sgr{\`o}}, C. and {Sikora}, M. and {Simone}, D. and {Siskind}, E.~J. and {Spada}, F. and {Spinelli}, P. and {Stawarz}, L. and {Thayer}, J.~B. and {Thompson}, D.~J. and {Torres}, D.~F. and {Troja}, E. and {Uchiyama}, Y. and {Yuan}, Y. and {Zimmer}, S.},
        title = "{Minute-timescale >100 MeV {\ensuremath{\gamma}}-Ray Variability during the Giant Outburst of Quasar 3C 279 Observed by Fermi-LAT in 2015 June}",
      journal = {\apjl},
         year = 2016,
        month = jun,
       volume = {824},
       number = {2},
          eid = {L20},
        pages = {L20},
          doi = {10.3847/2041-8205/824/2/L20},
archivePrefix = {arXiv},
       eprint = {1605.05324},
 primaryClass = {astro-ph.HE},
       adsurl = {https://ui.adsabs.harvard.edu/abs/2016ApJ...824L..20A}
}

@ARTICLE{Abdo2011ApJ733,
       author = {{Abdo}, A.~A. and {Ackermann}, M. and {Ajello}, M. and {Allafort}, A. and {Baldini}, L. and {Ballet}, J. and {Barbiellini}, G. and {Bastieri}, D. and {Bellazzini}, R. and {Berenji}, B. and {Blandford}, R.~D. and {Bloom}, E.~D. and {Bonamente}, E. and {Borgland}, A.~W. and {Bouvier}, A. and {Bregeon}, J. and {Brigida}, M. and {Bruel}, P. and {Buehler}, R. and {Buson}, S. and {Caliandro}, G.~A. and {Cameron}, R.~A. and {Caraveo}, P.~A. and {Casandjian}, J.~M. and {Cavazzuti}, E. and {Cecchi}, C. and {Charles}, E. and {Chekhtman}, A. and {Cheung}, C.~C. and {Chiang}, J. and {Ciprini}, S. and {Claus}, R. and {Conrad}, J. and {Cutini}, S. and {D'Ammando}, F. and {de Angelis}, A. and {de Palma}, F. and {Dermer}, C.~D. and {Digel}, S.~W. and {Silva}, E. do Couto e. and {Drell}, P.~S. and {Dubois}, R. and {Dumora}, D. and {Escande}, L. and {Favuzzi}, C. and {Fegan}, S.~J. and {Ferrara}, E.~C. and {Fortin}, P. and {Fukazawa}, Y. and {Fusco}, P. and {Gargano}, F. and {Gasparrini}, D. and {Gehrels}, N. and {Germani}, S. and {Giglietto}, N. and {Giommi}, P. and {Giordano}, F. and {Giroletti}, M. and {Glanzman}, T. and {Godfrey}, G. and {Grenier}, I.~A. and {Grove}, J.~E. and {Guiriec}, S. and {Hadasch}, D. and {Hayashida}, M. and {Hays}, E. and {Horan}, D. and {Itoh}, R. and {J{\'o}hannesson}, G. and {Johnson}, A.~S. and {Kamae}, T. and {Katagiri}, H. and {Kataoka}, J. and {Kn{\"o}dlseder}, J. and {Kuss}, M. and {Lande}, J. and {Larsson}, S. and {Latronico}, L. and {Lee}, S.-H. and {Longo}, F. and {Loparco}, F. and {Lott}, B. and {Lovellette}, M.~N. and {Lubrano}, P. and {Madejski}, G.~M. and {Makeev}, A. and {Mazziotta}, M.~N. and {McConville}, W. and {McEnery}, J.~E. and {Michelson}, P.~F. and {Mitthumsiri}, W. and {Mizuno}, T. and {Moiseev}, A.~A. and {Monte}, C. and {Monzani}, M.~E. and {Morselli}, A. and {Moskalenko}, I.~V. and {Murgia}, S. and {Naumann-Godo}, M. and {Nishino}, S. and {Nolan}, P.~L. and {Norris}, J.~P. and {Nuss}, E. and {Ohsugi}, T. and {Okumura}, A. and {Orlando}, E. and {Ormes}, J.~F. and {Paneque}, D. and {Pelassa}, V. and {Pesce-Rollins}, M. and {Pierbattista}, M. and {Piron}, F. and {Porter}, T.~A. and {Rain{\`o}}, S. and {Rando}, R. and {Razzaque}, S. and {Reimer}, A. and {Reimer}, O. and {Ritz}, S. and {Roth}, M. and {Sadrozinski}, H.~F.-W. and {Sanchez}, D. and {Scargle}, J.~D. and {Schalk}, T.~L. and {Sgr{\`o}}, C. and {Siskind}, E.~J. and {Smith}, P.~D. and {Spandre}, G. and {Spinelli}, P. and {Strickman}, M.~S. and {Takahashi}, H. and {Takahashi}, T. and {Tanaka}, T. and {Tanaka}, Y. and {Thayer}, J.~G. and {Thayer}, J.~B. and {Thompson}, D.~J. and {Tibaldo}, L. and {Torres}, D.~F. and {Tosti}, G. and {Tramacere}, A. and {Troja}, E. and {Vandenbroucke}, J. and {Vasileiou}, V. and {Vianello}, G. and {Vilchez}, N. and {Vitale}, V. and {Waite}, A.~P. and {Wang}, P. and {Winer}, B.~L. and {Wood}, K.~S. and {Yang}, Z. and {Ziegler}, M.},
        title = "{Fermi Gamma-ray Space Telescope Observations of the Gamma-ray Outburst from 3C454.3 in November 2010}",
      journal = {\apjl},
         year = 2011,
        month = jun,
       volume = {733},
       number = {2},
        pages = {L26},
          doi = {10.1088/2041-8205/733/2/L26},
archivePrefix = {arXiv},
       eprint = {1102.0277},
 primaryClass = {astro-ph.HE},
       adsurl = {https://ui.adsabs.harvard.edu/abs/2011ApJ...733L..26A}
}

@ARTICLE{Aharonian2007ApJ664,
       author = {{Aharonian}, F. and {Akhperjanian}, A.~G. and {Bazer-Bachi}, A.~R. and {Behera}, B. and {Beilicke}, M. and {Benbow}, W. and {Berge}, D. and {Bernl{\"o}hr}, K. and {Boisson}, C. and {Bolz}, O. and {Borrel}, V. and {Boutelier}, T. and {Braun}, I. and {Brion}, E. and {Brown}, A.~M. and {B{\"u}hler}, R. and {B{\"u}sching}, I. and {Bulik}, T. and {Carrigan}, S. and {Chadwick}, P.~M. and {Clapson}, A.~C. and {Chounet}, L.-M. and {Coignet}, G. and {Cornils}, R. and {Costamante}, L. and {Degrange}, B. and {Dickinson}, H.~J. and {Djannati-Ata{\"\i}}, A. and {Domainko}, W. and {Drury}, L. O'C. and {Dubus}, G. and {Dyks}, J. and {Egberts}, K. and {Emmanoulopoulos}, D. and {Espigat}, P. and {Farnier}, C. and {Feinstein}, F. and {Fiasson}, A. and {F{\"o}rster}, A. and {Fontaine}, G. and {Funk}, Seb. and {Funk}, S. and {F{\"u}{\ss}ling}, M. and {Gallant}, Y.~A. and {Giebels}, B. and {Glicenstein}, J.~F. and {Gl{\"u}ck}, B. and {Goret}, P. and {Hadjichristidis}, C. and {Hauser}, D. and {Hauser}, M. and {Heinzelmann}, G. and {Henri}, G. and {Hermann}, G. and {Hinton}, J.~A. and {Hoffmann}, A. and {Hofmann}, W. and {Holleran}, M. and {Hoppe}, S. and {Horns}, D. and {Jacholkowska}, A. and {de Jager}, O.~C. and {Kendziorra}, E. and {Kerschhaggl}, M. and {Kh{\'e}lifi}, B. and {Komin}, Nu. and {Kosack}, K. and {Lamanna}, G. and {Latham}, I.~J. and {Le Gallou}, R. and {Lemi{\`e}re}, A. and {Lemoine-Goumard}, M. and {Lenain}, J.-P. and {Lohse}, T. and {Martin}, J.~M. and {Martineau-Huynh}, O. and {Marcowith}, A. and {Masterson}, C. and {Maurin}, G. and {McComb}, T.~J.~L. and {Moderski}, R. and {Moulin}, E. and {de Naurois}, M. and {Nedbal}, D. and {Nolan}, S.~J. and {Olive}, J.-P. and {Orford}, K.~J. and {Osborne}, J.~L. and {Ostrowski}, M. and {Panter}, M. and {Pedaletti}, G. and {Pelletier}, G. and {Petrucci}, P.-O. and {Pita}, S. and {P{\"u}hlhofer}, G. and {Punch}, M. and {Ranchon}, S. and {Raubenheimer}, B.~C. and {Raue}, M. and {Rayner}, S.~M. and {Renaud}, M. and {Ripken}, J. and {Rob}, L. and {Rolland}, L. and {Rosier-Lees}, S. and {Rowell}, G. and {Rudak}, B. and {Ruppel}, J. and {Sahakian}, V. and {Santangelo}, A. and {Saug{\'e}}, L. and {Schlenker}, S. and {Schlickeiser}, R. and {Schr{\"o}der}, R. and {Schwanke}, U. and {Schwarzburg}, S. and {Schwemmer}, S. and {Shalchi}, A. and {Sol}, H. and {Spangler}, D. and {Stawarz}, {\L}. and {Steenkamp}, R. and {Stegmann}, C. and {Superina}, G. and {Tam}, P.~H. and {Tavernet}, J.-P. and {Terrier}, R. and {van Eldik}, C. and {Vasileiadis}, G. and {Venter}, C. and {Vialle}, J.~P. and {Vincent}, P. and {Vivier}, M. and {V{\"o}lk}, H.~J. and {Volpe}, F. and {Wagner}, S.~J. and {Ward}, M. and {Zdziarski}, A.~A.},
        title = "{An Exceptional Very High Energy Gamma-Ray Flare of PKS 2155-304}",
      journal = {\apjl},
         year = 2007,
        month = aug,
       volume = {664},
       number = {2},
        pages = {L71-L74},
          doi = {10.1086/520635},
archivePrefix = {arXiv},
       eprint = {0706.0797},
 primaryClass = {astro-ph},
       adsurl = {https://ui.adsabs.harvard.edu/abs/2007ApJ...664L..71A}
}

@ARTICLE{Raiteri2023MNRAS526,
       author = {{Raiteri}, C.~M. and {Villata}, M. and {Carnerero}, M.~I. and {Savchenko}, S.~S. and {Kurtanidze}, S.~O. and {Vlasyuk}, V.~V. and {Marchini}, A. and {Matsumoto}, K. and {Lorey}, C. and {Joner}, M.~D. and {Gazeas}, K. and {Carosati}, D. and {Mirzaqulov}, D.~O. and {Acosta Pulido}, J.~A. and {Agudo}, I. and {Bachev}, R. and {Ben{\'\i}tez}, E. and {Borman}, G.~A. and {Calcidese}, P. and {Chen}, W.~P. and {Damljanovic}, G. and {Ehgamberdiev}, S.~A. and {Els{\"a}sser}, D. and {Feige}, M. and {Frasca}, A. and {Gaur}, H. and {Grishina}, T.~S. and {Gupta}, A.~C. and {Hiriart}, D. and {Holland}, M. and {Horst}, B. and {Ibryamov}, S. and {Ivanidze}, R.~Z. and {Jensen}, J. and {Jithesh}, V. and {Jovanovic}, M.~D. and {Kiehlmann}, S. and {Kimeridze}, G.~N. and {Kishore}, S. and {Kopatskaya}, E.~N. and {Kurtanidze}, O.~M. and {Larionova}, E.~G. and {Lin}, H.~C. and {Mannheim}, K. and {Marinelli}, C. and {Moreira Reyes}, J. and {Morozova}, D.~A. and {Nikolashvili}, M.~G. and {Reinhart}, D. and {Romanov}, F.~D. and {Semkov}, E. and {Seufert}, J. and {Shishkina}, E.~V. and {Sigua}, L.~A. and {Skalidis}, R. and {Spiridonova}, O.~I. and {Stojanovic}, M. and {Strigachev}, A. and {Troitskaya}, Y.~V. and {Troitskiy}, I.~S. and {Tsai}, A. and {Vasilyev}, A.~A. and {Vince}, O. and {Vrontaki}, K. and {Wani}, K. and {Watts}, D. and {Zhovtan}, A.~V.},
        title = "{Extreme photometric and polarimetric variability of blazar S4 0954+65 at its maximum optical and {\ensuremath{\gamma}}-ray brightness levels}",
      journal = {\mnras},
         year = 2023,
        month = dec,
       volume = {526},
       number = {3},
        pages = {4502-4513},
          doi = {10.1093/mnras/stad3064},
archivePrefix = {arXiv},
       eprint = {2310.11108},
 primaryClass = {astro-ph.HE},
       adsurl = {https://ui.adsabs.harvard.edu/abs/2023MNRAS.526.4502R}
}

@ARTICLE{Blandford2019ARA&A57,
       author = {{Blandford}, Roger and {Meier}, David and {Readhead}, Anthony},
        title = "{Relativistic Jets from Active Galactic Nuclei}",
      journal = {\araa},
         year = 2019,
        month = aug,
       volume = {57},
        pages = {467-509},
          doi = {10.1146/annurev-astro-081817-051948},
archivePrefix = {arXiv},
       eprint = {1812.06025},
 primaryClass = {astro-ph.HE},
       adsurl = {https://ui.adsabs.harvard.edu/abs/2019ARA&A..57..467B}
}

@ARTICLE{Padovani2017A&ARv25,
       author = {{Padovani}, P. and {Alexander}, D.~M. and {Assef}, R.~J. and {De Marco}, B. and {Giommi}, P. and {Hickox}, R.~C. and {Richards}, G.~T. and {Smol{\v{c}}i{\'c}}, V. and {Hatziminaoglou}, E. and {Mainieri}, V. and {Salvato}, M.},
        title = "{Active galactic nuclei: what's in a name?}",
      journal = {\aapr},
         year = 2017,
        month = aug,
       volume = {25},
       number = {1},
          eid = {2},
        pages = {2},
          doi = {10.1007/s00159-017-0102-9},
archivePrefix = {arXiv},
       eprint = {1707.07134},
 primaryClass = {astro-ph.GA},
       adsurl = {https://ui.adsabs.harvard.edu/abs/2017A&ARv..25....2P}
}

@ARTICLE{Krawczynski2012ApJ744,
       author = {{Krawczynski}, H.},
        title = "{The Polarization Properties of Inverse Compton Emission and Implications for Blazar Observations with the GEMS X-Ray Polarimeter}",
      journal = {\apj},
         year = 2012,
        month = jan,
       volume = {744},
       number = {1},
          eid = {30},
        pages = {30},
          doi = {10.1088/0004-637X/744/1/30},
archivePrefix = {arXiv},
       eprint = {1109.2186},
 primaryClass = {astro-ph.HE},
       adsurl = {https://ui.adsabs.harvard.edu/abs/2012ApJ...744...30K}
}

@ARTICLE{Aiello2024APh16202990,
       author = {{Aiello}, S. and {Albert}, A. and {Alshamsi}, M. and {Alves Garre}, S. and {Aly}, Z. and {Ambrosone}, A. and {Ameli}, F. and {Andre}, M. and {Androutsou}, E. and {Anguita}, M. and {Aphecetche}, L. and {Ardid}, M. and {Ardid}, S. and {Atmani}, H. and {Aublin}, J. and {Badaracco}, F. and {Bailly-Salins}, L. and {Barda{\v{c}}ov{\'a}}, Z. and {Baret}, B. and {Bariego-Quintana}, A. and {Basegmez du Pree}, S. and {Becherini}, Y. and {Bendahman}, M. and {Benfenati}, F. and {Benhassi}, M. and {Benoit}, D.~M. and {Berbee}, E. and {Bertin}, V. and {Biagi}, S. and {Boettcher}, M. and {Bonanno}, D. and {Boumaaza}, J. and {Bouta}, M. and {Bouwhuis}, M. and {Bozza}, C. and {Bozza}, R.~M. and {Br{\^a}nza{\textcommabelow s}}, H. and {Bretaudeau}, F. and {Breuhaus}, M. and {Bruijn}, R. and {Brunner}, J. and {Bruno}, R. and {Buis}, E. and {Buompane}, R. and {Busto}, J. and {Caiffi}, B. and {Calvo}, D. and {Campion}, S. and {Capone}, A. and {Carenini}, F. and {Carretero}, V. and {Cartraud}, T. and {Castaldi}, P. and {Cecchini}, V. and {Celli}, S. and {Cerisy}, L. and {Chabab}, M. and {Chadolias}, M. and {Chen}, A. and {Cherubini}, S. and {Chiarusi}, T. and {Circella}, M. and {Cocimano}, R. and {Coelho}, J.~A.~B. and {Coleiro}, A. and {Coniglione}, R. and {Coyle}, P. and {Creusot}, A. and {Cuttone}, G. and {Dallier}, R. and {Darras}, Y. and {De Benedittis}, A. and {De Martino}, B. and {Decoene}, V. and {Del Burgo}, R. and {Del Rosso}, I. and {Di Mauro}, L.~S. and {Di Palma}, I. and {D{\'\i}az}, A.~F. and {Diaz}, C. and {Diego-Tortosa}, D. and {Distefano}, C. and {Domi}, A. and {Donzaud}, C. and {Dornic}, D. and {D{\"o}rr}, M. and {Drakopoulou}, E. and {Drouhin}, D. and {Dvornick{\'y}}, R. and {Eberl}, T. and {Eckerov{\'a}}, E. and {Eddymaoui}, A. and {van Eeden}, T. and {Eff}, M. and {van Eijk}, D. and {El Bojaddaini}, I. and {El Hedri}, S. and {Enzenh{\"o}fer}, A. and {Ferrara}, G. and {Filipovi{\'c}}, M.~D. and {Filippini}, F. and {Franciotti}, D. and {Fusco}, L.~A. and {Gabriel}, J. and {Gagliardini}, S. and {Gal}, T. and {Garc{\'\i}a M{\'e}ndez}, J. and {Garcia Soto}, A. and {Gatius Oliver}, C. and {Gei{\ss}elbrecht}, N. and {Ghaddari}, H. and {Gialanella}, L. and {Gibson}, B.~K. and {Giorgio}, E. and {Goos}, I. and {Goswami}, P. and {Goupilliere}, D. and {Gozzini}, S.~R. and {Gracia}, R. and {Graf}, K. and {Guidi}, C. and {Guillon}, B. and {Guti{\'e}rrez}, M. and {van Haren}, H. and {Heijboer}, A. and {Hekalo}, A. and {Hennig}, L. and {Hern{\'a}ndez-Rey}, J.~J. and {Ibnsalih}, W. Idrissi and {Illuminati}, G. and {de Jong}, M. and {de Jong}, P. and {Jung}, B.~J. and {Kalaczy{\'n}ski}, P. and {Kalekin}, O. and {Katz}, U.~F. and {Kistauri}, G. and {Kopper}, C. and {Kouchner}, A. and {Kueviakoe}, V. and {Kulikovskiy}, V. and {Kvatadze}, R. and {Labalme}, M. and {Lahmann}, R. and {Larosa}, G. and {Lastoria}, C. and {Lazo}, A. and {Le Stum}, S. and {Lehaut}, G. and {Leonora}, E. and {Lessing}, N. and {Levi}, G. and {Clark}, M. Lindsey and {Longhitano}, F. and {Magnani}, F. and {Majumdar}, J. and {Malerba}, L. and {Mamedov}, F. and {Ma{\'n}czak}, J. and {Manfreda}, A. and {Manzaneda}, M. and {Marconi}, M. and {Margiotta}, A. and {Marinelli}, A. and {Markou}, C. and {Martin}, L. and {Mart{\'\i}nez-Mora}, J.~A. and {Marzaioli}, F. and {Mastrodicasa}, M. and {Mastroianni}, S. and {Miccich{\`e}}, S. and {Miele}, G. and {Migliozzi}, P. and {Migneco}, E. and {Mitsou}, M.~L. and {Mollo}, C.~M. and {Morales-Gallegos}, L. and {Morga}, M. and {Moussa}, A. and {Mateo}, I. Mozun and {Muller}, R. and {Musone}, M.~R. and {Musumeci}, M. and {Navas}, S. and {Nayerhoda}, A. and {Nicolau}, C.~A. and {Nkosi}, B. and {Fearraigh}, B. {\'O}. and {Oliviero}, V. and {Orlando}, A. and {Oukacha}, E. and {Paesani}, D. and {Palacios Gonz{\'a}lez}, J. and {Papalashvili}, G. and {Parisi}, V. and {Gomez}, E.~J. Pastor and {P{\u{a}}un}, A.~M. and {P{\u{a}}v{\u{a}}la{\textcommabelow s}}, G.~E. and {Pe{\~n}a Mart{\'\i}nez}, S. and {Perrin-Terrin}, M.},
        title = "{Differential Sensitivity of the KM3NeT/ARCA detector to a diffuse neutrino flux and to point-like source emission: Exploring the case of the Starburst Galaxies}",
      journal = {Astroparticle Physics},
         year = 2024,
        month = oct,
       volume = {162},
          eid = {102990},
        pages = {102990},
          doi = {10.1016/j.astropartphys.2024.102990},
archivePrefix = {arXiv},
       eprint = {2402.09088},
 primaryClass = {astro-ph.HE},
       adsurl = {https://ui.adsabs.harvard.edu/abs/2024APh...16202990A}
}

@ARTICLE{AartsenIceCube2019EPJC79,
       author = {{Aartsen}, M.~G. and {Ackermann}, M. and {Adams}, J. and {Aguilar}, J.~A. and {Ahlers}, M. and {Ahrens}, M. and {Altmann}, D. and {Andeen}, K. and {Anderson}, T. and {Ansseau}, I. and {Anton}, G. and {Arg{\"u}elles}, C. and {Auffenberg}, J. and {Axani}, S. and {Backes}, P. and {Bagherpour}, H. and {Bai}, X. and {Barbano}, A. and {Barron}, J.~P. and {Barwick}, S.~W. and {Baum}, V. and {Bay}, R. and {Beatty}, J.~J. and {Becker Tjus}, J. and {Becker}, K.-H. and {BenZvi}, S. and {Berley}, D. and {Bernardini}, E. and {Besson}, D.~Z. and {Binder}, G. and {Bindig}, D. and {Blaufuss}, E. and {Blot}, S. and {Bohm}, C. and {B{\"o}rner}, M. and {Bos}, F. and {B{\"o}ser}, S. and {Botner}, O. and {Bourbeau}, E. and {Bourbeau}, J. and {Bradascio}, F. and {Braun}, J. and {Bretz}, H.-P. and {Bron}, S. and {Brostean-Kaiser}, J. and {Burgman}, A. and {Busse}, R.~S. and {Carver}, T. and {Chen}, C. and {Cheung}, E. and {Chirkin}, D. and {Clark}, K. and {Classen}, L. and {Collin}, G.~H. and {Conrad}, J.~M. and {Coppin}, P. and {Correa}, P. and {Cowen}, D.~F. and {Cross}, R. and {Dave}, P. and {Day}, M. and {de Andr{\'e}}, J.~P.~A.~M. and {De Clercq}, C. and {DeLaunay}, J.~J. and {Dembinski}, H. and {Deoskar}, K. and {De Ridder}, S. and {Desiati}, P. and {de Vries}, K.~D. and {de Wasseige}, G. and {de With}, M. and {DeYoung}, T. and {D{\'\i}az-V{\'e}lez}, J.~C. and {Dujmovic}, H. and {Dunkman}, M. and {Dvorak}, E. and {Eberhardt}, B. and {Ehrhardt}, T. and {Eichmann}, B. and {Eller}, P. and {Evenson}, P.~A. and {Fahey}, S. and {Fazely}, A.~R. and {Felde}, J. and {Filimonov}, K. and {Finley}, C. and {Franckowiak}, A. and {Friedman}, E. and {Fritz}, A. and {Gaisser}, T.~K. and {Gallagher}, J. and {Ganster}, E. and {Garrappa}, S. and {Gerhardt}, L. and {Ghorbani}, K. and {Giang}, W. and {Glauch}, T. and {Gl{\"u}senkamp}, T. and {Goldschmidt}, A. and {Gonzalez}, J.~G. and {Grant}, D. and {Griffith}, Z. and {Haack}, C. and {Hallgren}, A. and {Halve}, L. and {Halzen}, F. and {Hanson}, K. and {Hebecker}, D. and {Heereman}, D. and {Helbing}, K. and {Hellauer}, R. and {Hickford}, S. and {Hignight}, J. and {Hill}, G.~C. and {Hoffman}, K.~D. and {Hoffmann}, R. and {Hoinka}, T. and {Hokanson-Fasig}, B. and {Hoshina}, K. and {Huang}, F. and {Huber}, M. and {Hultqvist}, K. and {H{\"u}nnefeld}, M. and {Hussain}, R. and {In}, S. and {Iovine}, N. and {Ishihara}, A. and {Jacobi}, E. and {Japaridze}, G.~S. and {Jeong}, M. and {Jero}, K. and {Jones}, B.~J.~P. and {Kalaczynski}, P. and {Kang}, W. and {Kappes}, A. and {Kappesser}, D. and {Karg}, T. and {Karle}, A. and {Katz}, U. and {Kauer}, M. and {Keivani}, A. and {Kelley}, J.~L. and {Kheirandish}, A. and {Kim}, J. and {Kintscher}, T. and {Kiryluk}, J. and {Kittler}, T. and {Klein}, S.~R. and {Koirala}, R. and {Kolanoski}, H. and {K{\"o}pke}, L. and {Kopper}, C. and {Kopper}, S. and {Koskinen}, D.~J. and {Kowalski}, M. and {Krings}, K. and {Kroll}, M. and {Kr{\"u}ckl}, G. and {Kunwar}, S. and {Kurahashi}, N. and {Kyriacou}, A. and {Labare}, M. and {Lanfranchi}, J.~L. and {Larson}, M.~J. and {Lauber}, F. and {Leonard}, K. and {Leuermann}, M. and {Liu}, Q.~R. and {Lohfink}, E. and {Mariscal}, C.~J. Lozano and {Lu}, L. and {L{\"u}nemann}, J. and {Luszczak}, W. and {Madsen}, J. and {Maggi}, G. and {Mahn}, K.~B.~M. and {Makino}, Y. and {Mancina}, S. and {Mari{\textcommabelow s}}, I.~C. and {Maruyama}, R. and {Mase}, K. and {Maunu}, R. and {Meagher}, K. and {Medici}, M. and {Meier}, M. and {Menne}, T. and {Merino}, G. and {Meures}, T. and {Miarecki}, S. and {Micallef}, J. and {Moment{\'e}}, G. and {Montaruli}, T. and {Moore}, R.~W. and {Moulai}, M. and {Nagai}, R. and {Nahnhauer}, R. and {Nakarmi}, P. and {Naumann}, U. and {Neer}, G. and {Niederhausen}, H.},
        title = "{Search for steady point-like sources in the astrophysical muon neutrino flux with 8 years of IceCube data}",
      journal = {European Physical Journal C},
         year = 2019,
        month = mar,
       volume = {79},
       number = {3},
          eid = {234},
        pages = {234},
          doi = {10.1140/epjc/s10052-019-6680-0},
archivePrefix = {arXiv},
       eprint = {1811.07979},
 primaryClass = {hep-ph},
       adsurl = {https://ui.adsabs.harvard.edu/abs/2019EPJC...79..234A}
}

@ARTICLE{Poutanen2008MNRAS389,
       author = {{Poutanen}, Juri and {Zdziarski}, Andrzej A. and {Ibragimov}, Askar},
        title = "{Superorbital variability of X-ray and radio emission of Cyg X-1 - II. Dependence of the orbital modulation and spectral hardness on the superorbital phase}",
      journal = {\mnras},
         year = 2008,
        month = sep,
       volume = {389},
       number = {3},
        pages = {1427-1438},
          doi = {10.1111/j.1365-2966.2008.13666.x},
archivePrefix = {arXiv},
       eprint = {0802.1391},
 primaryClass = {astro-ph},
       adsurl = {https://ui.adsabs.harvard.edu/abs/2008MNRAS.389.1427P}
}

@INPROCEEDINGS{Tomsick2022icrcconfE,
       author = {{Tomsick}, J. and {COSI Collaboration}},
        title = "{The Compton Spectrometer and Imager Project for MeV Astronomy}",
    booktitle = {37th International Cosmic Ray Conference},
         year = 2022,
        month = mar,
          eid = {652},
        pages = {652},
          doi = {10.22323/1.395.0652},
archivePrefix = {arXiv},
       eprint = {2109.10403},
 primaryClass = {astro-ph.IM},
       adsurl = {https://ui.adsabs.harvard.edu/abs/2022icrc.confE.652T}
}

@ARTICLE{Kim2022ApJ925,
       author = {{Kim}, Dae-Won and {Kravchenko}, Evgeniya V. and {Kutkin}, Alexander M. and {B{\"o}ttcher}, Markus and {G{\'o}mez}, Jos{\'e} L. and {Gurwell}, Mark and {Jorstad}, Svetlana G. and {L{\"a}hteenm{\"a}ki}, Anne and {Marscher}, Alan P. and {Ramakrishnan}, Venkatessh and {Tornikoski}, Merja and {Trippe}, Sascha and {Weaver}, Zachary and {Williamson}, Karen E.},
        title = "{Radio and {\ensuremath{\gamma}}-Ray Activity in the Jet of the Blazar S5 0716+714}",
      journal = {\apj},
         year = 2022,
        month = jan,
       volume = {925},
       number = {1},
          eid = {64},
        pages = {64},
          doi = {10.3847/1538-4357/ac31b4},
archivePrefix = {arXiv},
       eprint = {2111.03006},
 primaryClass = {astro-ph.HE},
       adsurl = {https://ui.adsabs.harvard.edu/abs/2022ApJ...925...64K}
}

@ARTICLE{MAGIC2014A&A569,
       author = {{Aleksi{\'c}}, J. and {Ansoldi}, S. and {Antonelli}, L.~A. and {Antoranz}, P. and {Babic}, A. and {Bangale}, P. and {Barres de Almeida}, U. and {Barrio}, J.~A. and {Becerra Gonz{\'a}lez}, J. and {Bednarek}, W. and {Bernardini}, E. and {Biland}, A. and {Blanch}, O. and {Bonnefoy}, S. and {Bonnoli}, G. and {Borracci}, F. and {Bretz}, T. and {Carmona}, E. and {Carosi}, A. and {Carreto Fidalgo}, D. and {Colin}, P. and {Colombo}, E. and {Contreras}, J.~L. and {Cortina}, J. and {Covino}, S. and {Da Vela}, P. and {Dazzi}, F. and {De Angelis}, A. and {De Caneva}, G. and {De Lotto}, B. and {Delgado Mendez}, C. and {Doert}, M. and {Dom{\'\i}nguez}, A. and {Dominis Prester}, D. and {Dorner}, D. and {Doro}, M. and {Einecke}, S. and {Eisenacher}, D. and {Elsaesser}, D. and {Farina}, E. and {Ferenc}, D. and {Fonseca}, M.~V. and {Font}, L. and {Frantzen}, K. and {Fruck}, C. and {Garc{\'\i}a L{\'o}pez}, R.~J. and {Garczarczyk}, M. and {Garrido Terrats}, D. and {Gaug}, M. and {Godinovi{\'c}}, N. and {Gonz{\'a}lez Mu{\~n}oz}, A. and {Gozzini}, S.~R. and {Hadasch}, D. and {Hayashida}, M. and {Herrera}, J. and {Herrero}, A. and {Hildebrand}, D. and {Hose}, J. and {Hrupec}, D. and {Idec}, W. and {Kadenius}, V. and {Kellermann}, H. and {Kodani}, K. and {Konno}, Y. and {Krause}, J. and {Kubo}, H. and {Kushida}, J. and {La Barbera}, A. and {Lelas}, D. and {Lewandowska}, N. and {Lindfors}, E. and {Lombardi}, S. and {L{\'o}pez}, M. and {L{\'o}pez-Coto}, R. and {L{\'o}pez-Oramas}, A. and {Lorenz}, E. and {Lozano}, I. and {Makariev}, M. and {Mallot}, K. and {Maneva}, G. and {Mankuzhiyil}, N. and {Mannheim}, K. and {Maraschi}, L. and {Marcote}, B. and {Mariotti}, M. and {Mart{\'\i}nez}, M. and {Mazin}, D. and {Menzel}, U. and {Meucci}, M. and {Miranda}, J.~M. and {Mirzoyan}, R. and {Moralejo}, A. and {Munar-Adrover}, P. and {Nakajima}, D. and {Niedzwiecki}, A. and {Nilsson}, K. and {Nishijima}, K. and {Noda}, K. and {Nowak}, N. and {Orito}, R. and {Overkemping}, A. and {Paiano}, S. and {Palatiello}, M. and {Paneque}, D. and {Paoletti}, R. and {Paredes}, J.~M. and {Paredes-Fortuny}, X. and {Partini}, S. and {Persic}, M. and {Prada}, F. and {Prada Moroni}, P.~G. and {Prandini}, E. and {Preziuso}, S. and {Puljak}, I. and {Reinthal}, R. and {Rhode}, W. and {Rib{\'o}}, M. and {Rico}, J. and {Rodriguez Garcia}, J. and {R{\"u}gamer}, S. and {Saggion}, A. and {Saito}, T. and {Saito}, K. and {Satalecka}, K. and {Scalzotto}, V. and {Scapin}, V. and {Schultz}, C. and {Schweizer}, T. and {Shore}, S.~N. and {Sillanp{\"a}{\"a}}, A. and {Sitarek}, J. and {Snidaric}, I. and {Sobczynska}, D. and {Spanier}, F. and {Stamatescu}, V. and {Stamerra}, A. and {Steinbring}, T. and {Storz}, J. and {Strzys}, M. and {Sun}, S. and {Suri{\'c}}, T. and {Takalo}, L. and {Takami}, H. and {Tavecchio}, F. and {Temnikov}, P. and {Terzi{\'c}}, T. and {Tescaro}, D. and {Teshima}, M. and {Thaele}, J. and {Tibolla}, O. and {Torres}, D.~F. and {Toyama}, T. and {Treves}, A. and {Uellenbeck}, M. and {Vogler}, P. and {Wagner}, R.~M. and {Zandanel}, F. and {Zanin}, R. and {MAGIC Collaboration} and {Lucarelli}, F. and {Pittori}, C. and {Vercellone}, S. and {Verrecchia}, F. and {AGILE Collaboration} and {Buson}, S. and {D'Ammando}, F. and {Stawarz}, L. and {Giroletti}, M. and {Orienti}, M. and {Fermi-LAT Collaboration} and {Mundell}, C. and {Steele}, I. and {Zarpudin}, B. and {Raiteri}, C.~M. and {Villata}, M. and {Sandrinelli}, A. and {L{\"a}hteenm{\"a}ki}, A. and {Tammi}, J. and {Tornikoski}, M. and {Hovatta}, T. and {Readhead}, A.~C.~S. and {Max-Moerbeck}, W. and {Richards}, J.~L. and {Jorstad}, S. and {Marscher}, A. and {Gurwell}, M.~A. and {Larionov}, V.~M. and {Blinov}, D.~A. and {Konstantinova}, T.~S. and {Kopatskaya}, E.~N. and {Larionova}, L.~V. and {Larionova}, E.~G. and {Morozova}, D.~A. and {Troitsky}, I.~S. and {Mokrushina}, A.~A. and {Pavlova}, Yu. V. and {Chen}, W.~P. and {Lin}, H.~C. and {Panwar}, N. and {Agudo}, I.},
        title = "{MAGIC gamma-ray and multi-frequency observations of flat spectrum radio quasar PKS 1510-089 in early 2012}",
      journal = {\aap},
         year = 2014,
        month = sep,
       volume = {569},
          eid = {A46},
        pages = {A46},
          doi = {10.1051/0004-6361/201423484},
archivePrefix = {arXiv},
       eprint = {1401.5646},
 primaryClass = {astro-ph.HE},
       adsurl = {https://ui.adsabs.harvard.edu/abs/2014A&A...569A..46A}
}

@ARTICLE{Zhang2013ApJ774,
       author = {{Zhang}, H. and {B{\"o}ttcher}, M.},
        title = "{X-Ray and Gamma-Ray Polarization in Leptonic and Hadronic Jet Models of Blazars}",
      journal = {\apj},
         year = 2013,
        month = sep,
       volume = {774},
       number = {1},
          eid = {18},
        pages = {18},
          doi = {10.1088/0004-637X/774/1/18},
archivePrefix = {arXiv},
       eprint = {1307.4187},
 primaryClass = {astro-ph.HE},
       adsurl = {https://ui.adsabs.harvard.edu/abs/2013ApJ...774...18Z}
}

@ARTICLE{Peirson2019ApJ885,
       author = {{Peirson}, A.~L. and {Romani}, Roger W.},
        title = "{The Polarization Behavior of Relativistic Synchrotron Self-Compton Jets}",
      journal = {\apj},
         year = 2019,
        month = nov,
       volume = {885},
       number = {1},
          eid = {76},
        pages = {76},
          doi = {10.3847/1538-4357/ab46b1},
archivePrefix = {arXiv},
       eprint = {1909.10563},
 primaryClass = {astro-ph.HE},
       adsurl = {https://ui.adsabs.harvard.edu/abs/2019ApJ...885...76P}
}

@ARTICLE{Kouch2025A&A695,
       author = {{Kouch}, Pouya M. and {Liodakis}, Ioannis and {Fenu}, Francesco and {Zhang}, Haocheng and {Boula}, Stella and {Middei}, Riccardo and {Di Gesu}, Laura and {Paraschos}, Georgios F. and {Agudo}, Iv{\'a}n and {Jorstad}, Svetlana G. and {Lindfors}, Elina and {Marscher}, Alan P. and {Krawczynski}, Henric and {Negro}, Michela and {Hu}, Kun and {Kim}, Dawoon E. and {Cavazzuti}, Elisabetta and {Errando}, Manel and {Blinov}, Dmitry and {Gourni}, Anastasia and {Kiehlmann}, Sebastian and {Kourtidis}, Angelos and {Mandarakas}, Nikos and {Triantafyllou}, Nikolaos and {Vervelaki}, Anna and {Borman}, George A. and {Kopatskaya}, Evgenia N. and {Larionova}, Elena G. and {Morozova}, Daria A. and {Savchenko}, Sergey S. and {Vasilyev}, Andrey A. and {Troitskiy}, Ivan S. and {Grishina}, Tatiana S. and {Shishkina}, Ekaterina V. and {Zhovtan}, Alexey V. and {Aceituno}, Francisco Jos{\'e} and {Bonnoli}, Giacomo and {Casanova}, V{\'\i}ctor and {Escudero}, Juan and {Ag{\'\i}s-Gonz{\'a}lez}, Beatriz and {Husillos}, C{\'e}sar and {Otero-Santos}, Jorge and {Piirola}, Vilppu and {Sota}, Alfredo and {Myserlis}, Ioannis and {Gurwell}, Mark and {Keating}, Garrett K. and {Rao}, Ramprasad and {Angelakis}, Emmanouil and {Kraus}, Alexander and {Antonelli}, Lucio Angelo and {Bachetti}, Matteo and {Baldini}, Luca and {Baumgartner}, Wayne H. and {Bellazzini}, Ronaldo and {Bianchi}, Stefano and {Bongiorno}, Stephen D. and {Bonino}, Raffaella and {Brez}, Alessandro and {Bucciantini}, Niccol{\`o} and {Capitanio}, Fiamma and {Castellano}, Simone and {Chen}, Chien-Ting and {Ciprini}, Stefano and {Costa}, Enrico and {De Rosa}, Alessandra and {Del Monte}, Ettore and {Di Lalla}, Niccol{\`o} and {Di Marco}, Alessandro and {Donnarumma}, Immacolata and {Doroshenko}, Victor and {Dov{\v{c}}iak}, Michal and {Ehlert}, Steven R. and {Enoto}, Teruaki and {Evangelista}, Yuri and {Fabiani}, Sergio and {Ferrazzoli}, Riccardo and {Garcia}, Javier A. and {Gunji}, Shuichi and {Hayashida}, Kiyoshi and {Heyl}, Jeremy and {Iwakiri}, Wataru and {Kaaret}, Philip and {Karas}, Vladimir and {Kislat}, Fabian and {Kitaguchi}, Takao and {Kolodziejczak}, Jeffery J. and {La Monaca}, Fabio and {Latronico}, Luca and {Maldera}, Simone and {Manfreda}, Alberto and {Marin}, Fr{\'e}d{\'e}ric and {Marinucci}, Andrea and {Marshall}, Herman L. and {Massaro}, Francesco and {Matt}, Giorgio and {Mitsuishi}, Ikuyuki and {Mizuno}, Tsunefumi and {Muleri}, Fabio and {Ng}, Chi-Yung and {O'Dell}, Stephen L. and {Omodei}, Nicola and {Oppedisano}, Chiara and {Papitto}, Alessandro and {Pavlov}, George G. and {Peirson}, Abel Lawrence and {Perri}, Matteo and {Pesce-Rollins}, Melissa and {Petrucci}, Pierre-Olivier and {Pilia}, Maura and {Possenti}, Andrea and {Poutanen}, Juri and {Puccetti}, Simonetta and {Ramsey}, Brian D. and {Rankin}, John and {Ratheesh}, Ajay and {Roberts}, Oliver J. and {Sgr{\`o}}, Carmelo and {Slane}, Patrick and {Soffitta}, Paolo and {Spandre}, Gloria and {Swartz}, Douglas A. and {Tamagawa}, Toru and {Tavecchio}, Fabrizio and {Taverna}, Roberto and {Tawara}, Yuzuru and {Tennant}, Allyn F. and {Thomas}, Nicholas E. and {Tombesi}, Francesco and {Trois}, Alessio and {Tsygankov}, Sergey S. and {Turolla}, Roberto and {Romani}, Roger W. and {Vink}, Jacco and {Weisskopf}, Martin C. and {Wu}, Kinwah and {Xie}, Fei and {Zane}, Silvia},
        title = "{IXPE observation of the low-synchrotron peaked blazar S4 0954+65 during an optical-X-ray flare}",
      journal = {\aap},
         year = 2025,
        month = mar,
       volume = {695},
          eid = {A99},
        pages = {A99},
          doi = {10.1051/0004-6361/202453127},
archivePrefix = {arXiv},
       eprint = {2411.16868},
 primaryClass = {astro-ph.HE},
       adsurl = {https://ui.adsabs.harvard.edu/abs/2025A&A...695A..99K}
}

@ARTICLE{Kouch2026A&A708_A382,
       author = {{Kouch}, Pouya M. and {Lindfors}, Elina and {Hovatta}, Talvikki and {Liodakis}, Ioannis and {Koljonen}, Karri I.~I. and {Paggi}, Alessandro and {Nilsson}, Kari and {Jormanainen}, Jenni and {Fallah Ramazani}, Vandad and {Kankkunen}, Sofia and {Wierda}, Folkert and {Wagner}, Sarah M. and {Graham}, Matthew J.},
        title = "{CAZ catalog and optical light curves of 7918 blazar-selected active galactic nuclei}",
      journal = {\aap},
         year = 2026,
        month = apr,
       volume = {708},
          eid = {A382},
        pages = {A382},
          doi = {10.1051/0004-6361/202557582},
archivePrefix = {arXiv},
       eprint = {2510.16584},
 primaryClass = {astro-ph.HE},
       adsurl = {https://ui.adsabs.harvard.edu/abs/2026A&A...708A.382K}
}

@ARTICLE{Kouch2026A&A708_A383,
       author = {{Kouch}, Pouya M. and {Hovatta}, Talvikki and {Lindfors}, Elina and {Liodakis}, Ioannis and {Koljonen}, Karri I.~I. and {Paggi}, Alessandro},
        title = "{Association of the IceCube neutrinos with CAZ blazar light curves}",
      journal = {\aap},
         year = 2026,
        month = apr,
       volume = {708},
          eid = {A383},
        pages = {A383},
          doi = {10.1051/0004-6361/202557584},
archivePrefix = {arXiv},
       eprint = {2510.16585},
 primaryClass = {astro-ph.HE},
       adsurl = {https://ui.adsabs.harvard.edu/abs/2026A&A...708A.383K}
}

@ARTICLE{Giommi2020MNRAS497,
       author = {{Giommi}, P. and {Glauch}, T. and {Padovani}, P. and {Resconi}, E. and {Turcati}, A. and {Chang}, Y.~L.},
        title = "{Dissecting the regions around IceCube high-energy neutrinos: growing evidence for the blazar connection}",
      journal = {\mnras},
         year = 2020,
        month = sep,
       volume = {497},
       number = {1},
        pages = {865-878},
          doi = {10.1093/mnras/staa2082},
archivePrefix = {arXiv},
       eprint = {2001.09355},
 primaryClass = {astro-ph.HE},
       adsurl = {https://ui.adsabs.harvard.edu/abs/2020MNRAS.497..865G}
}

@ARTICLE{Shah2021MNRAS504,
       author = {{Shah}, Zahir and {Jithesh}, V. and {Sahayanathan}, S. and {Iqbal}, Naseer},
        title = "{Unveiling the broad-band spectral and temporal properties of PKS 0903-57 during its brightest flare}",
      journal = {\mnras},
         year = 2021,
        month = jun,
       volume = {504},
       number = {1},
        pages = {416-427},
          doi = {10.1093/mnras/stab834},
archivePrefix = {arXiv},
       eprint = {2103.13657},
 primaryClass = {astro-ph.HE},
       adsurl = {https://ui.adsabs.harvard.edu/abs/2021MNRAS.504..416S}
}

@ARTICLE{Abdo2011ApJ726,
       author = {{Abdo}, A.~A. and {Ackermann}, M. and {Ajello}, M. and {Baldini}, L. and {Ballet}, J. and {Barbiellini}, G. and {Bastieri}, D. and {Bechtol}, K. and {Bellazzini}, R. and {Berenji}, B. and {Blandford}, R.~D. and {Bonamente}, E. and {Borgland}, A.~W. and {Bouvier}, A. and {Bregeon}, J. and {Brez}, A. and {Brigida}, M. and {Bruel}, P. and {Buehler}, R. and {Buson}, S. and {Caliandro}, G.~A. and {Cameron}, R.~A. and {Caraveo}, P.~A. and {Carrigan}, S. and {Casandjian}, J.~M. and {Cavazzuti}, E. and {Cecchi}, C. and {{\c{C}}elik}, {\"O}. and {Charles}, E. and {Chekhtman}, A. and {Cheung}, C.~C. and {Chiang}, J. and {Ciprini}, S. and {Claus}, R. and {Cohen-Tanugi}, J. and {Conrad}, J. and {Costamante}, L. and {Cutini}, S. and {Davis}, D.~S. and {Dermer}, C.~D. and {de Palma}, F. and {Digel}, S.~W. and {do Couto e Silva}, E. and {Drell}, P.~S. and {Dubois}, R. and {Dumora}, D. and {Favuzzi}, C. and {Fegan}, S.~J. and {Fortin}, P. and {Frailis}, M. and {Fuhrmann}, L. and {Fukazawa}, Y. and {Funk}, S. and {Fusco}, P. and {Gargano}, F. and {Gasparrini}, D. and {Gehrels}, N. and {Germani}, S. and {Giglietto}, N. and {Giommi}, P. and {Giordano}, F. and {Giroletti}, M. and {Glanzman}, T. and {Godfrey}, G. and {Grenier}, I.~A. and {Grove}, J.~E. and {Guillemot}, L. and {Guiriec}, S. and {Hadasch}, D. and {Hayashida}, M. and {Hays}, E. and {Horan}, D. and {Hughes}, R.~E. and {Itoh}, R. and {J{\'o}hannesson}, G. and {Johnson}, A.~S. and {Johnson}, T.~J. and {Johnson}, W.~N. and {Kamae}, T. and {Katagiri}, H. and {Kataoka}, J. and {Kn{\"o}dlseder}, J. and {Kuss}, M. and {Lande}, J. and {Latronico}, L. and {Lee}, S.-H. and {Longo}, F. and {Loparco}, F. and {Lott}, B. and {Lovellette}, M.~N. and {Lubrano}, P. and {Makeev}, A. and {Mazziotta}, M.~N. and {McEnery}, J.~E. and {Mehault}, J. and {Michelson}, P.~F. and {Mizuno}, T. and {Moiseev}, A.~A. and {Monte}, C. and {Monzani}, M.~E. and {Morselli}, A. and {Moskalenko}, I.~V. and {Murgia}, S. and {Nakamori}, T. and {Naumann-Godo}, M. and {Nestoras}, I. and {Nolan}, P.~L. and {Norris}, J.~P. and {Nuss}, E. and {Ohsugi}, T. and {Okumura}, A. and {Omodei}, N. and {Orlando}, E. and {Ormes}, J.~F. and {Ozaki}, M. and {Paneque}, D. and {Panetta}, J.~H. and {Parent}, D. and {Pelassa}, V. and {Pepe}, M. and {Pesce-Rollins}, M. and {Piron}, F. and {Porter}, T.~A. and {Rain{\`o}}, S. and {Rando}, R. and {Razzano}, M. and {Reimer}, A. and {Reimer}, O. and {Reyes}, L.~C. and {Ripken}, J. and {Ritz}, S. and {Romani}, R.~W. and {Roth}, M. and {Sadrozinski}, H.~F.-W. and {Sanchez}, D. and {Sander}, A. and {Scargle}, J.~D. and {Sgr{\`o}}, C. and {Shaw}, M.~S. and {Smith}, P.~D. and {Spandre}, G. and {Spinelli}, P. and {Strickman}, M.~S. and {Suson}, D.~J. and {Takahashi}, H. and {Tanaka}, T. and {Thayer}, J.~B. and {Thayer}, J.~G. and {Thompson}, D.~J. and {Tibaldo}, L. and {Torres}, D.~F. and {Tosti}, G. and {Tramacere}, A. and {Usher}, T.~L. and {Vandenbroucke}, J. and {Vasileiou}, V. and {Vilchez}, N. and {Vitale}, V. and {Waite}, A.~P. and {Wang}, P. and {Winer}, B.~L. and {Wood}, K.~S. and {Yang}, Z. and {Ylinen}, T. and {Ziegler}, M. and {Acciari}, V.~A. and {Aliu}, E. and {Arlen}, T. and {Aune}, T. and {Beilicke}, M. and {Benbow}, W. and {B{\"o}ttcher}, M. and {Boltuch}, D. and {Bradbury}, S.~M. and {Buckley}, J.~H. and {Bugaev}, V. and {Byrum}, K. and {Cannon}, A. and {Cesarini}, A. and {Christiansen}, J.~L. and {Ciupik}, L. and {Cui}, W. and {de la Calle Perez}, I. and {Dickherber}, R. and {Errando}, M. and {Falcone}, A. and {Finley}, J.~P. and {Finnegan}, G. and {Fortson}, L. and {Furniss}, A. and {Galante}, N. and {Gall}, D. and {Gillanders}, G.~H. and {Godambe}, S. and {Grube}, J. and {Guenette}, R. and {Gyuk}, G. and {Hanna}, D. and {Holder}, J. and {Hui}, C.~M.},
        title = "{Multi-wavelength Observations of the Flaring Gamma-ray Blazar 3C 66A in 2008 October}",
      journal = {\apj},
         year = 2011,
        month = jan,
       volume = {726},
       number = {1},
          eid = {43},
        pages = {43},
          doi = {10.1088/0004-637X/726/1/43},
archivePrefix = {arXiv},
       eprint = {1011.1053},
 primaryClass = {astro-ph.HE},
       adsurl = {https://ui.adsabs.harvard.edu/abs/2011ApJ...726...43A}
}

@ARTICLE{Acciari2009ApJ707,
       author = {{Acciari}, V.~A. and {Aliu}, E. and {Aune}, T. and {Beilicke}, M. and {Benbow}, W. and {B{\"o}ttcher}, M. and {Boltuch}, D. and {Buckley}, J.~H. and {Bradbury}, S.~M. and {Bugaev}, V. and {Byrum}, K. and {Cannon}, A. and {Cesarini}, A. and {Ciupik}, L. and {Cogan}, P. and {Cui}, W. and {Dickherber}, R. and {Duke}, C. and {Falcone}, A. and {Finley}, J.~P. and {Fortin}, P. and {Fortson}, L. and {Furniss}, A. and {Galante}, N. and {Gall}, D. and {Gibbs}, K. and {Gillanders}, G.~H. and {Grube}, J. and {Guenette}, R. and {Gyuk}, G. and {Hanna}, D. and {Holder}, J. and {Hui}, C.~M. and {Humensky}, T.~B. and {Kaaret}, P. and {Karlsson}, N. and {Kertzman}, M. and {Kieda}, D. and {Konopelko}, A. and {Krawczynski}, H. and {Krennrich}, F. and {Lang}, M.~J. and {Le Bohec}, S. and {Maier}, G. and {McArthur}, S. and {McCann}, A. and {McCutcheon}, M. and {Millis}, J. and {Moriarty}, P. and {Ong}, R.~A. and {Otte}, A.~N. and {Pandel}, D. and {Perkins}, J.~S. and {Pichel}, A. and {Pohl}, M. and {Quinn}, J. and {Ragan}, K. and {Reyes}, L.~C. and {Reynolds}, P.~T. and {Roache}, E. and {Rose}, H.~J. and {Sembroski}, G.~H. and {Smith}, A.~W. and {Steele}, D. and {Theiling}, M. and {Thibadeau}, S. and {Varlotta}, A. and {Vassiliev}, V.~V. and {Vincent}, S. and {Wakely}, S.~P. and {Ward}, J.~E. and {Weekes}, T.~C. and {Weinstein}, A. and {Weisgarber}, T. and {Williams}, D.~A. and {Wissel}, S. and {Wood}, M. and {Pian}, E. and {Vercellone}, S. and {Donnarumma}, I. and {D'Ammando}, F. and {Bulgarelli}, A. and {Chen}, A.~W. and {Giuliani}, A. and {Longo}, F. and {Pacciani}, L. and {Pucella}, G. and {Vittorini}, V. and {Tavani}, M. and {Argan}, A. and {Barbiellini}, G. and {Caraveo}, P. and {Cattaneo}, P.~W. and {Cocco}, V. and {Costa}, E. and {Del Monte}, E. and {De Paris}, G. and {Di Cocco}, G. and {Evangelista}, Y. and {Feroci}, M. and {Fiorini}, M. and {Froysland}, T. and {Frutti}, M. and {Fuschino}, F. and {Galli}, M. and {Gianotti}, F. and {Labanti}, C. and {Lapshov}, I. and {Lazzarotto}, F. and {Lipari}, P. and {Marisaldi}, M. and {Mastropietro}, M. and {Mereghetti}, S. and {Morelli}, E. and {Morselli}, A. and {Pellizzoni}, A. and {Perotti}, F. and {Piano}, G. and {Picozza}, P. and {Pilia}, M. and {Porrovecchio}, G. and {Prest}, M. and {Rapisarda}, M. and {Rappoldi}, A. and {Rubini}, A. and {Sabatini}, S. and {Soffitta}, P. and {Trifoglio}, M. and {Trois}, A. and {Vallazza}, E. and {Zambra}, A. and {Zanello}, D. and {Pittori}, C. and {Santolamazza}, P. and {Verrecchia}, F. and {Giommi}, P. and {Colafrancesco}, S. and {Salotti}, L. and {Villata}, M. and {Raiteri}, C.~M. and {Aller}, H.~D. and {Aller}, M.~F. and {Arkharov}, A.~A. and {Efimova}, N.~V. and {Larionov}, V.~M. and {Leto}, P. and {Ligustri}, R. and {Lindfors}, E. and {Pasanen}, M. and {Kurtanidze}, O.~M. and {Tetradze}, S.~D. and {Lahteenmaki}, A. and {Kotiranta}, M. and {Cucchiara}, A. and {Romano}, P. and {Nesci}, R. and {Pursimo}, T. and {Heidt}, J. and {Benitez}, E. and {Hiriart}, D. and {Nilsson}, K. and {Berdyugin}, A. and {Mujica}, R. and {Dultzin}, D. and {Lopez}, J.~M. and {Mommert}, M. and {Sorcia}, M. and {de la Calle Perez}, I.},
        title = "{Multiwavelength Observations of a TeV-Flare from W Comae}",
      journal = {\apj},
         year = 2009,
        month = dec,
       volume = {707},
       number = {1},
        pages = {612-620},
          doi = {10.1088/0004-637X/707/1/612},
archivePrefix = {arXiv},
       eprint = {0910.3750},
 primaryClass = {astro-ph.HE},
       adsurl = {https://ui.adsabs.harvard.edu/abs/2009ApJ...707..612A}
}

@ARTICLE{Petropoulou2015MNRAS448,
       author = {{Petropoulou}, M. and {Dimitrakoudis}, S. and {Padovani}, P. and {Mastichiadis}, A. and {Resconi}, E.},
        title = "{Photohadronic origin of {\ensuremath{\gamma}} -ray BL Lac emission: implications for IceCube neutrinos}",
      journal = {\mnras},
         year = 2015,
        month = apr,
       volume = {448},
       number = {3},
        pages = {2412-2429},
          doi = {10.1093/mnras/stv179},
archivePrefix = {arXiv},
       eprint = {1501.07115},
 primaryClass = {astro-ph.HE},
       adsurl = {https://ui.adsabs.harvard.edu/abs/2015MNRAS.448.2412P}
}

@ARTICLE{Bottcher2013ApJ768,
       author = {{B{\"o}ttcher}, M. and {Reimer}, A. and {Sweeney}, K. and {Prakash}, A.},
        title = "{Leptonic and Hadronic Modeling of Fermi-detected Blazars}",
      journal = {\apj},
         year = 2013,
        month = may,
       volume = {768},
       number = {1},
          eid = {54},
        pages = {54},
          doi = {10.1088/0004-637X/768/1/54},
archivePrefix = {arXiv},
       eprint = {1304.0605},
 primaryClass = {astro-ph.HE},
       adsurl = {https://ui.adsabs.harvard.edu/abs/2013ApJ...768...54B}
}

@ARTICLE{Marscher2010ApJ710,
       author = {{Marscher}, Alan P. and {Jorstad}, Svetlana G. and {Larionov}, Valeri M. and {Aller}, Margo F. and {Aller}, Hugh D. and {L{\"a}hteenm{\"a}ki}, Anne and {Agudo}, Iv{\'a}n and {Smith}, Paul S. and {Gurwell}, Mark and {Hagen-Thorn}, Vladimir A. and {Konstantinova}, Tatiana S. and {Larionova}, Elena G. and {Larionova}, Liudmila V. and {Melnichuk}, Daria A. and {Blinov}, Dmitry A. and {Kopatskaya}, Evgenia N. and {Troitsky}, Ivan S. and {Tornikoski}, Merja and {Hovatta}, Talvikki and {Schmidt}, Gary D. and {D'Arcangelo}, Francesca D. and {Bhattarai}, Dipesh and {Taylor}, Brian and {Olmstead}, Alice R. and {Manne-Nicholas}, Emily and {Roca-Sogorb}, Mar and {G{\'o}mez}, Jos{\'e} L. and {McHardy}, Ian M. and {Kurtanidze}, Omar and {Nikolashvili}, Maria G. and {Kimeridze}, Givi N. and {Sigua}, Lorand A.},
        title = "{Probing the Inner Jet of the Quasar PKS 1510-089 with Multi-Waveband Monitoring During Strong Gamma-Ray Activity}",
      journal = {\apjl},
         year = 2010,
        month = feb,
       volume = {710},
       number = {2},
        pages = {L126-L131},
          doi = {10.1088/2041-8205/710/2/L126},
archivePrefix = {arXiv},
       eprint = {1001.2574},
 primaryClass = {astro-ph.CO},
       adsurl = {https://ui.adsabs.harvard.edu/abs/2010ApJ...710L.126M}
}

@ARTICLE{Marscher1985ApJ298,
       author = {{Marscher}, A.~P. and {Gear}, W.~K.},
        title = "{Models for high-frequency radio outbursts in extragalactic sources, with application to the early 1983 millimeter-to-infrared flare of 3C 273.}",
      journal = {\apj},
         year = 1985,
        month = nov,
       volume = {298},
        pages = {114-127},
          doi = {10.1086/163592},
       adsurl = {https://ui.adsabs.harvard.edu/abs/1985ApJ...298..114M}
}

@ARTICLE{Chatterjee2008ApJ689,
       author = {{Chatterjee}, Ritaban and {Jorstad}, Svetlana G. and {Marscher}, Alan P. and {Oh}, Haruki and {McHardy}, Ian M. and {Aller}, Margo F. and {Aller}, Hugh D. and {Balonek}, Thomas J. and {Miller}, H. Richard and {Ryle}, Wesley T. and {Tosti}, Gino and {Kurtanidze}, Omar and {Nikolashvili}, Maria and {Larionov}, Valeri M. and {Hagen-Thorn}, Vladimir A.},
        title = "{Correlated Multi-Wave Band Variability in the Blazar 3C 279 from 1996 to 2007}",
      journal = {\apj},
         year = 2008,
        month = dec,
       volume = {689},
       number = {1},
        pages = {79-94},
          doi = {10.1086/592598},
archivePrefix = {arXiv},
       eprint = {0808.2194},
 primaryClass = {astro-ph},
       adsurl = {https://ui.adsabs.harvard.edu/abs/2008ApJ...689...79C}
}

@BOOK{Longair2011hea_book,
       author = {{Longair}, Malcolm S.},
        title = "{High Energy Astrophysics}",
         year = 2011,
       adsurl = {https://ui.adsabs.harvard.edu/abs/2011hea..book.....L}
}

@INCOLLECTION{Yuan2022,
       author = {{Yuan}, Weimin and {Zhang}, Chen and {Chen}, Yong and {Ling}, Zhixing},
        title = "{The Einstein Probe Mission}",
    booktitle = {Handbook of X-ray and Gamma-ray Astrophysics},
         year = 2022,
       editor = {{Bambi}, Cosimo and {Sangangelo}, Andrea},
          eid = {86},
        pages = {86},
          doi = {10.1007/978-981-16-4544-0_151-1},
       adsurl = {https://ui.adsabs.harvard.edu/abs/2022hxga.book...86Y},
    publisher = {Springer}
}

@article{Yuan2025,
Author = {Yuan, Weimin and Dai, Lixin and Feng, Hua and Jin, Chichuan and Jonker,
   Peter and Kuulkers, Erik and Liu, Yuan and Nandra, Kirpal and O'Brien,
   Paul and Piro, Luigi and Rau, Arne and Rea, Nanda and Sanders, Jeremy
   and Tao, Lian and Wang, Junfeng and Wu, Xuefeng and Zhang, Bing and
   Zhang, Shuangnan and Ai, Shunke and Buchner, Johannes and Bulbul, Esra
   and Chen, Hechao and Chen, Minghua and Chen, Yong and Chen, Yu-Peng and
   Coleiro, Alexis and Zelati, Francesco Coti and Dai, Zigao and Fan,
   Xilong and Fan, Zhou and Friedrich, Susanne and Gao, He and Ge, Chong
   and Ge, Mingyu and Geng, Jinjun and Ghirlanda, Giancarlo and Gianfagna,
   Giulia and Gou, Lijun and Guillot, Sebastien and Hou, Xian and Hu,
   Jingwei and Huang, Yongfeng and Ji, Long and Jia, Shumei and Komossa, S.
   and Kong, Albert K. H. and Lan, Lin and Li, An and Li, Ang and Li,
   Chengkui and Li, Dongyue and Li, Jian and Li, Zhaosheng and Ling,
   Zhixing and Liu, Ang and Liu, Jinzhong and Liu, Liangduan and Liu, Zhu
   and Luo, Jiawei and Ma, Ruican and Maggi, Pierre and Maitra, Chandreyee
   and Marino, Alessio and Ng, Stephen Chi-Yung and Pan, Haiwu and Rukdee,
   Surangkhana and Soria, Roberto and Sun, Hui and Tam, Pak-Hin Thomas and
   Thakur, Aishwarya Linesh and Tian, Hui and Troja, Eleonora and Wang, Wei
   and Wang, Xiangyu and Wang, Yanan and Wei, Junjie and Wen, Sixiang and
   Wu, Jianfeng and Wu, Ting and Xiao, Di and Xu, Dong and Xu, Renxin and
   Xu, Yanjun and Xu, Yu and Yang, Haonan and You, Bei and Yu, Heng and Yu,
   Yunwei and Zhang, Binbin and Zhang, Chen and Zhang, Guobao and Zhang,
   Liang and Zhang, Wenda and Zhang, Yu and Zhou, Ping and Zou, Zecheng},
Title = {Science objectives of the Einstein Probe mission},
Journal = {SCIENCE CHINA-PHYSICS MECHANICS \& ASTRONOMY},
Year = {2025},
Volume = {68},
Number = {3},
Month = {MAR},
DOI = {10.1007/s11433-024-2600-3},
Article-Number = {239501},
ISSN = {1674-7348},
EISSN = {1869-1927},
ResearcherID-Numbers = {Wu, Jianfeng/KCY-4997-2024
   xu, yanjun/LZE-9536-2025
   wang, yiran/IAP-0414-2023
   fan, xilong/B-1555-2009
   Zhang, Bing/AAG-2824-2019
   Luo, Jiawei/GYV-3644-2022
   LIU, JIALIN/JXN-8034-2024
   Dai, Lixin/Q-7569-2019
   Zelati, Francesco/A-5362-2018
   Ge, Chong/JFJ-3512-2023
   Yu, Heng/I-3485-2016
   yu, yun-wei/X-2088-2019
   Hou, Xian/AAL-5645-2021
   Chen, Minghua/JDW-6758-2023
   hb, l/KIA-5795-2024
   wang, sijia/IYJ-0510-2023
   You, Bei/HKF-8297-2023},
Unique-ID = {WOS:001407830200001},
}

@ARTICLE{Cheng2025,
       author = {{Cheng}, Huaqing and {Zhang}, Chen and {Ling}, Zhixing and {Sun}, Xiaojin and {Sun}, Shengli and {Liu}, Yuan and {Dai}, Yanfeng and {Jia}, Zhenqing and {Pan}, Haiwu and {Wang}, Wenxin and {Zhao}, Donghua and {Chen}, Yifan and {Cheng}, Zhiwei and {Fu}, Wei and {Han}, Yixiao and {Li}, Junfei and {Li}, Zhengda and {Ma}, Xiaohao and {Xue}, Yulong and {Yan}, Ailiang and {Zhang}, Qiang and {Wang}, Yusa and {Yang}, Xiongtao and {Zhao}, Zijian and {Li}, Longhui and {Jin}, Ge and {Yuan}, Weimin},
        title = "{Ground calibration result of the wide-field X-ray telescope (WXT) onboard the Einstein probe}",
      journal = {Experimental Astronomy},
         year = 2025,
        month = oct,
       volume = {60},
       number = {2},
          eid = {15},
        pages = {15},
          doi = {10.1007/s10686-025-10025-9},
archivePrefix = {arXiv},
       eprint = {2505.18939},
 primaryClass = {astro-ph.IM},
       adsurl = {https://ui.adsabs.harvard.edu/abs/2025ExA....60...15C}
}

@INPROCEEDINGS{Arnaud1996,
       author = {{Arnaud}, K.~A.},
        title = "{XSPEC: The First Ten Years}",
    booktitle = {Astronomical Data Analysis Software and Systems V},
         year = 1996,
       editor = {{Jacoby}, George H. and {Barnes}, Jeannette},
       series = {Astronomical Society of the Pacific Conference Series},
       volume = {101},
        month = jan,
        pages = {17},
       adsurl = {https://ui.adsabs.harvard.edu/abs/1996ASPC..101...17A}
}

@ARTICLE{Blumenthal1970PRD1,
       author = {{Blumenthal}, George R.},
        title = "{Energy Loss of High-Energy Cosmic Rays in Pair-Producing Collisions with Ambient Photons}",
      journal = {\prd},
         year = 1970,
        month = mar,
       volume = {1},
       number = {6},
        pages = {1596-1602},
          doi = {10.1103/PhysRevD.1.1596},
       adsurl = {https://ui.adsabs.harvard.edu/abs/1970PhRvD...1.1596B}
}

@ARTICLE{Chodorowski1992ApJ400,
       author = {{Chodorowski}, Michal J. and {Zdziarski}, Andrzej A. and {Sikora}, Marek},
        title = "{Reaction Rate and Energy-Loss Rate for Photopair Production by Relativistic Nuclei}",
      journal = {\apj},
         year = 1992,
        month = nov,
       volume = {400},
        pages = {181},
          doi = {10.1086/171984},
       adsurl = {https://ui.adsabs.harvard.edu/abs/1992ApJ...400..181C}
}

@ARTICLE{Begelman1990ApJ362,
       author = {{Begelman}, Mitchell C. and {Rudak}, Bronislaw and {Sikora}, Marek},
        title = "{Consequences of Relativistic Proton Injection in Active Galactic Nuclei}",
      journal = {\apj},
         year = 1990,
        month = oct,
       volume = {362},
        pages = {38},
          doi = {10.1086/169241},
       adsurl = {https://ui.adsabs.harvard.edu/abs/1990ApJ...362...38B}
}

@ARTICLE{Kelner2008PhRvD78,
       author = {{Kelner}, S.~R. and {Aharonian}, F.~A.},
        title = "{Energy spectra of gamma rays, electrons, and neutrinos produced at interactions of relativistic protons with low energy radiation}",
      journal = {\prd},
         year = 2008,
        month = aug,
       volume = {78},
       number = {3},
          eid = {034013},
        pages = {034013},
          doi = {10.1103/PhysRevD.78.034013},
archivePrefix = {arXiv},
       eprint = {0803.0688},
 primaryClass = {astro-ph},
       adsurl = {https://ui.adsabs.harvard.edu/abs/2008PhRvD..78c4013K}
}

@ARTICLE{Reimer2019ApJ881,
       author = {{Reimer}, Anita and {B{\"o}ttcher}, Markus and {Buson}, Sara},
        title = "{Cascading Constraints from Neutrino-emitting Blazars: The Case of TXS 0506+056}",
      journal = {\apj},
         year = 2019,
        month = aug,
       volume = {881},
       number = {1},
          eid = {46},
        pages = {46},
          doi = {10.3847/1538-4357/ab2bff},
archivePrefix = {arXiv},
       eprint = {1812.05654},
 primaryClass = {astro-ph.HE},
       adsurl = {https://ui.adsabs.harvard.edu/abs/2019ApJ...881...46R}
}

@ARTICLE{IceCube2018Sci361,
       author = {{IceCube Collaboration} and {Aartsen}, M.~G. and {Ackermann}, M. and {Adams}, J. and {Aguilar}, J.~A. and {Ahlers}, M. and {Ahrens}, M. and {Al Samarai}, I. and {Altmann}, D. and {Andeen}, K. and {Anderson}, T. and {Ansseau}, I. and {Anton}, G. and {Arg{\"u}elles}, C. and {Auffenberg}, J. and {Axani}, S. and {Bagherpour}, H. and {Bai}, X. and {Barron}, J.~P. and {Barwick}, S.~W. and {Baum}, V. and {Bay}, R. and {Beatty}, J.~J. and {Becker Tjus}, J. and {Becker}, K.-H. and {BenZvi}, S. and {Berley}, D. and {Bernardini}, E. and {Besson}, D.~Z. and {Binder}, G. and {Bindig}, D. and {Blaufuss}, E. and {Blot}, S. and {Bohm}, C. and {B{\"o}rner}, M. and {Bos}, F. and {B{\"o}ser}, S. and {Botner}, O. and {Bourbeau}, E. and {Bourbeau}, J. and {Bradascio}, F. and {Braun}, J. and {Brenzke}, M. and {Bretz}, H.-P. and {Bron}, S. and {Brostean-Kaiser}, J. and {Burgman}, A. and {Busse}, R.~S. and {Carver}, T. and {Cheung}, E. and {Chirkin}, D. and {Christov}, A. and {Clark}, K. and {Classen}, L. and {Coenders}, S. and {Collin}, G.~H. and {Conrad}, J.~M. and {Coppin}, P. and {Correa}, P. and {Cowen}, D.~F. and {Cross}, R. and {Dave}, P. and {Day}, M. and {de Andr{\'e}}, J.~P.~A.~M. and {De Clercq}, C. and {DeLaunay}, J.~J. and {Dembinski}, H. and {De Ridder}, S. and {Desiati}, P. and {de Vries}, K.~D. and {de Wasseige}, G. and {de With}, M. and {DeYoung}, T. and {D{\'\i}az-V{\'e}lez}, J.~C. and {di Lorenzo}, V. and {Dujmovic}, H. and {Dumm}, J.~P. and {Dunkman}, M. and {Dvorak}, E. and {Eberhardt}, B. and {Ehrhardt}, T. and {Eichmann}, B. and {Eller}, P. and {Evenson}, P.~A. and {Fahey}, S. and {Fazely}, A.~R. and {Felde}, J. and {Filimonov}, K. and {Finley}, C. and {Flis}, S. and {Franckowiak}, A. and {Friedman}, E. and {Fritz}, A. and {Gaisser}, T.~K. and {Gallagher}, J. and {Gerhardt}, L. and {Ghorbani}, K. and {Glauch}, T. and {Gl{\"u}senkamp}, T. and {Goldschmidt}, A. and {Gonzalez}, J.~G. and {Grant}, D. and {Griffith}, Z. and {Haack}, C. and {Hallgren}, A. and {Halzen}, F. and {Hanson}, K. and {Hebecker}, D. and {Heereman}, D. and {Helbing}, K. and {Hellauer}, R. and {Hickford}, S. and {Hignight}, J. and {Hill}, G.~C. and {Hoffman}, K.~D. and {Hoffmann}, R. and {Hoinka}, T. and {Hokanson-Fasig}, B. and {Hoshina}, K. and {Huang}, F. and {Huber}, M. and {Hultqvist}, K. and {H{\"u}nnefeld}, M. and {Hussain}, R. and {In}, S. and {Iovine}, N. and {Ishihara}, A. and {Jacobi}, E. and {Japaridze}, G.~S. and {Jeong}, M. and {Jero}, K. and {Jones}, B.~J.~P. and {Kalaczynski}, P. and {Kang}, W. and {Kappes}, A. and {Kappesser}, D. and {Karg}, T. and {Karle}, A. and {Katz}, U. and {Kauer}, M. and {Keivani}, A. and {Kelley}, J.~L. and {Kheirandish}, A. and {Kim}, J. and {Kim}, M. and {Kintscher}, T. and {Kiryluk}, J. and {Kittler}, T. and {Klein}, S.~R. and {Koirala}, R. and {Kolanoski}, H. and {K{\"o}pke}, L. and {Kopper}, C. and {Kopper}, S. and {Koschinsky}, J.~P. and {Koskinen}, D.~J. and {Kowalski}, M. and {Krings}, K. and {Kroll}, M. and {Kr{\"u}ckl}, G. and {Kunwar}, S. and {Kurahashi}, N. and {Kuwabara}, T. and {Kyriacou}, A. and {Labare}, M. and {Lanfranchi}, J.~L. and {Larson}, M.~J. and {Lauber}, F. and {Leonard}, K. and {Lesiak-Bzdak}, M. and {Leuermann}, M. and {Liu}, Q.~R. and {Lozano Mariscal}, C.~J. and {Lu}, L. and {L{\"u}nemann}, J. and {Luszczak}, W. and {Madsen}, J. and {Maggi}, G. and {Mahn}, K.~B.~M. and {Mancina}, S. and {Maruyama}, R. and {Mase}, K. and {Maunu}, R. and {Meagher}, K. and {Medici}, M. and {Meier}, M. and {Menne}, T. and {Merino}, G. and {Meures}, T. and {Miarecki}, S. and {Micallef}, J. and {Moment{\'e}}, G. and {Montaruli}, T. and {Moore}, R.~W. and {Morse}, R. and {Moulai}, M. and {Nahnhauer}, R. and {Nakarmi}, P. and {Naumann}, U. and {Neer}, G.},
        title = "{Multimessenger observations of a flaring blazar coincident with high-energy neutrino IceCube-170922A}",
      journal = {Science},
         year = 2018,
        month = jul,
       volume = {361},
       number = {6398},
          eid = {eaat1378},
        pages = {eaat1378},
          doi = {10.1126/science.aat1378},
archivePrefix = {arXiv},
       eprint = {1807.08816},
 primaryClass = {astro-ph.HE},
       adsurl = {https://ui.adsabs.harvard.edu/abs/2018Sci...361.1378I}
}

@ARTICLE{Cerruti2021MNRAS502,
       author = {{Cerruti}, M. and {Zech}, A. and {Boisson}, C. and {Emery}, G. and {Inoue}, S. and {Lenain}, J.-P.},
        title = "{Erratum: Lepto-hadronic single-zone models for the electromagnetic and neutrino emission of TXS 0506+056}",
      journal = {\mnras},
         year = 2021,
        month = mar,
       volume = {502},
       number = {1},
        pages = {L21-L22},
          doi = {10.1093/mnrasl/slaa188},
       adsurl = {https://ui.adsabs.harvard.edu/abs/2021MNRAS.502L..21C}
}

@ARTICLE{Maraschi1992ApJ397,
       author = {{Maraschi}, L. and {Ghisellini}, G. and {Celotti}, A.},
        title = "{A Jet Model for the Gamma-Ray--emitting Blazar 3C 279}",
      journal = {\apjl},
         year = 1992,
        month = sep,
       volume = {397},
        pages = {L5},
          doi = {10.1086/186531},
       adsurl = {https://ui.adsabs.harvard.edu/abs/1992ApJ...397L...5M}
}

@ARTICLE{Ghisellini2009MNRAS399,
       author = {{Ghisellini}, G. and {Tavecchio}, F. and {Ghirlanda}, G.},
        title = "{Jet and accretion power in the most powerful Fermi blazars}",
      journal = {\mnras},
         year = 2009,
        month = nov,
       volume = {399},
       number = {4},
        pages = {2041-2054},
          doi = {10.1111/j.1365-2966.2009.15397.x},
archivePrefix = {arXiv},
       eprint = {0906.2195},
 primaryClass = {astro-ph.HE},
       adsurl = {https://ui.adsabs.harvard.edu/abs/2009MNRAS.399.2041G}
}

@ARTICLE{Diwan2023MNRAS524,
       author = {{Diwan}, Rishank and {Prince}, Raj and {Agarwal}, Aditi and {Bose}, Debanjan and {Majumdar}, Pratik and {{\"O}zd{\"o}nmez}, Aykut and {Chandra}, Sunil and {Khatoon}, Rukaiya and {Ege}, Erg{\"u}n},
        title = "{Multiwavelength study of TeV blazar 1ES 1218+304 using gamma-ray, X-ray and optical observations}",
      journal = {\mnras},
         year = 2023,
        month = sep,
       volume = {524},
       number = {3},
        pages = {4333-4345},
          doi = {10.1093/mnras/stad2088},
archivePrefix = {arXiv},
       eprint = {2301.00991},
 primaryClass = {astro-ph.HE},
       adsurl = {https://ui.adsabs.harvard.edu/abs/2023MNRAS.524.4333D}
}

@ARTICLE{Xiao2025ApJ991,
       author = {{Xiao}, Hubing and {Cao}, Haitao and {Xue}, Rui and {Ouyang}, Zhihao and {Zhang}, Shaohua and {Chen}, Junping and {Luo}, Zhijian and {Yang}, Jianghe and {Fan}, Junhui},
        title = "{The Averaged Broadband Spectral Energy Distribution Study of Fermi Bright BL Lac Objects}",
      journal = {\apj},
         year = 2025,
        month = sep,
       volume = {991},
       number = {1},
          eid = {78},
        pages = {78},
          doi = {10.3847/1538-4357/adf7ad},
archivePrefix = {arXiv},
       eprint = {2509.11820},
 primaryClass = {astro-ph.HE},
       adsurl = {https://ui.adsabs.harvard.edu/abs/2025ApJ...991...78X}
}

@ARTICLE{MAGIC2025MNRAS540_OT081,
       author = {{Abe}, H. and {Abe}, S. and {Acciari}, V.~A. and {Agudo}, I. and {Aniello}, T. and {Ansoldi}, S. and {Antonelli}, L.~A. and {Arbet Engels}, A. and {Arcaro}, C. and {Artero}, M. and {Asano}, K. and {Baack}, D. and {Babi{\'c}}, A. and {Baquero}, A. and {Barres de Almeida}, U. and {Batkovi{\'c}}, I. and {Baxter}, J. and {Bernardini}, E. and {Bernardos}, M. and {Bernete}, J. and {Berti}, A. and {Bigongiari}, C. and {Biland}, A. and {Blanch}, O. and {Bonnoli}, G. and {Bo{\v{s}}njak}, {\v{Z}}. and {Burelli}, I. and {Busetto}, G. and {Campoy-Ordaz}, A. and {Carosi}, A. and {Carosi}, R. and {Carretero-Castrillo}, M. and {Castro-Tirado}, A.~J. and {Chai}, Y. and {Cifuentes}, A. and {Cikota}, S. and {Colombo}, E. and {Contreras}, J.~L. and {Cortina}, J. and {Covino}, S. and {D'Amico}, G. and {D'Elia}, V. and {Da Vela}, P. and {Dazzi}, F. and {De Angelis}, A. and {De Lotto}, B. and {Del Popolo}, A. and {Delfino}, M. and {Delgado}, J. and {Delgado Mendez}, C. and {Depaoli}, D. and {Di Pierro}, F. and {Di Venere}, L. and {Dominis Prester}, D. and {Donini}, A. and {Dorner}, D. and {Doro}, M. and {Elsaesser}, D. and {Emery}, G. and {Escudero}, J. and {Fari{\~n}a}, L. and {Fattorini}, A. and {Foffano}, L. and {Font}, L. and {Fukami}, S. and {Fukazawa}, Y. and {Garc{\'\i}a L{\'o}pez}, R.~J. and {Gasparyan}, S. and {Gaug}, M. and {Giesbrecht Paiva}, J.~G. and {Giglietto}, N. and {Giordano}, F. and {Gliwny}, P. and {Grau}, R. and {Green}, J.~G. and {Hadasch}, D. and {Hahn}, A. and {Heckmann}, L. and {Herrera}, J. and {Hrupec}, D. and {H{\"u}tten}, M. and {Imazawa}, R. and {Inada}, T. and {Iotov}, R. and {Ishio}, K. and {Jim{\'e}nez Mart{\'\i}nez}, I. and {Jormanainen}, J. and {Kerszberg}, D. and {Kluge}, G.~W. and {Kobayashi}, Y. and {Kubo}, H. and {Kushida}, J. and {L{\'a}inez Lez{\'a}un}, M. and {Lamastra}, A. and {Leone}, F. and {Lindfors}, E. and {Linhoff}, L. and {Lombardi}, S. and {Longo}, F. and {L{\'o}pez-Moya}, M. and {L{\'o}pez-Oramas}, A. and {Loporchio}, S. and {Lorini}, A. and {Machado de Oliveira Fraga}, B. and {Majumdar}, P. and {Makariev}, M. and {Maneva}, G. and {Mang}, N. and {Manganaro}, M. and {Mangano}, S. and {Mannheim}, K. and {Mariotti}, M. and {Mart{\'\i}nez}, M. and {Mas-Aguilar}, A. and {Mazin}, D. and {Menchiari}, S. and {Mender}, S. and {Mi{\'c}anovi{\'c}}, S. and {Miceli}, D. and {Miranda}, J.~M. and {Mirzoyan}, R. and {Molina}, E. and {Mondal}, H.~A. and {Morcuende}, D. and {Nanci}, C. and {Neustroev}, V. and {Nigro}, C. and {Nishijima}, K. and {Njoh Ekoume}, T. and {Noda}, K. and {Nozaki}, S. and {Ohtani}, Y. and {Otero-Santos}, J. and {Paiano}, S. and {Palatiello}, M. and {Paneque}, D. and {Paoletti}, R. and {Paredes}, J.~M. and {Pavleti{\'c}}, L. and {Persic}, M. and {Pihet}, M. and {Pirola}, G. and {Podobnik}, F. and {Prada Moroni}, P.~G. and {Prandini}, E. and {Principe}, G. and {Priyadarshi}, C. and {Rhode}, W. and {Rib{\'o}}, M. and {Rico}, J. and {Righi}, C. and {Sahakyan}, N. and {Saito}, T. and {Satalecka}, K. and {Saturni}, F.~G. and {Schleicher}, B. and {Schmidt}, K. and {Schmuckermaier}, F. and {Schubert}, J.~L. and {Schweizer}, T. and {Sitarek}, J. and {Spolon}, A. and {Stamerra}, A. and {Stri{\v{s}}kovi{\'c}}, J. and {Strom}, D. and {Suda}, Y. and {Suri{\'c}}, T. and {Suutarinen}, S. and {Tajima}, H. and {Takahashi}, M. and {Takeishi}, R. and {Tavecchio}, F. and {Temnikov}, P. and {Terzi{\'c}}, T. and {Teshima}, M. and {Tosti}, L. and {Truzzi}, S. and {Ubach}, S. and {van Scherpenberg}, J. and {Ventura}, S. and {Verguilov}, V. and {Viale}, I. and {Vigorito}, C.~F. and {Vitale}, V. and {Walter}, R. and {Yamamoto}, T. and {Ait Benkhali}, F. and {Becherini}, Y. and {Bi}, B. and {B{\"o}ttcher}, M. and {Bolmont}, J. and {Brown}, A. and {Bulik}, T. and {Casanova}, S. and {Chand}, T. and {Chandra}, S. and {Chibueze}, J. and {Chibueze}, O. and {Egberts}, K. and {Einecke}, S.},
        title = "{Multiwavelength study of OT 081: broadband modelling of a transitional blazar}",
      journal = {\mnras},
         year = 2025,
        month = jun,
       volume = {540},
       number = {1},
        pages = {364-384},
          doi = {10.1093/mnras/stae2469},
archivePrefix = {arXiv},
       eprint = {2410.22557},
 primaryClass = {astro-ph.HE},
       adsurl = {https://ui.adsabs.harvard.edu/abs/2025MNRAS.540..364A}
}

@ARTICLE{Planck2020A&A641,
       author = {{Planck Collaboration} and {Aghanim}, N. and {Akrami}, Y. and {Ashdown}, M. and {Aumont}, J. and {Baccigalupi}, C. and {Ballardini}, M. and {Banday}, A.~J. and {Barreiro}, R.~B. and {Bartolo}, N. and {Basak}, S. and {Battye}, R. and {Benabed}, K. and {Bernard}, J.-P. and {Bersanelli}, M. and {Bielewicz}, P. and {Bock}, J.~J. and {Bond}, J.~R. and {Borrill}, J. and {Bouchet}, F.~R. and {Boulanger}, F. and {Bucher}, M. and {Burigana}, C. and {Butler}, R.~C. and {Calabrese}, E. and {Cardoso}, J.-F. and {Carron}, J. and {Challinor}, A. and {Chiang}, H.~C. and {Chluba}, J. and {Colombo}, L.~P.~L. and {Combet}, C. and {Contreras}, D. and {Crill}, B.~P. and {Cuttaia}, F. and {de Bernardis}, P. and {de Zotti}, G. and {Delabrouille}, J. and {Delouis}, J.-M. and {Di Valentino}, E. and {Diego}, J.~M. and {Dor{\'e}}, O. and {Douspis}, M. and {Ducout}, A. and {Dupac}, X. and {Dusini}, S. and {Efstathiou}, G. and {Elsner}, F. and {En{\ss}lin}, T.~A. and {Eriksen}, H.~K. and {Fantaye}, Y. and {Farhang}, M. and {Fergusson}, J. and {Fernandez-Cobos}, R. and {Finelli}, F. and {Forastieri}, F. and {Frailis}, M. and {Fraisse}, A.~A. and {Franceschi}, E. and {Frolov}, A. and {Galeotta}, S. and {Galli}, S. and {Ganga}, K. and {G{\'e}nova-Santos}, R.~T. and {Gerbino}, M. and {Ghosh}, T. and {Gonz{\'a}lez-Nuevo}, J. and {G{\'o}rski}, K.~M. and {Gratton}, S. and {Gruppuso}, A. and {Gudmundsson}, J.~E. and {Hamann}, J. and {Handley}, W. and {Hansen}, F.~K. and {Herranz}, D. and {Hildebrandt}, S.~R. and {Hivon}, E. and {Huang}, Z. and {Jaffe}, A.~H. and {Jones}, W.~C. and {Karakci}, A. and {Keih{\"a}nen}, E. and {Keskitalo}, R. and {Kiiveri}, K. and {Kim}, J. and {Kisner}, T.~S. and {Knox}, L. and {Krachmalnicoff}, N. and {Kunz}, M. and {Kurki-Suonio}, H. and {Lagache}, G. and {Lamarre}, J.-M. and {Lasenby}, A. and {Lattanzi}, M. and {Lawrence}, C.~R. and {Le Jeune}, M. and {Lemos}, P. and {Lesgourgues}, J. and {Levrier}, F. and {Lewis}, A. and {Liguori}, M. and {Lilje}, P.~B. and {Lilley}, M. and {Lindholm}, V. and {L{\'o}pez-Caniego}, M. and {Lubin}, P.~M. and {Ma}, Y.-Z. and {Mac{\'\i}as-P{\'e}rez}, J.~F. and {Maggio}, G. and {Maino}, D. and {Mandolesi}, N. and {Mangilli}, A. and {Marcos-Caballero}, A. and {Maris}, M. and {Martin}, P.~G. and {Martinelli}, M. and {Mart{\'\i}nez-Gonz{\'a}lez}, E. and {Matarrese}, S. and {Mauri}, N. and {McEwen}, J.~D. and {Meinhold}, P.~R. and {Melchiorri}, A. and {Mennella}, A. and {Migliaccio}, M. and {Millea}, M. and {Mitra}, S. and {Miville-Desch{\^e}nes}, M.-A. and {Molinari}, D. and {Montier}, L. and {Morgante}, G. and {Moss}, A. and {Natoli}, P. and {N{\o}rgaard-Nielsen}, H.~U. and {Pagano}, L. and {Paoletti}, D. and {Partridge}, B. and {Patanchon}, G. and {Peiris}, H.~V. and {Perrotta}, F. and {Pettorino}, V. and {Piacentini}, F. and {Polastri}, L. and {Polenta}, G. and {Puget}, J.-L. and {Rachen}, J.~P. and {Reinecke}, M. and {Remazeilles}, M. and {Renzi}, A. and {Rocha}, G. and {Rosset}, C. and {Roudier}, G. and {Rubi{\~n}o-Mart{\'\i}n}, J.~A. and {Ruiz-Granados}, B. and {Salvati}, L. and {Sandri}, M. and {Savelainen}, M. and {Scott}, D. and {Shellard}, E.~P.~S. and {Sirignano}, C. and {Sirri}, G. and {Spencer}, L.~D. and {Sunyaev}, R. and {Suur-Uski}, A.-S. and {Tauber}, J.~A. and {Tavagnacco}, D. and {Tenti}, M. and {Toffolatti}, L. and {Tomasi}, M. and {Trombetti}, T. and {Valenziano}, L. and {Valiviita}, J. and {Van Tent}, B. and {Vibert}, L. and {Vielva}, P. and {Villa}, F. and {Vittorio}, N. and {Wandelt}, B.~D. and {Wehus}, I.~K. and {White}, M. and {White}, S.~D.~M. and {Zacchei}, A. and {Zonca}, A.},
        title = "{Planck 2018 results. VI. Cosmological parameters}",
      journal = {\aap},
         year = 2020,
        month = sep,
       volume = {641},
          eid = {A6},
        pages = {A6},
          doi = {10.1051/0004-6361/201833910},
archivePrefix = {arXiv},
       eprint = {1807.06209},
 primaryClass = {astro-ph.CO},
       adsurl = {https://ui.adsabs.harvard.edu/abs/2020A&A...641A...6P}
}

@ARTICLE{Sokolov2004ApJ613,
       author = {{Sokolov}, Andrei and {Marscher}, Alan P. and {McHardy}, Ian M.},
        title = "{Synchrotron Self-Compton Model for Rapid Nonthermal Flares in Blazars with Frequency-dependent Time Lags}",
      journal = {\apj},
         year = 2004,
        month = oct,
       volume = {613},
       number = {2},
        pages = {725-746},
          doi = {10.1086/423165},
archivePrefix = {arXiv},
       eprint = {astro-ph/0406235},
 primaryClass = {astro-ph},
       adsurl = {https://ui.adsabs.harvard.edu/abs/2004ApJ...613..725S}
}

@ARTICLE{Kirk1998A&A333,
       author = {{Kirk}, J.~G. and {Rieger}, F.~M. and {Mastichiadis}, A.},
        title = "{Particle acceleration and synchrotron emission in blazar jets}",
      journal = {\aap},
         year = 1998,
        month = may,
       volume = {333},
        pages = {452-458},
          doi = {10.48550/arXiv.astro-ph/9801265},
archivePrefix = {arXiv},
       eprint = {astro-ph/9801265},
 primaryClass = {astro-ph},
       adsurl = {https://ui.adsabs.harvard.edu/abs/1998A&A...333..452K}
}

@ARTICLE{Kraub2016A&A591,
       author = {{Krau{\ss}}, F. and {Wilms}, J. and {Kadler}, M. and {Ojha}, R. and {Schulz}, R. and {Tr{\"u}stedt}, J. and {Edwards}, P.~G. and {Stevens}, J. and {Ros}, E. and {Baumgartner}, W. and {Beuchert}, T. and {Blanchard}, J. and {Buson}, S. and {Carpenter}, B. and {Dauser}, T. and {Falkner}, S. and {Gehrels}, N. and {Gr{\"a}fe}, C. and {Gulyaev}, S. and {Hase}, H. and {Horiuchi}, S. and {Kreikenbohm}, A. and {Kreykenbohm}, I. and {Langejahn}, M. and {Leiter}, K. and {Lovell}, J.~E.~J. and {M{\"u}ller}, C. and {Natusch}, T. and {Nesci}, R. and {Pursimo}, T. and {Phillips}, C. and {Pl{\"o}tz}, C. and {Quick}, J. and {Tzioumis}, A.~K. and {Weston}, S.},
        title = "{The TANAMI Multiwavelength Program: Dynamic spectral energy distributions of southern blazars}",
      journal = {\aap},
         year = 2016,
        month = jun,
       volume = {591},
          eid = {A130},
        pages = {A130},
          doi = {10.1051/0004-6361/201628595},
archivePrefix = {arXiv},
       eprint = {1605.00841},
 primaryClass = {astro-ph.HE},
       adsurl = {https://ui.adsabs.harvard.edu/abs/2016A&A...591A.130K}
}

@ARTICLE{Middei2022MNRAS514,
       author = {{Middei}, R. and {Giommi}, P. and {Perri}, M. and {Turriziani}, S. and {Sahakyan}, N. and {Chang}, Y.~L. and {Leto}, C. and {Verrecchia}, F.},
        title = "{The first hard X-ray spectral catalogue of Blazars observed by NuSTAR}",
      journal = {\mnras},
         year = 2022,
        month = aug,
       volume = {514},
       number = {3},
        pages = {3179-3190},
          doi = {10.1093/mnras/stac1185},
archivePrefix = {arXiv},
       eprint = {2205.05089},
 primaryClass = {astro-ph.HE},
       adsurl = {https://ui.adsabs.harvard.edu/abs/2022MNRAS.514.3179M}
}

@ARTICLE{Taylor2026A&A706,
       author = {{Taylor}, Gabrielle L. and {Wagner}, Stefan J. and {Wierzcholska}, Alicja and {Zacharias}, Michael},
        title = "{Not so Swift: 20 years of multi-wavelength observations of Mrk 421 and Mrk 501}",
      journal = {\aap},
         year = 2026,
        month = feb,
       volume = {706},
          eid = {A366},
        pages = {A366},
          doi = {10.1051/0004-6361/202557635},
archivePrefix = {arXiv},
       eprint = {2601.08344},
 primaryClass = {astro-ph.HE},
       adsurl = {https://ui.adsabs.harvard.edu/abs/2026A&A...706A.366T}
}

@ARTICLE{MAGIC2025A&A697_B2_1811+31,
       author = {{Abe}, K. and {Abe}, S. and {Abhir}, J. and {Abhishek}, A. and {Acciari}, V.~A. and {Aguasca-Cabot}, A. and {Agudo}, I. and {Aniello}, T. and {Ansoldi}, S. and {Antonelli}, L.~A. and {Arbet Engels}, A. and {Arcaro}, C. and {Asano}, K. and {Babi{\'c}}, A. and {Barres de Almeida}, U. and {Barrio}, J.~A. and {Barrios-Jim{\'e}nez}, L. and {Batkovi{\'c}}, I. and {Baxter}, J. and {Becerra Gonz{\'a}lez}, J. and {Bednarek}, W. and {Bernardini}, E. and {Bernete}, J. and {Berti}, A. and {Besenrieder}, J. and {Bigongiari}, C. and {Biland}, A. and {Blanch}, O. and {Bonnoli}, G. and {Bo{\v{s}}njak}, {\v{Z}}. and {Bronzini}, E. and {Burelli}, I. and {Campoy-Ordaz}, A. and {Carosi}, A. and {Carosi}, R. and {Carretero-Castrillo}, M. and {Castro-Tirado}, A.~J. and {Cerasole}, D. and {Ceribella}, G. and {Chilingarian}, A. and {Cifuentes}, A. and {Colombo}, E. and {Contreras}, J.~L. and {Cortina}, J. and {Covino}, S. and {D'Ammando}, F. and {D'Amico}, G. and {Da Vela}, P. and {Dazzi}, F. and {De Angelis}, A. and {De Lotto}, B. and {de Menezes}, R. and {Delfino}, M. and {Delgado}, J. and {Delgado Mendez}, C. and {Di Pierro}, F. and {Di Tria}, R. and {Di Venere}, L. and {Dinesh}, A. and {Dominis Prester}, D. and {Donini}, A. and {Dorner}, D. and {Doro}, M. and {Eisenberger}, L. and {Elsaesser}, D. and {Escudero}, J. and {Fari{\~n}a}, L. and {Foffano}, L. and {Font}, L. and {Fr{\"o}se}, S. and {Fukazawa}, Y. and {Garc{\'\i}a L{\'o}pez}, R.~J. and {Garczarczyk}, M. and {Gasparyan}, S. and {Gaug}, M. and {Giesbrecht Paiva}, J.~G. and {Giglietto}, N. and {Giordano}, F. and {Gliwny}, P. and {Gradetzke}, T. and {Grau}, R. and {Green}, D. and {Green}, J.~G. and {G{\"u}nther}, P. and {Hadasch}, D. and {Hahn}, A. and {Hassan}, T. and {Heckmann}, L. and {Herrera Llorente}, J. and {Hrupec}, D. and {Imazawa}, R. and {Israyelyan}, D. and {Itokawa}, T. and {Jim{\'e}nez Mart{\'\i}nez}, I. and {Jim{\'e}nez Quiles}, J. and {Jormanainen}, J. and {Kankkunen}, S. and {Kayanoki}, T. and {Kerszberg}, D. and {Khachatryan}, M. and {Kluge}, G.~W. and {Kobayashi}, Y. and {Konrad}, J. and {Kouch}, P.~M. and {Kubo}, H. and {Kushida}, J. and {L{\'a}inez}, M. and {Lamastra}, A. and {Lindfors}, E. and {Lombardi}, S. and {Longo}, F. and {L{\'o}pez-Coto}, R. and {L{\'o}pez-Moya}, M. and {L{\'o}pez-Oramas}, A. and {Loporchio}, S. and {Lorini}, A. and {Luli{\'c}}, L. and {Lyard}, E. and {Majumdar}, P. and {Makariev}, M. and {Maneva}, G. and {Manganaro}, M. and {Mangano}, S. and {Mannheim}, K. and {Mariotti}, M. and {Mart{\'\i}nez}, M. and {Maru{\v{s}}evec}, P. and {Mas-Aguilar}, A. and {Mazin}, D. and {Menchiari}, S. and {Mender}, S. and {Miceli}, D. and {Miranda}, J.~M. and {Mirzoyan}, R. and {Molero Gonz{\'a}lez}, M. and {Molina}, E. and {Mondal}, H.~A. and {Moralejo}, A. and {Nakamori}, T. and {Nanci}, C. and {Neustroev}, V. and {Nickel}, L. and {Nievas Rosillo}, M. and {Nigro}, C. and {Nikoli{\'c}}, L. and {Nilsson}, K. and {Nishijima}, K. and {Njoh Ekoume}, T. and {Noda}, K. and {Nozaki}, S. and {Okumura}, A. and {Paiano}, S. and {Paneque}, D. and {Paoletti}, R. and {Paredes}, J.~M. and {Pavleti{\'c}}, L. and {Peresano}, M. and {Persic}, M. and {Pihet}, M. and {Pirola}, G. and {Podobnik}, F. and {Prada Moroni}, P.~G. and {Prandini}, E. and {Principe}, G. and {Rhode}, W. and {Rib{\'o}}, M. and {Rico}, J. and {Righi}, C. and {Sahakyan}, N. and {Saito}, T. and {Saturni}, F.~G. and {Schmitz}, K. and {Schmuckermaier}, F. and {Schubert}, J.~L. and {Sciaccaluga}, A. and {Silvestri}, G. and {Sitarek}, J. and {Sliusar}, V. and {Sobczynska}, D. and {Stamerra}, A. and {Stri{\v{s}}kovi{\'c}}, J. and {Strom}, D. and {Strzys}, M. and {Suda}, Y. and {Tajima}, H. and {Takahashi}, M. and {Takeishi}, R. and {Temnikov}, P. and {Terauchi}, K. and {Terzi{\'c}}, T. and {Teshima}, M. and {Tutone}, A. and {Ubach}, S. and {van Scherpenberg}, J. and {Vazquez Acosta}, M. and {Ventura}, S. and {Verna}, G. and {Viale}, I. and {Vigliano}, A. and {Vigorito}, C.~F.},
        title = "{Very high-energy gamma-ray detection and long-term multiwavelength view of the flaring blazar B2 1811+31}",
      journal = {\aap},
         year = 2025,
        month = may,
       volume = {697},
          eid = {A172},
        pages = {A172},
          doi = {10.1051/0004-6361/202452942},
archivePrefix = {arXiv},
       eprint = {2503.19054},
 primaryClass = {astro-ph.HE},
       adsurl = {https://ui.adsabs.harvard.edu/abs/2025A&A...697A.172A}
}

@ARTICLE{KrishnaMohana2024MNRAS527,
       author = {{Krishna Mohana}, A. and {Gupta}, Alok C. and {Marscher}, Alan P. and {Sotnikova}, Yulia V. and {Jorstad}, S.~G. and {Wiita}, Paul J. and {Cui}, Lang and {Aller}, Margo F. and {Aller}, Hugh D. and {Kovalev}, Yu A. and {Kovalev}, Y.~Y. and {Liu}, Xiang and {Mufakharov}, T.~V. and {Popkov}, A.~V. and {Mingaliev}, M.~G. and {Erkenov}, A.~K. and {Nizhelsky}, N.~A. and {Tsybulev}, P.~G. and {Zhao}, Wei and {Weaver}, Z.~R. and {Morozova}, D.~A.},
        title = "{Multiband cross-correlated radio variability of the blazar 3C 279}",
      journal = {\mnras},
         year = 2024,
        month = jan,
       volume = {527},
       number = {3},
        pages = {6970-6980},
          doi = {10.1093/mnras/stad3583},
archivePrefix = {arXiv},
       eprint = {2311.02395},
 primaryClass = {astro-ph.HE},
       adsurl = {https://ui.adsabs.harvard.edu/abs/2024MNRAS.527.6970K}
}

@ARTICLE{MAGIC2023A&A670_1ES_0647+250,
       author = {{MAGIC Collaboration} and {Acciari}, V.~A. and {Aniello}, T. and {Ansoldi}, S. and {Antonelli}, L.~A. and {Arbet Engels}, A. and {Arcaro}, C. and {Artero}, M. and {Asano}, K. and {Baack}, D. and {Babi{\'c}}, A. and {Baquero}, A. and {Barres de Almeida}, U. and {Barrio}, J.~A. and {Batkovi{\'c}}, I. and {Becerra Gonz{\'a}lez}, J. and {Bednarek}, W. and {Bernardini}, E. and {Bernardos}, M. and {Berti}, A. and {Besenrieder}, J. and {Bhattacharyya}, W. and {Bigongiari}, C. and {Biland}, A. and {Blanch}, O. and {B{\"o}kenkamp}, H. and {Bonnoli}, G. and {Bo{\v{s}}njak}, {\v{Z}}. and {Burelli}, I. and {Busetto}, G. and {Carosi}, R. and {Carretero-Castrillo}, M. and {Ceribella}, G. and {Chai}, Y. and {Chilingarian}, A. and {Cikota}, S. and {Colombo}, E. and {Contreras}, J.~L. and {Cortina}, J. and {Covino}, S. and {D'Amico}, G. and {D'Elia}, V. and {da Vela}, P. and {Dazzi}, F. and {de Angelis}, A. and {de Lotto}, B. and {Del Popolo}, A. and {Delfino}, M. and {Delgado}, J. and {Delgado Mendez}, C. and {Depaoli}, D. and {di Pierro}, F. and {di Venere}, L. and {Do Souto Espi{\~n}eira}, E. and {Dominis Prester}, D. and {Donini}, A. and {Dorner}, D. and {Doro}, M. and {Elsaesser}, D. and {Emery}, G. and {Fallah Ramazani}, V. and {Fari{\~n}a}, L. and {Fattorini}, A. and {Font}, L. and {Fruck}, C. and {Fukami}, S. and {Fukazawa}, Y. and {Garc{\'\i}a L{\'o}pez}, R.~J. and {Garczarczyk}, M. and {Gasparyan}, S. and {Gaug}, M. and {Giesbrecht Paiva}, J.~G. and {Giglietto}, N. and {Giordano}, F. and {Gliwny}, P. and {Godinovi{\'c}}, N. and {Green}, J.~G. and {Green}, D. and {Hadasch}, D. and {Hahn}, A. and {Hassan}, T. and {Heckmann}, L. and {Herrera}, J. and {Hrupec}, D. and {H{\"u}tten}, M. and {Inada}, T. and {Iotov}, R. and {Ishio}, K. and {Iwamura}, Y. and {Jim{\'e}nez Mart{\'\i}nez}, I. and {Jormanainen}, J. and {Kerszberg}, D. and {Kobayashi}, Y. and {Kubo}, H. and {Kushida}, J. and {Lamastra}, A. and {Lelas}, D. and {Leone}, F. and {Lindfors}, E. and {Linhoff}, L. and {Lombardi}, S. and {Longo}, F. and {L{\'o}pez-Coto}, R. and {L{\'o}pez-Moya}, M. and {L{\'o}pez-Oramas}, A. and {Loporchio}, S. and {Lorini}, A. and {Lyard}, E. and {Machado de Oliveira Fraga}, B. and {Majumdar}, P. and {Makariev}, M. and {Maneva}, G. and {Manganaro}, M. and {Mangano}, S. and {Mannheim}, K. and {Mariotti}, M. and {Mart{\'\i}nez}, M. and {Mas Aguilar}, A. and {Mazin}, D. and {Menchiari}, S. and {Mender}, S. and {Mi{\'c}anovi{\'c}}, S. and {Miceli}, D. and {Miener}, T. and {Miranda}, J.~M. and {Mirzoyan}, R. and {Molina}, E. and {Mondal}, H.~A. and {Moralejo}, A. and {Morcuende}, D. and {Moreno}, V. and {Nakamori}, T. and {Nanci}, C. and {Nava}, L. and {Neustroev}, V. and {Nievas Rosillo}, M. and {Nigro}, C. and {Nilsson}, K. and {Nishijima}, K. and {Njoh Ekoume}, T. and {Noda}, K. and {Nozaki}, S. and {Ohtani}, Y. and {Oka}, T. and {Otero-Santos}, J. and {Paiano}, S. and {Palatiello}, M. and {Paneque}, D. and {Paoletti}, R. and {Paredes}, J.~M. and {Pavleti{\'c}}, L. and {Persic}, M. and {Pihet}, M. and {Podobnik}, F. and {Prada Moroni}, P.~G. and {Prandini}, E. and {Principe}, G. and {Priyadarshi}, C. and {Puljak}, I. and {Rhode}, W. and {Rib{\'o}}, M. and {Rico}, J. and {Righi}, C. and {Rugliancich}, A. and {Sahakyan}, N. and {Saito}, T. and {Sakurai}, S. and {Satalecka}, K. and {Saturni}, F.~G. and {Schleicher}, B. and {Schmidt}, K. and {Schmuckermaier}, F. and {Schubert}, J.~L. and {Schweizer}, T. and {Sitarek}, J. and {Sliusar}, V. and {Sobczynska}, D. and {Spolon}, A. and {Stamerra}, A. and {Stri{\v{s}}kovi{\'c}}, J. and {Strom}, D. and {Strzys}, M. and {Suda}, Y. and {Suri{\'c}}, T. and {Takahashi}, M. and {Takeishi}, R. and {Tavecchio}, F. and {Temnikov}, P. and {Terzi{\'c}}, T. and {Teshima}, M. and {Tosti}, L. and {Truzzi}, S. and {Tutone}, A. and {Ubach}, S. and {van Scherpenberg}, J. and {Vanzo}, G. and {Vazquez Acosta}, M. and {Ventura}, S. and {Verguilov}, V. and {Viale}, I.},
        title = "{Long-term multi-wavelength study of 1ES 0647+250}",
      journal = {\aap},
         year = 2023,
        month = feb,
       volume = {670},
          eid = {A49},
        pages = {A49},
          doi = {10.1051/0004-6361/202244477},
archivePrefix = {arXiv},
       eprint = {2211.13268},
 primaryClass = {astro-ph.HE},
       adsurl = {https://ui.adsabs.harvard.edu/abs/2023A&A...670A..49M}
}

@ARTICLE{Morozova2014AJ148,
       author = {{Morozova}, D.~A. and {Larionov}, V.~M. and {Troitsky}, I.~S. and {Jorstad}, S.~G. and {Marscher}, A.~P. and {G{\'o}mez}, J.~L. and {Blinov}, D.~A. and {Efimova}, N.~V. and {Hagen-Thorn}, V.~A. and {Hagen-Thorn}, E.~I. and {Joshi}, M. and {Konstantinova}, T.~S. and {Kopatskaya}, E.~N. and {Larionova}, L.~V. and {Larionova}, E.~G. and {L{\"a}hteenm{\"a}ki}, A. and {Tammi}, J. and {Rastorgueva-Foi}, E. and {McHardy}, I. and {Tornikoski}, M. and {Agudo}, I. and {Casadio}, C. and {Molina}, S.~N. and {Volvach}, A.~E. and {Volvach}, L.~N.},
        title = "{The Outburst of the Blazar S4 0954+658 in 2011 March-April}",
      journal = {\aj},
         year = 2014,
        month = sep,
       volume = {148},
       number = {3},
          eid = {42},
        pages = {42},
          doi = {10.1088/0004-6256/148/3/42},
archivePrefix = {arXiv},
       eprint = {1406.2286},
 primaryClass = {astro-ph.HE},
       adsurl = {https://ui.adsabs.harvard.edu/abs/2014AJ....148...42M}
}

@INPROCEEDINGS{Alexander1997ASSL218,
       author = {{Alexander}, Tal},
        title = "{Is AGN Variability Correlated with Other AGN Properties? ZDCF Analysis of Small Samples of Sparse Light Curves}",
    booktitle = {Astronomical Time Series},
         year = 1997,
       editor = {{Maoz}, D. and {Sternberg}, A. and {Leibowitz}, E.~M.},
       series = {Astrophysics and Space Science Library},
       volume = {218},
        month = jan,
        pages = {163},
          doi = {10.1007/978-94-015-8941-3_14},
       adsurl = {https://ui.adsabs.harvard.edu/abs/1997ASSL..218..163A}
}

@ARTICLE{Alexander2013arXiv1302_1508,
       author = {{Alexander}, Tal},
        title = "{Improved AGN light curve analysis with the z-transformed discrete correlation function}",
      journal = {arXiv e-prints},
         year = 2013,
        month = feb,
          eid = {arXiv:1302.1508},
        pages = {arXiv:1302.1508},
          doi = {10.48550/arXiv.1302.1508},
archivePrefix = {arXiv},
       eprint = {1302.1508},
 primaryClass = {astro-ph.IM},
       adsurl = {https://ui.adsabs.harvard.edu/abs/2013arXiv1302.1508A}
}

@ARTICLE{Mukherjee1995ApJ445,
       author = {{Mukherjee}, R. and {Aller}, H.~D. and {Aller}, M.~F. and {Bertsch}, D.~L. and {Collmar}, W. and {Digel}, S.~W. and {Dingus}, B.~L. and {Esposito}, J.~A. and {Fichtel}, C.~E. and {Hartman}, R.~C. and {Hunter}, S.~D. and {Johnson}, W.~N. and {Kanbach}, G. and {Kniffen}, D.~A. and {Lin}, Y.~C. and {Mattox}, J.~R. and {Mayer-Hasselwander}, H.~A. and {Michelson}, P.~F. and {von Montigny}, C. and {Nolan}, P.~L. and {Ramanamurthy}, P.~V. and {Schneid}, E. and {Sreekumar}, P. and {Teraesranta}, H. and {Thompson}, D.~J. and {Wagner}, S.~J.},
        title = "{EGRET Gamma-Ray Sources: GRO J0744+54 and GRO J0957+65 (= BL Lacertae Object 0954+658)}",
      journal = {\apj},
         year = 1995,
        month = may,
       volume = {445},
        pages = {189},
          doi = {10.1086/175685},
       adsurl = {https://ui.adsabs.harvard.edu/abs/1995ApJ...445..189M}
}

@ARTICLE{Cohen1977MmRAS84,
       author = {{Cohen}, A.~M. and {Porcas}, R.~W. and {Browne}, I.~W.~A. and {Daintree}, E.~J. and {Walsh}, D.},
        title = "{Accurate position measurements and optical identification for radio sources selected at 966 MHz}",
      journal = {\memras},
         year = 1977,
        month = jan,
       volume = {84},
        pages = {1},
       adsurl = {https://ui.adsabs.harvard.edu/abs/1977MmRAS..84....1C}
}

@ARTICLE{Walsh1984MNRAS211,
       author = {{Walsh}, D. and {Beckers}, J.~M. and {Carswell}, R.~F. and {Weymann}, R.~J.},
        title = "{Spectroscopy of 26 QSO candidates form the Jodrell Bank 966-MHz survey.}",
      journal = {\mnras},
         year = 1984,
        month = nov,
       volume = {211},
        pages = {105-109},
          doi = {10.1093/mnras/211.1.105},
       adsurl = {https://ui.adsabs.harvard.edu/abs/1984MNRAS.211..105W}
}

@ARTICLE{Rani2022ATel15375,
       author = {{Rani}, B. and {Valverde}, J. and {La Mura}, G.},
        title = "{Fermi-LAT detection of renewed gamma-ray activity from the BL Lac S4 0954+65}",
      journal = {The Astronomer's Telegram},
         year = 2022,
        month = may,
       volume = {15375},
        pages = {1},
       adsurl = {https://ui.adsabs.harvard.edu/abs/2022ATel15375....1R}
}

@ARTICLE{Ojha2015ATel7093,
       author = {{Ojha}, Roopesh and {Carpenter}, Bryce and {Tanaka}, Yasuyuki},
        title = "{Fermi LAT Detection of a Hard Spectrum Gamma-ray Flare from the FSRQ S4 0954+65}",
      journal = {The Astronomer's Telegram},
         year = 2015,
        month = feb,
       volume = {7093},
        pages = {1},
       adsurl = {https://ui.adsabs.harvard.edu/abs/2015ATel.7093....1O}
}

@ARTICLE{Ghisellini2011MNRAS414,
       author = {{Ghisellini}, G. and {Tavecchio}, F. and {Foschini}, L. and {Ghirlanda}, G.},
        title = "{The transition between BL Lac objects and flat spectrum radio quasars}",
      journal = {\mnras},
         year = 2011,
        month = jul,
       volume = {414},
       number = {3},
        pages = {2674-2689},
          doi = {10.1111/j.1365-2966.2011.18578.x},
archivePrefix = {arXiv},
       eprint = {1012.0308},
 primaryClass = {astro-ph.CO},
       adsurl = {https://ui.adsabs.harvard.edu/abs/2011MNRAS.414.2674G}
}

@ARTICLE{Xue2025EPJC85,
       author = {{Xue}, Rui and {Wang}, Ze-Rui and {Hu}, Hai-Bin},
        title = "{Time-dependent leptohadronic modeling of Markarian 421}",
      journal = {European Physical Journal C},
         year = 2025,
        month = jul,
       volume = {85},
       number = {7},
          eid = {779},
        pages = {779},
          doi = {10.1140/epjc/s10052-025-14468-4},
archivePrefix = {arXiv},
       eprint = {2507.01265},
 primaryClass = {astro-ph.HE},
       adsurl = {https://ui.adsabs.harvard.edu/abs/2025EPJC...85..779X}
}

@ARTICLE{Zhao-Xie2026ChPhL43,
       author = {{Zhao-Xie}, Hua and {Du}, Lei-Ming and {Xu}, Yun-Bing and {Xue}, Rui and {Wang}, Ze-Rui},
        title = "{On the Hadronuclear Origin of the TeV Emission of Ap Librae}",
      journal = {Chinese Physics Letters},
         year = 2026,
        month = feb,
       volume = {43},
       number = {2},
          eid = {021101},
        pages = {021101},
          doi = {10.1088/0256-307X/43/2/021101},
       adsurl = {https://ui.adsabs.harvard.edu/abs/2026ChPhL..43b1101Z}
}

@ARTICLE{Bernard2025ATel16994,
       author = {{Bernard}, Denis},
        title = "{Fermi-LAT detection of renewed gamma-ray activity from the BL Lac S4 0954+65 and from the FSRQ PKS 0235-618}",
      journal = {The Astronomer's Telegram},
         year = 2025,
        month = jan,
       volume = {16994},
        pages = {1},
       adsurl = {https://ui.adsabs.harvard.edu/abs/2025ATel16994....1B}
}

@ARTICLE{Bartolini2025ATel17042,
       author = {{Bartolini}, Chiara},
        title = "{Fermi-LAT detection of renewed gamma-ray activity from the BL Lac S4 0954+65}",
      journal = {The Astronomer's Telegram},
         year = 2025,
        month = feb,
       volume = {17042},
        pages = {1},
       adsurl = {https://ui.adsabs.harvard.edu/abs/2025ATel17042....1B}
}

@ARTICLE{Weaver2022ApJS260,
       author = {{Weaver}, Zachary R. and {Jorstad}, Svetlana G. and {Marscher}, Alan P. and {Morozova}, Daria A. and {Troitsky}, Ivan S. and {Agudo}, Iv{\'a}n and {G{\'o}mez}, Jos{\'e} L. and {L{\"a}hteenm{\"a}ki}, Anne and {Tammi}, Joni and {Tornikoski}, Merja},
        title = "{Kinematics of Parsec-scale Jets of Gamma-Ray Blazars at 43 GHz during 10 yr of the VLBA-BU-BLAZAR Program}",
      journal = {\apjs},
         year = 2022,
        month = may,
       volume = {260},
       number = {1},
          eid = {12},
        pages = {12},
          doi = {10.3847/1538-4365/ac589c},
archivePrefix = {arXiv},
       eprint = {2202.12290},
 primaryClass = {astro-ph.HE},
       adsurl = {https://ui.adsabs.harvard.edu/abs/2022ApJS..260...12W}
}

@ARTICLE{Jorstad2017ApJ846,
       author = {{Jorstad}, Svetlana G. and {Marscher}, Alan P. and {Morozova}, Daria A. and {Troitsky}, Ivan S. and {Agudo}, Iv{\'a}n and {Casadio}, Carolina and {Foord}, Adi and {G{\'o}mez}, Jos{\'e} L. and {MacDonald}, Nicholas R. and {Molina}, Sol N. and {L{\"a}hteenm{\"a}ki}, Anne and {Tammi}, Joni and {Tornikoski}, Merja},
        title = "{Kinematics of Parsec-scale Jets of Gamma-Ray Blazars at 43 GHz within the VLBA-BU-BLAZAR Program}",
      journal = {\apj},
         year = 2017,
        month = sep,
       volume = {846},
       number = {2},
          eid = {98},
        pages = {98},
          doi = {10.3847/1538-4357/aa8407},
archivePrefix = {arXiv},
       eprint = {1711.03983},
 primaryClass = {astro-ph.GA},
       adsurl = {https://ui.adsabs.harvard.edu/abs/2017ApJ...846...98J}
}

@ARTICLE{Jorstad2005AJ130,
       author = {{Jorstad}, Svetlana G. and {Marscher}, Alan P. and {Lister}, Matthew L. and {Stirling}, Alastair M. and {Cawthorne}, Timothy V. and {Gear}, Walter K. and {G{\'o}mez}, Jos{\'e} L. and {Stevens}, Jason A. and {Smith}, Paul S. and {Forster}, James R. and {Robson}, E. Ian},
        title = "{Polarimetric Observations of 15 Active Galactic Nuclei at High Frequencies: Jet Kinematics from Bimonthly Monitoring with the Very Long Baseline Array}",
      journal = {\aj},
         year = 2005,
        month = oct,
       volume = {130},
       number = {4},
        pages = {1418-1465},
          doi = {10.1086/444593},
archivePrefix = {arXiv},
       eprint = {astro-ph/0502501},
 primaryClass = {astro-ph},
       adsurl = {https://ui.adsabs.harvard.edu/abs/2005AJ....130.1418J}
}

@ARTICLE{Jorstad2016Galax4,
       author = {{Jorstad}, Svetlana and {Marscher}, Alan},
        title = "{The VLBA-BU-BLAZAR Multi-Wavelength Monitoring Program}",
      journal = {Galaxies},
         year = 2016,
        month = oct,
       volume = {4},
       number = {4},
          eid = {47},
        pages = {47},
          doi = {10.3390/galaxies4040047},
       adsurl = {https://ui.adsabs.harvard.edu/abs/2016Galax...4...47J}
}

@inproceedings{Habbema1974stepwise,
  author    = {Habbema, J. D. F. and Hermans, J. and van den Broek, K.},
  title     = {A Stepwise Discriminant Analysis Program Using Density Estimation},
  booktitle = {COMPSTAT 1974: Proceedings in Computational Statistics},
  editor    = {Bruckmann, G.},
  pages     = {101--110},
  year      = {1974},
  publisher = {Physica Verlag},
  address   = {Vienna}
}

@ARTICLE{BOLAT2024110775,
    title = {Stable training of probabilistic models using the leave-one-out maximum log-likelihood objective},
    journal = {Electric Power Systems Research},
    volume = {235},
    pages = {110775},
    year = {2024},
    issn = {0378-7796},
    doi = {https://doi.org/10.1016/j.epsr.2024.110775},
    url = {https://www.sciencedirect.com/science/article/pii/S0378779624006618},
    author = {Kutay Bölat and Simon H. Tindemans and Peter Palensky}
}

@ARTICLE{Wang2024ApJS270,
       author = {{Wang}, Gege and {Xiao}, Hubing and {Fan}, Junhui and {Zhang}, Xin},
        title = "{GeV Variability Properties of TeV Blazars Detected by Fermi-LAT}",
      journal = {\apjs},
         year = 2024,
        month = feb,
       volume = {270},
       number = {2},
          eid = {22},
        pages = {22},
          doi = {10.3847/1538-4365/ad0e08},
archivePrefix = {arXiv},
       eprint = {2312.01122},
 primaryClass = {astro-ph.HE},
       adsurl = {https://ui.adsabs.harvard.edu/abs/2024ApJS..270...22W}
}

@ARTICLE{Duda2021MNRAS508,
       author = {{Duda}, Jaros{\l}aw and {Bhatta}, Gopal},
        title = "{Gamma-ray blazar variability: new statistical methods of time-flux distributions}",
      journal = {\mnras},
         year = 2021,
        month = nov,
       volume = {508},
       number = {1},
        pages = {1446-1458},
          doi = {10.1093/mnras/stab2574},
archivePrefix = {arXiv},
       eprint = {2005.14040},
 primaryClass = {astro-ph.HE},
       adsurl = {https://ui.adsabs.harvard.edu/abs/2021MNRAS.508.1446D}
}

@ARTICLE{Sinha2018MNRAS480,
       author = {{Sinha}, Atreyee and {Khatoon}, Rukaiya and {Misra}, Ranjeev and {Sahayanathan}, Sunder and {Mandal}, Soma and {Gogoi}, Rupjyoti and {Bhatt}, Nilay},
        title = "{The flux distribution of individual blazars as a key to understand the dynamics of particle acceleration}",
      journal = {\mnras},
         year = 2018,
        month = oct,
       volume = {480},
       number = {1},
        pages = {L116-L120},
          doi = {10.1093/mnrasl/sly136},
archivePrefix = {arXiv},
       eprint = {1807.09073},
 primaryClass = {astro-ph.HE},
       adsurl = {https://ui.adsabs.harvard.edu/abs/2018MNRAS.480L.116S}
}

@ARTICLE{Emmanoulopoulos2013MNRAS433,
       author = {{Emmanoulopoulos}, D. and {McHardy}, I.~M. and {Papadakis}, I.~E.},
        title = "{Generating artificial light curves: revisited and updated}",
      journal = {\mnras},
         year = 2013,
        month = aug,
       volume = {433},
       number = {2},
        pages = {907-927},
          doi = {10.1093/mnras/stt764},
archivePrefix = {arXiv},
       eprint = {1305.0304},
 primaryClass = {astro-ph.IM},
       adsurl = {https://ui.adsabs.harvard.edu/abs/2013MNRAS.433..907E}
}

@ARTICLE{MAGIC2021A&A655A,
       author = {{MAGIC Collaboration} and {Acciari}, V.~A. and {Ansoldi}, S. and {Antonelli}, L.~A. and {Arbet Engels}, A. and {Artero}, M. and {Asano}, K. and {Babi{\'c}}, A. and {Baquero}, A. and {Barres de Almeida}, U. and {Barrio}, J.~A. and {Batkovi{\'c}}, I. and {Becerra Gonz{\'a}lez}, J. and {Bednarek}, W. and {Bellizzi}, L. and {Bernardini}, E. and {Bernardos}, M. and {Berti}, A. and {Besenrieder}, J. and {Bhattacharyya}, W. and {Bigongiari}, C. and {Blanch}, O. and {Bo{\v{s}}njak}, {\v{Z}}. and {Busetto}, G. and {Carosi}, R. and {Ceribella}, G. and {Cerruti}, M. and {Chai}, Y. and {Chilingarian}, A. and {Cikota}, S. and {Colak}, S.~M. and {Colombo}, E. and {Contreras}, J.~L. and {Cortina}, J. and {Covino}, S. and {D'Amico}, G. and {D'Elia}, V. and {da Vela}, P. and {Dazzi}, F. and {de Angelis}, A. and {de Lotto}, B. and {Delfino}, M. and {Delgado}, J. and {Delgado Mendez}, C. and {Depaoli}, D. and {di Pierro}, F. and {di Venere}, L. and {Do Souto Espi{\~n}eira}, E. and {Dominis Prester}, D. and {Donini}, A. and {Doro}, M. and {Fallah Ramazani}, V. and {Fattorini}, A. and {Ferrara}, G. and {Fonseca}, M.~V. and {Font}, L. and {Fruck}, C. and {Fukami}, S. and {Garc{\'\i}a L{\'o}pez}, R.~J. and {Garczarczyk}, M. and {Gasparyan}, S. and {Gaug}, M. and {Giglietto}, N. and {Giordano}, F. and {Gliwny}, P. and {Godinovi{\'c}}, N. and {Green}, J.~G. and {Green}, D. and {Hadasch}, D. and {Hahn}, A. and {Heckmann}, L. and {Herrera}, J. and {Hoang}, J. and {Hrupec}, D. and {H{\"u}tten}, M. and {Inada}, T. and {Inoue}, S. and {Ishio}, K. and {Iwamura}, Y. and {Jim{\'e}nez}, I. and {Jormanainen}, J. and {Jouvin}, L. and {Kajiwara}, Y. and {Karjalainen}, M. and {Kerszberg}, D. and {Kobayashi}, Y. and {Kubo}, H. and {Kushida}, J. and {Lamastra}, A. and {Lelas}, D. and {Leone}, F. and {Lindfors}, E. and {Lombardi}, S. and {Longo}, F. and {L{\'o}pez-Coto}, R. and {L{\'o}pez-Moya}, M. and {L{\'o}pez-Oramas}, A. and {Loporchio}, S. and {Machado de Oliveira Fraga}, B. and {Maggio}, C. and {Majumdar}, P. and {Makariev}, M. and {Mallamaci}, M. and {Maneva}, G. and {Manganaro}, M. and {Maraschi}, L. and {Mariotti}, M. and {Mart{\'\i}nez}, M. and {Mazin}, D. and {Menchiari}, S. and {Mender}, S. and {Mi{\'c}anovi{\'c}}, S. and {Miceli}, D. and {Miener}, T. and {Minev}, M. and {Miranda}, J.~M. and {Mirzoyan}, R. and {Molina}, E. and {Moralejo}, A. and {Morcuende}, D. and {Moreno}, V. and {Moretti}, E. and {Neustroev}, V. and {Nigro}, C. and {Nilsson}, K. and {Nishijima}, K. and {Noda}, K. and {Nozaki}, S. and {Ohtani}, Y. and {Oka}, T. and {Otero-Santos}, J. and {Paiano}, S. and {Palatiello}, M. and {Paneque}, D. and {Paoletti}, R. and {Paredes}, J.~M. and {Pavleti{\'c}}, L. and {Pe{\~n}il}, P. and {Perennes}, C. and {Persic}, M. and {Prada Moroni}, P.~G. and {Prandini}, E. and {Priyadarshi}, C. and {Puljak}, I. and {Rib{\'o}}, M. and {Rico}, J. and {Righi}, C. and {Rugliancich}, A. and {Saha}, L. and {Sahakyan}, N. and {Saito}, T. and {Sakurai}, S. and {Satalecka}, K. and {Saturni}, F.~G. and {Schmidt}, K. and {Schweizer}, T. and {Sitarek}, J. and {{\v{S}}nidari{\'c}}, I. and {Sobczynska}, D. and {Spolon}, A. and {Stamerra}, A. and {Strom}, D. and {Strzys}, M. and {Suda}, Y. and {Suri{\'c}}, T. and {Takahashi}, M. and {Tavecchio}, F. and {Temnikov}, P. and {Terzi{\'c}}, T. and {Teshima}, M. and {Tosti}, L. and {Truzzi}, S. and {Tutone}, A. and {Ubach}, S. and {van Scherpenberg}, J. and {Vanzo}, G. and {Vazquez Acosta}, M. and {Ventura}, S. and {Verguilov}, V. and {Vigorito}, C.~F. and {Vitale}, V. and {Vovk}, I. and {Will}, M. and {Wunderlich}, C. and {Zari{\'c}}, D. and {FACT Collaboration} and {Baack}, D. and {Balbo}, M. and {Biederbeck}, N. and {Biland}, A. and {Bretz}, T. and {Buss}, J. and {Dorner}, D. and {Eisenberger}, L. and {Elsaesser}, D. and {Hildebrand}, D. and {Iotov}, R. and {Mannheim}, K. and {Neise}, D. and {Noethe}, M.},
        title = "{Investigation of the correlation patterns and the Compton dominance variability of Mrk 421 in 2017}",
      journal = {\aap},
         year = 2021,
        month = nov,
       volume = {655},
          eid = {A89},
        pages = {A89},
          doi = {10.1051/0004-6361/202141004},
archivePrefix = {arXiv},
       eprint = {2106.05516},
 primaryClass = {astro-ph.HE},
       adsurl = {https://ui.adsabs.harvard.edu/abs/2021A&A...655A..89M}
}

\begin{appendix}





\onecolumn

\section{\emph{Swift} XRT spectral and UVOT photometric results}

Table~\ref{Tab:xrt_spec} summarizes the \emph{Swift}-XRT spectral fitting results, including the best-fit parameters and the 0.5--4.0 keV integrated energy fluxes.
Table~\ref{Tab:uvot_phot} gives the \emph{Swift}-UVOT flux densities in different filters, corrected for Galactic extinction.

\begin{table*}[ht!]
\centering
\renewcommand{\arraystretch}{1.3}
\setlength{\tabcolsep}{6.5pt}
\caption{\emph{Swift}-XRT spectral fitting results.}
\label{Tab:xrt_spec}
\begin{tabular}{cccccccc}
\hline\hline
ObsID & Exposure & Time & Photon Index & Norm & C-stat & d.o.f & Flux \\
      & (s) & (MJD) &  & (keV$^{-1}$ cm$^{-2}$ s$^{-1}$) &  &  & (erg cm$^{-2}$ s$^{-1}$) \\
\hline
00036326019 & 978.94  & 60709.74 & $1.56_{-0.11}^{+0.11}$ & $(1.31_{-0.11}^{+0.12}) \times 10^{-3}$ & 52.21  & 60  & $(5.29_{-0.55}^{+0.57}) \times 10^{-12}$ \\
00036326020 & 933.99  & 60712.74 & $1.46_{-0.09}^{+0.09}$ & $(1.76_{-0.13}^{+0.14}) \times 10^{-3}$ & 79.92  & 75  & $(7.48_{-0.67}^{+0.69}) \times 10^{-12}$ \\
00019580001 & 829.10  & 60739.35 & $1.14_{-0.12}^{+0.12}$ & $(1.75_{-0.19}^{+0.21}) \times 10^{-3}$ & 96.87  & 113 & $(8.94_{-1.21}^{+1.26}) \times 10^{-12}$ \\
00019580002 & 966.45  & 60746.20 & $1.26_{-0.13}^{+0.13}$ & $(8.23_{-0.94}^{+1.01}) \times 10^{-4}$ & 72.47  & 88  & $(3.90_{-0.55}^{+0.57}) \times 10^{-12}$ \\
\hline
\end{tabular}
\tablefoot{
For each observation ID (ObsID), we present the best-fit spectral parameters, including the photon index, the normalization at 1 keV, together with the corresponding Cash statistic and degrees of freedom. 
The integral energy flux in the 0.5--4.0 keV band is calculated accordingly.
}
\end{table*}

\begin{table*}[ht!]
\centering
\renewcommand{\arraystretch}{1.2}
\setlength{\tabcolsep}{8pt}
\caption{\emph{Swift}-UVOT photometric results.}
\label{Tab:uvot_phot}
\begin{tabular}{cccc}
\hline\hline
Time & Band (Filter) & Flux Density & Uncertainty \\
(MJD) &  & (mJy) & (mJy) \\
\hline
    60709.74 & B     & 4.05  & 0.15 \\
    60712.73 & B     & 2.93  & 0.12 \\
    60709.74 & U     & 2.99  & 0.12 \\
    60712.73 & U     & 2.00  & 0.09 \\
    60709.75 & UVM2  & 1.23  & 0.07 \\
    60712.74 & UVM2  & 0.87  & 0.05 \\
    60746.20 & UVM2  & 0.52  & 0.02 \\
    60709.74 & UVW1  & 1.57  & 0.09 \\
    60712.73 & UVW1  & 1.09  & 0.07 \\
    60739.35 & UVW1  & 1.91  & 0.09 \\
    60709.74 & UVW2  & 1.02  & 0.05 \\
    60712.74 & UVW2  & 0.70  & 0.04 \\
    60709.75 & V     & 5.37  & 0.22 \\
    60712.74 & V     & 3.87  & 0.19 \\
\hline
\end{tabular}
\tablefoot{
The table lists the \emph{Swift}-UVOT flux densities and the corresponding uncertainties in different filters. 
All flux densities have been corrected for Galactic extinction.
}
\end{table*}

\section{VLBA jet component parameters}

Table~\ref{Tab:VLBA_knot_params} lists the parameters of the VLBA jet components derived at each observing epoch. 
The table includes the flux density, radial distance, position angle, angular size, and observed brightness temperature, along with their associated uncertainties.

\longtab[1]{
\begin{landscape}
\renewcommand{\arraystretch}{1.3}
\setlength{\tabcolsep}{6pt}
\begin{longtable}{cccccccccccc}
\caption{Time evolution of the VLBA jet component parameters in S4~0954+65.}\\
\label{Tab:VLBA_knot_params} \\
\hline\hline
Epoch & Time & $\chi^2_{\rm red}$ & $S$ & $r$ & $\Theta$ & $a$ & $T_{\rm b, ~ obs}$ & Flag ${T_{\rm b, ~ obs}}$ & $x$ & $y$ & Knot \\
      & (MJD) &                    & (Jy) & (mas) & (deg) & (mas) & (K) &        & (mas) & (mas) & \\
\hline
\endfirsthead
\caption{continued.}\\
\hline\hline
Epoch & Time & $\chi^2_{\rm red}$ & $S$ & $r$ & $\Theta$ & $a$ & $T_{\rm b, ~ obs}$ & Flag ${T_{\rm b, ~ obs}}$ & $x$ & $y$ & Knot \\
      & (MJD) &                    & (Jy) & (mas) & (deg) & (mas) & (K) &        & (mas) & (mas) & \\
\hline
\endhead
\hline
\endfoot
2024.888 & 60635 & 0.76  & $1.918 \pm 0.096$ & 0 & 0 & $0.020 \pm 0.005$ & $3.60\times10^{12}$ & L & $0.000 \pm 0.005$ & $0.000 \pm 0.010$ & A0   \\
         &       &       & $0.671 \pm 0.034$ & $0.044 \pm 0.009$ & $-27.6 \pm 4.3$ & $0.034 \pm 0.008$ & $4.30\times10^{11}$ &    & $-0.020 \pm 0.005$ & $0.039 \pm 0.010$ & K24b \\
         &       &       & $0.064 \pm 0.010$ & $0.200 \pm 0.024$ & $-43.2 \pm 3.3$ & $0.073 \pm 0.021$ & $9.05\times10^{9}$  &    & $-0.137 \pm 0.015$ & $0.146 \pm 0.029$ & A1   \\
         &       &       & $0.306 \pm 0.018$ & $0.363 \pm 0.018$ & $-54.0 \pm 1.7$ & $0.137 \pm 0.020$ & $1.22\times10^{10}$ &    & $-0.293 \pm 0.013$ & $0.213 \pm 0.025$ & K24a \\
         &       &       & $0.041 \pm 0.013$ & $0.586 \pm 0.210$ & $-40.3 \pm 9.3$ & $0.370 \pm 0.053$ & $2.26\times10^{8}$  &    & $-0.379 \pm 0.127$ & $0.447 \pm 0.253$ & Unk  \\
\hline

2024.954 & 60659 & 0.79  & $1.789 \pm 0.090$ & 0 & 0 & $0.020 \pm 0.005$ & $3.35\times10^{12}$ & L & $0.000 \pm 0.005$ & $0.000 \pm 0.010$ & A0   \\
         &       &       & $0.500 \pm 0.026$ & $0.056 \pm 0.010$ & $-22.7 \pm 3.3$ & $0.036 \pm 0.009$ & $2.95\times10^{11}$ &    & $-0.021 \pm 0.005$ & $0.051 \pm 0.011$ & K24b \\
         &       &       & $0.168 \pm 0.012$ & $0.247 \pm 0.017$ & $-43.3 \pm 1.9$ & $0.085 \pm 0.018$ & $1.73\times10^{10}$ &    & $-0.170 \pm 0.011$ & $0.180 \pm 0.021$ & A1   \\
         &       &       & $0.256 \pm 0.016$ & $0.424 \pm 0.019$ & $-54.6 \pm 1.6$ & $0.135 \pm 0.020$ & $1.06\times10^{10}$ &    & $-0.345 \pm 0.014$ & $0.246 \pm 0.027$ & K24a \\
         &       &       & $0.035 \pm 0.014$ & $0.681 \pm 0.306$ & $-34.2 \pm 10.3$ & $0.447 \pm 0.061$ & $1.31\times10^{8}$  &    & $-0.382 \pm 0.175$ & $0.563 \pm 0.350$ & Unk  \\
\hline

2025.126 & 60722 & 0.783 & $1.504 \pm 0.075$ & 0 & 0 & $0.026 \pm 0.006$ & $1.73\times10^{12}$ &    & $0.000 \pm 0.005$ & $0.000 \pm 0.010$ & A0   \\
         &       &       & $0.147 \pm 0.010$ & $0.074 \pm 0.011$ & $-25.2 \pm 2.9$ & $0.034 \pm 0.012$ & $9.52\times10^{10}$ &    & $-0.032 \pm 0.006$ & $0.067 \pm 0.012$ & A2   \\
         &       &       & $0.094 \pm 0.010$ & $0.218 \pm 0.029$ & $-28.8 \pm 2.8$ & $0.097 \pm 0.022$ & $7.55\times10^{9}$  &    & $-0.105 \pm 0.016$ & $0.191 \pm 0.032$ & K24b \\
         &       &       & $0.144 \pm 0.013$ & $0.476 \pm 0.049$ & $-49.9 \pm 3.3$ & $0.222 \pm 0.030$ & $2.18\times10^{9}$  &    & $-0.364 \pm 0.033$ & $0.307 \pm 0.066$ & K24a \\
\hline

2025.184 & 60743 & 0.678 & $2.256 \pm 0.113$ & 0 & 0 & $0.020 \pm 0.005$ & $4.23\times10^{12}$ & L & $0.000 \pm 0.005$ & $0.000 \pm 0.010$ & A0   \\
         &       &       & $0.749 \pm 0.038$ & $0.047 \pm 0.010$ & $-20.8 \pm 3.7$ & $0.035 \pm 0.008$ & $4.47\times10^{11}$ &    & $-0.017 \pm 0.005$ & $0.044 \pm 0.010$ & A2  \\
         &       &       & $0.160 \pm 0.011$ & $0.243 \pm 0.015$ & $-29.1 \pm 1.3$ & $0.062 \pm 0.015$ & $3.13\times10^{10}$ &    & $-0.118 \pm 0.008$ & $0.212 \pm 0.016$ & K24b \\
         &       &       & $0.163 \pm 0.013$ & $0.481 \pm 0.032$ & $-47.5 \pm 2.0$ & $0.159 \pm 0.025$ & $4.85\times10^{9}$  &    & $-0.355 \pm 0.021$ & $0.325 \pm 0.041$ & K24a \\
         &       &       & $0.033 \pm 0.012$ & $0.636 \pm 0.105$ & $-58.1 \pm 6.2$ & $0.218 \pm 0.043$ & $5.13\times10^{8}$  &    & $-0.540 \pm 0.077$ & $0.336 \pm 0.155$ & Unk  \\
\hline

2025.222 & 60757 & 0.718 & $2.055 \pm 0.103$ & 0 & 0 & $0.020 \pm 0.005$ & $3.85\times10^{12}$ & L & $0.000 \pm 0.005$ & $0.000 \pm 0.010$ & A0   \\
         &       &       & $0.747 \pm 0.038$ & $0.062 \pm 0.010$ & $-23.0 \pm 2.9$ & $0.033 \pm 0.008$ & $5.13\times10^{11}$ &    & $-0.024 \pm 0.005$ & $0.057 \pm 0.010$ & A2  \\
         &       &       & $0.148 \pm 0.011$ & $0.262 \pm 0.017$ & $-28.9 \pm 1.3$ & $0.068 \pm 0.017$ & $2.39\times10^{10}$ &    & $-0.127 \pm 0.009$ & $0.229 \pm 0.018$ & K24b \\
         &       &       & $0.177 \pm 0.013$ & $0.491 \pm 0.030$ & $-47.3 \pm 1.8$ & $0.155 \pm 0.024$ & $5.48\times10^{9}$  &    & $-0.361 \pm 0.019$ & $0.333 \pm 0.039$ & K24a \\
         &       &       & $0.030 \pm 0.013$ & $0.681 \pm 0.129$ & $-58.9 \pm 7.2$ & $0.252 \pm 0.047$ & $3.59\times10^{8}$  &    & $-0.583 \pm 0.096$ & $0.352 \pm 0.192$ & Unk  \\
\hline

2025.282 & 60779 & 0.8   & $1.243 \pm 0.062$ & 0 & 0 & $0.020 \pm 0.005$ & $2.33\times10^{12}$ & L & $0.000 \pm 0.005$ & $0.000 \pm 0.010$ & A0   \\
         &       &       & $0.328 \pm 0.018$ & $0.073 \pm 0.011$ & $-25.4 \pm 2.9$ & $0.047 \pm 0.011$ & $1.13\times10^{11}$ &    & $-0.031 \pm 0.006$ & $0.066 \pm 0.012$ & K25  \\
         &       &       & $0.167 \pm 0.012$ & $0.243 \pm 0.015$ & $-30.5 \pm 1.3$ & $0.066 \pm 0.016$ & $2.86\times10^{10}$ &    & $-0.123 \pm 0.009$ & $0.210 \pm 0.017$ & K24b \\
         &       &       & $0.183 \pm 0.014$ & $0.494 \pm 0.044$ & $-44.9 \pm 2.6$ & $0.219 \pm 0.028$ & $2.85\times10^{9}$  &    & $-0.349 \pm 0.028$ & $0.349 \pm 0.056$ & K24a \\
         &       &       & $0.028 \pm 0.014$ & $0.763 \pm 0.238$ & $-55.0 \pm 11.0$ & $0.390 \pm 0.060$ & $1.39\times10^{8}$  &    & $-0.625 \pm 0.169$ & $0.438 \pm 0.338$ & Unk  \\
\hline

2025.356 & 60806 & 0.79  & $0.907 \pm 0.046$ & 0 & 0 & $0.020 \pm 0.006$ & $1.70\times10^{12}$ & L & $0.000 \pm 0.005$ & $0.000 \pm 0.010$ & A0   \\
         &       &       & $0.137 \pm 0.010$ & $0.100 \pm 0.015$ & $-21.6 \pm 2.6$ & $0.053 \pm 0.015$ & $3.71\times10^{10}$ &    & $-0.037 \pm 0.008$ & $0.093 \pm 0.016$ & K25  \\
         &       &       & $0.108 \pm 0.010$ & $0.285 \pm 0.021$ & $-27.1 \pm 1.5$ & $0.075 \pm 0.019$ & $1.44\times10^{10}$ &    & $-0.130 \pm 0.012$ & $0.254 \pm 0.023$ & K24b \\
         &       &       & $0.107 \pm 0.012$ & $0.504 \pm 0.052$ & $-41.8 \pm 2.8$ & $0.187 \pm 0.030$ & $2.29\times10^{9}$  &    & $-0.336 \pm 0.032$ & $0.376 \pm 0.064$ & K24a \\
         &       &       & $0.032 \pm 0.013$ & $0.694 \pm 0.175$ & $-55.2 \pm 9.0$ & $0.322 \pm 0.053$ & $2.31\times10^{8}$  &    & $-0.571 \pm 0.125$ & $0.396 \pm 0.249$ & Unk  \\
\hline

2025.54  & 60873 & 0.58  & $0.887 \pm 0.045$ & 0 & 0 & $0.020 \pm 0.006$ & $1.66\times10^{12}$ & L & $0.000 \pm 0.005$ & $0.000 \pm 0.010$ & A0   \\
         &       &       & $0.605 \pm 0.031$ & $0.053 \pm 0.009$ & $-33.8 \pm 3.8$ & $0.023 \pm 0.007$ & $8.64\times10^{11}$ &    & $-0.029 \pm 0.005$ & $0.044 \pm 0.010$ & A2   \\
         &       &       & $0.036 \pm 0.009$ & $0.198 \pm 0.020$ & $-29.1 \pm 2.1$ & $0.040 \pm 0.018$ & $1.64\times10^{10}$ &    & $-0.096 \pm 0.011$ & $0.173 \pm 0.022$ & K25  \\
         &       &       & $0.016 \pm 0.008$ & $0.344 \pm 0.016$ & $-20.9 \pm 0.8$ & $0.020 \pm 0.016$ & $2.95\times10^{10}$ & L & $-0.122 \pm 0.008$ & $0.321 \pm 0.017$ & K24b \\
         &       &       & $0.049 \pm 0.011$ & $0.521 \pm 0.080$ & $-33.3 \pm 3.4$ & $0.171 \pm 0.034$ & $1.26\times10^{9}$  &    & $-0.287 \pm 0.045$ & $0.435 \pm 0.091$ & K24a \\
         &       &       & $0.043 \pm 0.013$ & $0.641 \pm 0.207$ & $-44.6 \pm 9.2$ & $0.387 \pm 0.054$ & $2.15\times10^{8}$  &    & $-0.450 \pm 0.130$ & $0.456 \pm 0.261$ & Unk  \\
\hline

2025.655 & 60915 & 0.701 & $1.857 \pm 0.093$ & 0 & 0 & $0.020 \pm 0.005$ & $3.48\times10^{12}$ & L & $0.000 \pm 0.005$ & $0.000 \pm 0.010$ & A0   \\
         &       &       & $0.572 \pm 0.029$ & $0.052 \pm 0.010$ & $-24.6 \pm 3.5$ & $0.030 \pm 0.008$ & $4.83\times10^{11}$ &    & $-0.022 \pm 0.005$ & $0.047 \pm 0.010$ & A2   \\
         &       &       & $0.129 \pm 0.010$ & $0.240 \pm 0.012$ & $-34.5 \pm 1.2$ & $0.045 \pm 0.014$ & $4.74\times10^{10}$ &    & $-0.136 \pm 0.007$ & $0.198 \pm 0.014$ & K25  \\
         &       &       & $0.042 \pm 0.008$ & $0.368 \pm 0.015$ & $-28.1 \pm 0.8$ & $0.031 \pm 0.015$ & $3.28\times10^{10}$ &    & $-0.174 \pm 0.008$ & $0.325 \pm 0.016$ & K24b \\
         &       &       & $0.094 \pm 0.013$ & $0.602 \pm 0.112$ & $-33.6 \pm 4.2$ & $0.316 \pm 0.040$ & $7.06\times10^{8}$  &    & $-0.333 \pm 0.064$ & $0.501 \pm 0.128$ & K24a \\
\hline

2025.712 & 60936 & 0.762 & $1.252 \pm 0.063$ & 0 & 0 & $0.020 \pm 0.005$ & $2.35\times10^{12}$ & L & $0.000 \pm 0.005$ & $0.000 \pm 0.010$ & A0   \\
         &       &       & $0.385 \pm 0.020$ & $0.063 \pm 0.010$ & $-26.6 \pm 3.1$ & $0.033 \pm 0.009$ & $2.62\times10^{11}$ &    & $-0.028 \pm 0.005$ & $0.056 \pm 0.011$ & A2   \\
         &       &       & $0.139 \pm 0.011$ & $0.261 \pm 0.015$ & $-34.3 \pm 1.3$ & $0.063 \pm 0.016$ & $2.66\times10^{10}$ &    & $-0.147 \pm 0.009$ & $0.215 \pm 0.018$ & K25  \\
         &       &       & $0.037 \pm 0.010$ & $0.426 \pm 0.031$ & $-29.6 \pm 1.5$ & $0.063 \pm 0.023$ & $6.96\times10^{9}$  &    & $-0.211 \pm 0.017$ & $0.370 \pm 0.034$ & K24b \\
         &       &       & $0.063 \pm 0.013$ & $0.639 \pm 0.162$ & $-34.7 \pm 5.9$ & $0.354 \pm 0.047$ & $3.77\times10^{8}$  &    & $-0.364 \pm 0.093$ & $0.525 \pm 0.186$ & K24a \\

\end{longtable}
\tablefoot{
At each epoch, we report the flux density ($S$), distance from the core ($r$), position angle ($\Theta$), angular size ($a$), observed brightness temperature ($T_{\rm b, ~ obs}$, where ``L" denotes the lower limits), and position relative to the core ($x$ and $y$) for each identified component.
}
\end{landscape}
}

\section{Significance estimation of the correlations}\label{zdcf_significance}

We estimated the significance of the cross-correlation measured with the zDCF using Monte Carlo simulations.
For each observed light curve, $10^4$ artificial light curves were generated with the same probability distribution function (PDF) and power spectral density (PSD) as the real observed data using the algorithm of \citet{Emmanoulopoulos2013MNRAS433}. 
The simulated light curves were then rebinned to match the temporal sampling of the real data. 
For each simulated light-curve pair, the zDCF coefficients were calculated, and the significance at each time lag was determined from the percentile distribution of the simulated zDCF coefficients at that lag. 
In the following, we describe how the PDF and PSD were characterized.

\subsection{Probability distribution function fitting}

To characterize the observed light-curve distribution without imposing a specific parametric form on the flux distribution, we adopt a non-parametric kernel density estimation (KDE) approach. 
This method provides a flexible, data-driven representation of the underlying distribution and is particularly suitable for blazar variability studies, where the flux distribution may exhibit substantial skewness and heavy tails. 
In such cases, simple parametric models such as Gaussian or log-normal distributions may be insufficient to capture the full complexity of the observed variability, which likely reflects the superposition of multiple stochastic processes operating within relativistic jets \citep[e.g.,][]{Sinha2018MNRAS480, Duda2021MNRAS508, Wang2024ApJS270}.

We define the normalized flux as $x = F / \langle F \rangle$, and estimate its probability density function using a Gaussian kernel,
\begin{equation}
\hat{f}(x) = \frac{1}{n h} \sum_{i=1}^{n} K\left( \frac{x - x_i}{h} \right),
\end{equation}
where $K$ is the Gaussian kernel, $h$ is the bandwidth, and $n$ is the number of data points.
The optimal bandwidth is determined by maximizing the leave-one-out cross-validation log-likelihood \citep[e.g.,][]{Habbema1974stepwise, BOLAT2024110775},
\begin{equation}
\mathcal{L}_{\mathrm{LOO}} = \sum_{i=1}^{n} \ln \hat{f}_{-i}(x_i),
\end{equation}
where
\begin{equation}
\hat{f}_{-i}(x_i) = \frac{1}{(n-1)h} \sum_{j \ne i} K\left(\frac{x_i - x_j}{h}\right)
\end{equation}
is the kernel density estimate evaluated at $x_i$ with the $i$-th data point excluded.

The adequacy of the KDE model is assessed by comparing the cumulative distribution implied by the KDE with the empirical cumulative distribution function (ECDF), which is the cumulative distribution constructed directly from the observed data. 
To further evaluate the robustness of the inferred distribution, we perform bootstrap resampling with replacement. For each bootstrap realization, a KDE is constructed using the same bandwidth, and the resulting ensemble of density estimates is used to quantify the variability of the inferred distribution. 
The stability of key features, including the peak structure and tail behavior, is examined across the bootstrap realizations. 
Confidence intervals are then constructed from the empirical percentiles of the bootstrap KDE ensemble.

\begin{figure}
    \centering
    \includegraphics[width=0.8\hsize]{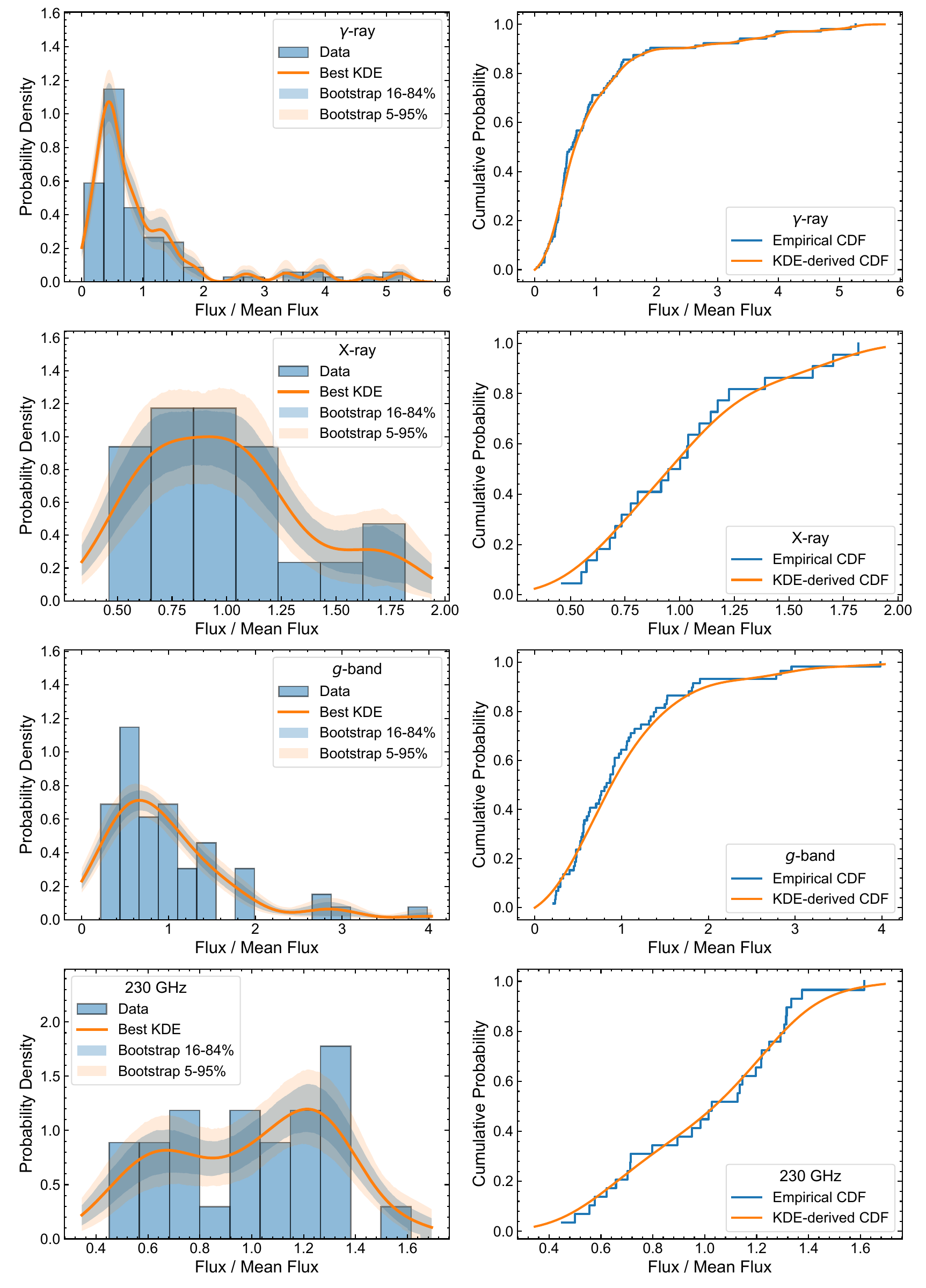}
    \caption{
    Left panel: Probability density distribution for each light curve. The orange curve represents the best-fit KDE, and the two shaded regions correspond to the 16--84\% and 5--95\% bootstrap confidence intervals.
    Right panel: Comparison of the empirical cumulative distribution function (ECDF) with the cumulative distribution function derived from the KDE, illustrating the quality of the fit in cumulative space.
    }
    \label{Fig:PDF_KDE}%
\end{figure}

The results are shown in Figure~\ref{Fig:PDF_KDE} and the resulting best-fit flux distribution is subsequently adopted as the input PDF for the simulations described above.

\subsection{Power spectral density fitting}

The intrinsic PSD cannot be directly inferred from observed light curves, since real data are typically unevenly sampled and affected by substantial gaps. 
In addition, the finite duration of the observations and the applied temporal binning can distort the variability power over frequency. 
These effects must therefore be taken into account when constraining the underlying PSD.

\begin{figure}
    \centering
    \includegraphics[width=0.8\hsize]{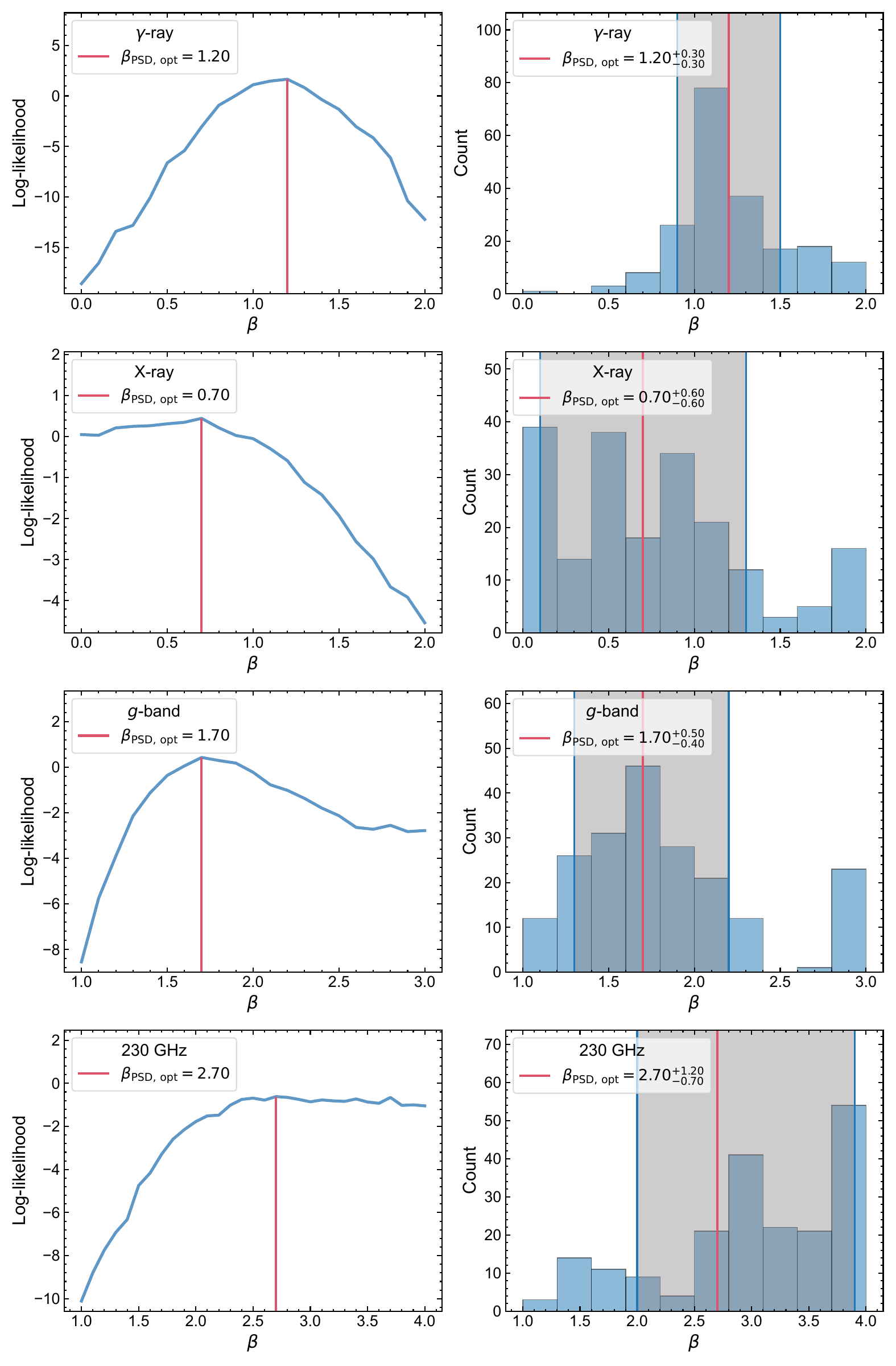}
    \caption{
    Left panel: Log-likelihood values for each light curve calculated by Equation \eqref{PSD_L}, with their corresponding best-fit PSD slope values indicated by vertical rose-red solid lines.
    Right panel: Histogram of the best-fit PSD slope values obtained from simulations for each light curve, where the input simulated light curve has the known PSD slope shown in the left panel.
    }
    \label{Fig:PSD}%
\end{figure}

We estimated the intrinsic PSD slope through a simulation-based forward-folding method, following \citet{MAGIC2021A&A655A} and \citet{MAGIC2025A&A694}. 
The PSD slope was scanned over the range 0.0--2.0 (1.0--3.0 for the optical g-band and 1.0--4.0 for the SMA 230 GHz band) in steps of 0.1. 
For each trial slope, 3000 artificial light curves were generated assuming a power-law PSD. 
The simulations were produced using the algorithm of \citet{Emmanoulopoulos2013MNRAS433}. 
To account for leakage effects, the simulated light curves were generated with lengths 100 times that of the observed ones and then rebinned to the same temporal sampling.
As a proxy for the PSD, we used the multiple fragments variance function \citep[MFVF;][]{Kastendieck2011A&A531}.
This quantity was obtained by recursively dividing the light curve into two equal segments and computing the variance in each segment, down to a minimum timescale of one day, which is a compromise between the limited number of measurements and the smallest timescale resolvable.
The resulting variance-timescale relation provides a robust characterization of the variability properties.
For each simulated light curve, the MFVF was calculated and used to derive, in each of the $N$ timescale bins, the probability density $p(\beta_{\rm PSD},f_i)$, i.e., the probability of obtaining a given MFVF value at frequency $f_i$ for an intrinsic PSD slope $\beta_{\rm PSD}$. 
The best-fit slope was then determined by maximizing the likelihood $\mathcal{L}(\beta_{\rm PSD})$ \citep{Kastendieck2011A&A531}
\begin{equation}\label{PSD_L}
\mathcal{L}(\beta_{\rm PSD})=\sum_{i=0}^{N}\ln p(\beta_{\rm PSD},f_i).
\end{equation}
The resulting likelihood profiles for each light curve are shown in the left panel of Figure~\ref{Fig:PSD}.
The best-fit PSD slope is subsequently adopted in the simulations described above.

Following \citet{MAGIC2021A&A655A} and \citet{MAGIC2025A&A694}, the uncertainty on the PSD slope was evaluated by generating 200 simulated light curves for each source using the corresponding best-fit slope. 
These simulated light curves were then refitted with the same procedure applied to the observed data. 
The resulting distribution of recovered slopes was used to estimate the uncertainty, which was defined as the 68\% confidence interval.
These distributions are shown in the right panel of Figure~\ref{Fig:PSD}.

\end{appendix}

\end{document}